\documentclass[12pt]{iopart}

\newcommand{\arccot}{\textmd{arccot}}

\usepackage{epsfig}\parskip 5pt plus 1pt
\usepackage{bbm}
\usepackage{amssymb}
\usepackage{graphicx}

\eqnobysec
\usepackage{float}
\usepackage{caption}
\usepackage{subfig}
\usepackage{cite}

\providecommand{\keywords}[1]{\textbf{Keywords:} #1}

\begin{document}

\title[Fermionic greybody factors in dilaton black holes]{Fermionic greybody factors in dilaton black holes}

\author{Jahed Abedi$^{1}$ and Hessamaddin Arfaei$^{1, 2}$}

\address{$^1$ Department of Physics, Sharif University of Technology,\\
P.O. Box 11155-9161, Tehran, Iran}
\address{$^2$ School of Particles and Accelerators,\\ Institute for Research in Fundamental Sciences (IPM), \\
 P.O. Box 19395-5531, Tehran, Iran}
\eads{\mailto{jahed$_-$abedi@physics.sharif.ir}, \mailto{arfaei@ipm.ir}}
\begin{abstract}
In this paper the question of emission of fermions in the process of
dilaton black hole evolution and its characters for different
dilaton coupling constants $\alpha$ is studied. The main quantity of
interest, the greybody factors are calculated both numerically and
in analytical approximation. The dependence of rates of evaporation
and behaviour on the dilaton coupling constant is analyzed. Having
calculated the greybody factors we are able to address  the
question of the final fate of the dilaton black hole.  For that we
also need to make  dynamical treatment of the solution by
considering the  backreaction which will show a crucial effect on
the final result. We find a transition line in $(Q/M, \alpha)$ plane
that separates the two regimes for the fate of the black hole, decay
regime and  extremal regime. In the decay regime the black hole
completely evaporates, while in the extremal regime the black hole
approaches  the extremal limit by radiation  and becomes stable.
\end{abstract}

\maketitle

\section{Introduction}
Gravitational systems coupled to Maxwell  and dilaton fields emerge
from several more fundamental theories. In particular the low energy
limit of (super) string theory or Kaluza-Klein compactifications
result in such systems, which  have been studied for long time
\cite{ Gibbons:1987ps, Garfinkle:1990qj, Holzhey:1991bx,
Koga:1995bs}. Corresponding black holes and their evaporation are
also studied previously. The exact black hole solution goes back to
\cite{Hawking:1974sw,Bardeen:1973gs} from 70's. The thermodynamics
of the black hole in this theory shows interesting properties which
depend on the dilaton coupling $\alpha$. For $\alpha<1$ one expects
similar properties as for the standard Reissner-Nordstr\"{o}m black
hole, although we find a range $1/\sqrt{3}\leq\alpha<1$ in which
some properties differ significantly. The behaviour of the theory in
the range of $\alpha>1$ is significantly different and shows
unexpected features some of which are addressed in this article. The
particle emission by dilaton black holes  studied in several
articles falls among them. Holzhey and Wilczek \cite{Holzhey:1991bx}
derived the potential barrier which for $\alpha>1$ strongly impedes
the particle radiation to the extent that may stop it.  In contrast,
Koga and Maeda \cite{Koga:1995bs} showed by numerical computation
that Hawking radiation wins over the barrier and the dilaton black
hole does not stop radiating, despite the fact that  the potential
barrier becomes infinitely high. All this is done for emission of
scalars and in semi-classical approximation. The question is which
of these results stay  valid once we consider the process for
fermions and consider next order correction arising from the
back-reaction.  As expected, fermionic emission show more or less
similar properties qualitatively as scalars, but  considering next
order of dynamical effect as back-reaction changes the scene and
becomes the key factor when the evolution of the black hole moves it
close to the extremal limit.

Shortly after the  discovery of Hawking radiation, it was noticed
that due to large value of  $\frac{e}{m} \simeq 2 \times 10^{21}$,
large black holes are unlikely to hold any charge and rapidly
radiate away their charges \cite{Gibbons:1975} and become neutral.
Hence in nature we must look for neutral black holes rather than
charged ones. However, considering the next order effect of the
dynamics of the dilaton black hole  we find that this need not be
valid for all ranges of parameters. We find a transition line in the
$(Q/M, \alpha)$ plane which designates the border between two
regions; one region specifies the parameters for black holes which
evaporate completely and the other for black holes that end up as
extremal condition. In the latter case the black hole stops
radiating and becomes stable.

The radiation rates of spin 1/2 particles and the fate of different
types of  Einstein Hilbert   black-holes in the semi-classical
approximation
\cite{Page:1977,Cvetic:1997ap,Kanti:2002ge,Creek:2007tw,Das:1999pt,Gubser:1997cm,alBinni:2009cu,Sampaio:2009ra,Casals:2006xp,Sampaio:2009tp}
using greybody factors   have been calculated both numerically and
analytically. Certain results on the scattering parameters of Dirac
field such as quasinormal frequencies or decay rates in the
background of dilaton \cite{Gibbons:2008rs,Nakonieczny:2011bs} and
other types of black holes are also presented
\cite{Chen:2007dja,Cho:2003qe,Cho:2004wj,Cho:2007zi,Chakrabarti:2008xz,Doran:2005vm,Dolan:2006vj,Jin:1998rg,Jing:2003wq,Jing:2004xv,Jing:2004zb,Jung:2004nh,LopezOrtega:2005ep,LopezOrtega:2010uu,LopezOrtega:2011sc,Moderski:2008nq,R:2008ub,Shu:2004fj,Wang:2009hr}.
But there are number of unsettled questions which we will consider
in this article.

For completeness and setting the notation the next section is
devoted to a quick review of the charged dilaton black hole and its
properties, such as general results on decay rates of mass and
charge, thermodynamics, etc.

In section \ref{Charged massive Dirac particle in the background
metric} we set up the Dirac equations for emission of charged
fermions in the background of a charged dilaton black hole and
derive the effective potentials.  We also solve the equations to
find the greybody factors that are the important factors in
calculation of the emission rates.

Our main results which are obtained by numerical computations are
presented in section \ref{Greybody factors}, but to get a better
view we also look at analytical approximation of  the solutions to
the Dirac equations and evaluation of the greybody factors using
Rosen-Morse potential \cite{Eckart:1930zza,Rosen:1932,Boonserm:2011}
and WKB approximation. The evolution of the charged dilaton black
hole and the rates of charge and mass emission are discussed and the
existence of a transition line  is demonstrated. We also compute the
transition line for different ranges of the parameters of the
problem at hand.

Section \ref{Results and discussion} is devoted to the analysis of
the results and their behaviour under change of various parameters
involved.

Finally in section \ref{Conclusion} we end with concluding remarks
and future plans.  In the appendix the details of the computation of
the effective potential for fermions in the background of most
general static black hole is presented.


\section{\label{A short review of dilaton black hole, greybody factors and Hawking radiation}A short review of dilaton black hole, greybody factors and Hawking radiation}
In this section we consider an Einstein-Maxwell gravity coupled to a
dilaton field $\phi$ with the dilaton coupling constant $\alpha$.
The action is
\begin{equation}
S=\int d^{4} x \sqrt{-g}[R-2(\nabla\phi)^{2}+e^{-2\alpha\phi}F^{2}] \label{eq2.1}.
\end{equation}

The signature of the metric is $(+ - - -)$. The parameter $\alpha$
is a dimensionless constant, and $F^{2}=F_{\mu\nu}F^{\mu\nu}$. The
behaviour of the theory shows non-trivial dependence on $\alpha$
that we will see in rest of the article. The equations of motion
are;

\noindent the Maxwell equations
\begin{equation}
\nabla_{\mu}(e^{-2\alpha\phi}F^{\mu\nu})=0,
\end{equation}
\begin{equation}
\partial_{[\rho}F_{\mu\nu]}=0,
\end{equation}
the Einstein equations
\begin{equation}
R_{\mu\nu}=e^{-2\alpha\phi}(-2F_{\mu\rho}F^{\rho}_{\nu}+\frac{1}{2}F^{2}g_{\mu\nu})+2\partial_{\mu}\phi\partial_{\nu}\phi,
\end{equation}
and the dilaton equation
\begin{equation}
g^{\mu\nu}\nabla_{\mu}\nabla_{\nu}\phi=\frac{1}{2}\alpha e^{-2\alpha\phi}F^{2} \label{eq2.5}.
\end{equation}

The spherically symmetric black hole solutions of this action are
well known and found long ago\cite{Gibbons:1987ps,Garfinkle:1990qj};

\begin{equation}
ds^{2}=f(r)^{2}dt^{2}-\frac{dr^{2}}{f(r)^{2}}-R(r)^{2}d\Omega^{2} \label{eq2.6},
\end{equation}
where
\begin{equation}
f(r)^{2}=(1-\frac{r_{+}}{r})(1-\frac{r_{-}}{r})^{\frac{1-\alpha^{2}}{1+\alpha^{2}}},
\end{equation}
and
\begin{equation}
R(r)^{2}=r^{2}(1-\frac{r_{-}}{r})^{\frac{2\alpha^{2}}{1+\alpha^{2}}}\label{eq2.8}.
\end{equation}

The Maxwell and dilaton fields of the solution are, $A_{\mu}=(A_{t},0,0,0)$, $A_{t}=-\frac{Q}{r}$,

\noindent with
\begin{equation}
F_{tr}=\frac{e^{2\alpha\phi}Q}{R(r)^{2}},
\end{equation}
and
\begin{equation}
e^{2\alpha\phi}=(1-\frac{r_{-}}{r})^{\frac{2\alpha^{2}}{1+\alpha^{2}}} \label{eq2.14}.
\end{equation}

The  two (inner and outer) horizons are located at
\begin{equation}
r_{+}=M+\sqrt{M^{2}-(1-\alpha^{2})Q^{2}},
\end{equation}
and
\begin{equation}
r_{-}=\frac{1+\alpha^{2}}{1-\alpha^{2}}(M-\sqrt{M^{2}-(1-\alpha^{2})Q^{2}}),
\end{equation}
where M and Q are ADM mass and charge of this black hole
respectively. Note that for $\alpha <1 $ in order to preserve
reality of the horizons one must have
$|Q/M|\leq\frac{1}{\sqrt{1-\alpha^2}}$, but for $\alpha> 1$ we do
not have such restriction. We shall see that the different behaviour
of the black hole for these two ranges of $\alpha $ occurs also in
several  places. To have $r_{+}>r_{-}$ we must also have
$\frac{Q^{2}}{M^{2}}<1+\alpha^{2} $ and in the extremal limit where
the two horizons coincide;
\begin{equation}
\frac{Q^{2}}{M^{2}}=1+\alpha^{2}.\label{eq13}
\end{equation}
The case of $r_{-}>r_{+}$ or equivalently
$\frac{Q^{2}}{M^{2}}>1+\alpha^{2}$ is not considered in detail in
this article. It requires particular attention since it behaves very
differently and is under study by the authors.

The Hawking temperature of this dilaton black hole is
\begin{equation}
T_{H}=\frac{1}{4\pi r_{+}}(1-\frac{r_{-}}{r_{+}})^{\frac{1-\alpha^{2}}{1+\alpha^{2}}}.
\end{equation}

This dilaton black hole demonstrates interesting thermodynamical
properties not present in non dilatonic
ones\cite{Holzhey:1991bx,Koga:1995bs,Preskill:1991tb,Koga:1994np}.
Obviously, the behaviour of the temperature is drastically different
from the normal Reissner-Nordstr\"{o}m black hole. For $\alpha<1$,
it  is much like that of RN black hole and approaches zero when the
black hole becomes extremal. The drastic difference occurs for
$\alpha> 1$ and $\alpha=1$. When $\alpha>1$, at the extremal limit
the temperature diverges, while for $\alpha=1$ it has a finite value
$T_{H}=1/4\pi r_{+}$. Such behaviour implies that the Hawking
radiation might be quite different with strong dependence  on  the
coupling constant $\alpha$.

The condition (\ref{eq13}) arise if $r_{+}$ is truly the outer
horizon, $r_{+}>r_{-}$. In this case the inner horizon $r_{-}$ of
this black hole has other interesting characteristics which is
unique among  black holes \cite{Holzhey:1991bx}. For non-zero
$\alpha$ and for extremal black holes the angular factor $R$ in the
metric (\ref{eq2.6}) vanishes at the event horizon and the geometry
becomes singular which must be resolved. However, there is no such a
singularity for Reissner-Nordstr\"{o}m black hole ($\alpha=0$). In
the string frame  ($\alpha=1$) this singularity completely
disappears by rescaling the metric with the  conformal factor. In
this frame which is obtained by removing the singular scale factor
$\left(1-\frac{r_{-}}{r}\right)$ from Einstein frame metric
(\ref{eq2.6}), and in the  extremal limit, and imposing the   the
asymptotic constant value of the dilaton $\phi_{0}=0$,  we have
\cite{Garfinkle:1990qj},
\begin{equation}
ds_{string}^{2}=dt^{2}-\left(1-\frac{r_{+}}{r}\right)^{-2}dr^{2}-r^{2}d\Omega^{2}.
\end{equation}

In the String frame the geometry of $t=cte.$ surfaces for this
metric is similar to that of Reissner-Nordstr\"{o}m for $t=cte.$
surfaces.

Contrary to Reissner-Nordstr\"{o}m ($\alpha=0$) where for $Q>M$ the
geometry becomes complex and exposes the naked singularity at $r=0$,
for dilaton black hole ($\alpha>0$) inner horizon can pass the outer
horizon $r_{-}>r_{+}$ or we can have
$1<\frac{Q}{M\sqrt{1+\alpha^{2}}}\leq \frac{1}{\sqrt{1-\alpha^{4}}}$
for $0<\alpha<1$ or $Q>M\sqrt{1+\alpha^{2}}$ for $\alpha \geq 1$ and
the geometry remains real \cite{Holzhey:1991bx}. This range of
parameters, as stated above requires its own analysis which is under
study and shall be reported separately.

Hawking radiation at the event horizon is exactly the black-body
radiation\cite{Hawking:1974sw}. However, before this radiation
reaches  the distant  observer, it  must pass the curved space-time
around the black hole horizon\cite{Maldacena:1996ix,Harmark:2007jy}
which modifies it to a large extent. Therefore an observer located
at far distance from the black hole observes a different spectrum
than pure black body radiation. The geometry outside the event
horizon apart from red-shifting the radiation also plays the role of
a potential barrier, thus filters the Hawking radiation. The portion
of the Hawking radiation passing the barrier, just goes under the
red shift to infinity whereas the remainder is reflected back into
the black hole. Hence, from viewpoint of infinite observer the
space-time around the black hole, acts like a potential barrier and
forces a deviation on blackbody spectrum. This deviation can be
calculated by obtaining greybody factors from the scattering
coefficients of the black hole.

Holzhey and Wilczek have obtained the potential for scalars,
$V_{\eta}$ that at the extremal limit is proportional to
$(1-r_{+}/r)^{2(1-\alpha^{2})/(1+\alpha^{2})}$
\cite{Holzhey:1991bx}. It is illuminating to write the potential as
a product of two factors as in the following,
\begin{equation}
V_{\eta}=V_{\eta 1}V_{\eta 2},\label{eq2.15}
\end{equation}
\begin{equation}
V_{\eta 1}=\left(1-\frac{r_{+}}{r}\right)\left(1-\frac{r_{-}}{r}\right)^{\frac{1-3 \alpha^{2}}{1+\alpha^{2}}},
\end{equation}
\begin{eqnarray}
\fl V_{\eta 2}=\frac{1}{r^{2}} \left( l(l+1)+\frac{r_{-} r + r_{+} r (1+\alpha^{2})^{2}-(2+\alpha^{2})r_{-}r_{+}}{(1+\alpha^{2})^{2} r^{2}}-\frac{\alpha^{4} r_{-} (1-\frac{r_{+}}{r})}{(1+\alpha^{2})^{2} r(1-\frac{r_{-}}{r})} \right).
\end{eqnarray}
Again the strong dependence on $\alpha$ with three distinct
behaviour for $\alpha<1$, $\alpha=1$ and $\alpha>1$ is visible. For
$\alpha<1$ it is qualitatively like RN black hole, i.e. $\alpha=0$:
it tends to zero at the event horizon,  increases to a maximum and
as $r$ becomes large tends to zero again. For the case $\alpha=1$,
the height of the potential barrier  near extremal limit remains
finite. For the class of black holes with $\alpha>1$ in the extremal
limit the height of the potential barrier diverges on the event
horizon. For non-extremal cases the height of potential barrier is
finite, but its peak grows as
$(r_{+}-r_{-})^{-2(\alpha^{2}-1)/(\alpha^{2}+1)}$ as one approaches
the extremal limit. The behavior of the potential can be better
understood in the tortoise coordinates. In this case and at the
extremal limit the tortoise coordinate in the event horizon
is finite and the height of potential barrier increases by decrease
in its width. Based on  the behaviour of effective potential for
$\alpha>1$ Holzhey and Wilczek \cite{Holzhey:1991bx} came to expect
that as one approaches the  extremal limit, the   emission rate of
the black hole tends to zero. However, later Koga and Maeda
\cite{Koga:1995bs}  under the assumption of conservation of the
Black hole charge, showed that numerical calculations  point to the
emission of the large amount of energy for the scalars  in the
extremal limit due to the afore-mentioned divergence of the
temperature.

Mass and charge evaporation rates of black hole in terms of
radiation spectrum are given by\cite{Sampaio:2009ra},
\begin{equation}
-\frac{dM}{dt}=\int_{m}^{\infty}\frac{d\omega}{2\pi}\sum_{mods\ n,\ charge\ q}\frac{\omega(1-|R_{n}(\omega)|^2)}{\exp((\omega-q\Phi_{H})/T_{H})\pm1},\label{eq16}
\end{equation}
\begin{equation}
-\frac{dQ}{dt}=\int_{m}^{\infty}\frac{d\omega}{2\pi}\sum_{mods\ n,\ charge\ q}\frac{q(1-|R_{n}(\omega)|^2)}{\exp((\omega-q\Phi_{H})/T_{H})\pm1}.\label{eq17}
\end{equation}
with the  minus sign is for bosons and the plus sign is for
fermions.  The electrical potential of black hole on the event
horizon $\Phi_{H}=Q/r_{+}$. For near extremal limit or for black
hole with small mass where emission of the quanta of energy and
charge alters the temperature of black hole significantly, one must
take into account backreaction effects in the Hawking radiation
spectrum \cite{Kraus:1995vr}. For this purpose, substituting
$\omega$ with $-dM$ and $q$ with $-dQ$ in above formula and using
first law of black hole thermodynamics\cite{Bardeen:1973gs} we
obtain the nonthermal spectrum of Hawking radiation,
\begin{equation}
-\frac{dM}{dt}=\int_{m}^{\infty}\frac{d\omega}{2\pi}\sum_{mods\ n,\ charge\ q}\frac{\omega(1-|R_{n}(\omega)|^2)}{\exp(-\triangle S_{BH}))\pm1},\label{eq18}
\end{equation}
\begin{equation}
-\frac{dQ}{dt}=\int_{m}^{\infty}\frac{d\omega}{2\pi}\sum_{mods\ n,\ charge\ q}\frac{q(1-|R_{n}(\omega)|^2)}{\exp(-\triangle S_{BH}))\pm1}.\label{eq19}
\end{equation}
$S_{BH}$, is the entropy of the black hole and $\triangle S_{BH}$ is
change of the entropy of black hole before and after radiation of
the quanta of energy and charge,
\begin{equation}
\triangle S_{BH}=S(M-\omega,Q-q)-S(M,Q),\label{eq2.18}
\end{equation}
If A stands for the surface area of the black hole (area of the
event horizon), then the black hole entropy, is given by
Bekenstein-Hawking Formula,
\begin{equation}
S_{BH}=\frac{1}{4}A=\pi r_{+}^{2} \left( 1-\frac{r_{-}}{r_{+}} \right)^{\frac{2\alpha^{2}}{1+\alpha^{2}}}.
\end{equation}
The first law of black hole thermodynamics\cite{Gibbons:1987ps,Bardeen:1973gs} states,
\begin{equation}
dM=T_{H}dS_{BH}+\Phi_{H} dQ.
\end{equation}
In the above formulae $R_{n}(\omega)$, is the reflection coefficient
of emitted  particle which can be obtained from the solution of wave
equation with appropriate boundary condition. $n$ is the angular
parameters of the emitted particle that  in this paper is replaced
by $\kappa$,  for spinors and $l$,  for scalars. $m,\ \omega$ and
$q$, are rest mass, energy and charge of the emitted particle. As we
will see in the next section the equation take a simple form in
tortoise coordinates defined as $r_{*}=\int dr/f(r)^{2}$. This
coordinate maps the location of event horizon $r=r_{+}$, to
$r_{*}=-\infty$. In this coordinate the boundary conditions or
asymptotic behavior of the wave functions for the particles leaving
the black hole horizon in terms of the transition and reflection
coefficients are,
\begin{equation}
\Psi = \left\{
  \begin{array}{l l}
     e^{+i(\omega-\frac{qQ}{r_{+}})r_{*}}+R_{n}(\omega)e^{-i(\omega-\frac{qQ}{r_{+}})r_{*}} & \quad r\rightarrow r_{+}\\
    T_{n}(\omega)e^{+i\omega r} & \quad r\rightarrow +\infty
  \end{array} \right.,
\end{equation}
where $\Psi$s' are the asymptotic solutions of wave equations for
outgoing modes.

The greybody factor defined as transition probability of wave in a
given mode through the black hole potential,   can be written in
terms of the reflection coefficient as follows,
\begin{equation}
\gamma_{n}(\omega)=1-|R_{n}(\omega)|^{2}.
\end{equation}
If we suppose the particles are coming from infinity into the black
hole, these factors will indicate absorption coefficients of black
hole. So, the greybody factors can be computed by obtaining the
scattering coefficients of black hole. In the next section we will
solve the corresponding equations to find these coefficients.

\section{\label{Charged massive Dirac particle in the background metric}Charged massive Dirac particle in the background metric}
In this section we address the main technical question of this
article, emission of charged massive spin 1/2 particle in the
background of dilaton black hole which is the key to our further
analysis. Details of this calculation is presented in the appendix.

The equation of motion for spin 1/2 particle with charge $q$ and
mass $m$ in the background metric (\ref{eq2.6}) is;
\begin{equation}
\left(i\gamma^{\mu}D_{\mu}-m \right) \Psi=0,\label{eq3.1}
\end{equation}
where
\begin{equation}
D_{\mu}=\partial_{\mu}+\Gamma_{\mu}-iqA_{\mu},
\end{equation}

$\Gamma_{\mu}$ is the spin connection defined by
\begin{equation}
\Gamma_{\mu}=\frac{1}{8}\left[\gamma^{a},\gamma^{b}\right] e_{a}^{\nu} e_{b\nu;\mu},
\end{equation}
$e^{a}_{\mu}$, the tetrad (vierbein) is
\begin{equation}
e^{a}_{\mu}=diag\left(f(r),f(r)^{-1},R(r),R(r)\sin\theta \right).
\end{equation}

We solve (\ref{eq3.1}) by separation of variables and taking
$\Psi=f(r)^{-\frac{1}{2}}(\sin{\theta})^{-\frac{1}{2}}\Phi$\cite{Chen:2007dja,R:2008ub}.

Let us define the operator $K$
\begin{eqnarray}
K=\gamma^{t} \gamma^{r} \gamma^{\theta} \frac{\partial}{\partial \theta} + \gamma^{t} \gamma^{r} \gamma^{\varphi} \frac{1}{\sin{\theta}} \frac{\partial}{\partial \varphi}.
\end{eqnarray}
with eigenvalues
\begin{eqnarray}
\kappa=
\left\{
  \begin{array}{cc}
    (j+\frac{1}{2}) & j=l+\frac{1}{2}, \\
    -(j+\frac{1}{2}) & j=l-\frac{1}{2}. \\
  \end{array}
\right.
\end{eqnarray}

Here $\kappa$ is a positive or negative integer
($\kappa=\kappa_{(\pm)}=\pm1, \pm2, ...$). Positive integers
indicate $(+)$ modes while negative integers indicate $(-)$ modes.

One can show that after separation of radial and angular variables $\Phi$ can be taken as;
\begin{eqnarray}
\Phi=
\left(
  \begin{array}{c}
    \frac{iG^{(\pm)}(r)}{R(r)} \phi^{(\pm)}_{jm}(\theta,\varphi) \\
    \frac{F^{(\pm)}(r)}{R(r)} \phi^{(\mp)}_{jm}(\theta,\varphi) \\
  \end{array}
\right) e^{-i\omega t},
\end{eqnarray}
with
\begin{eqnarray}
\phi^{+}_{jm}=
\left(
  \begin{array}{c}
    \sqrt{\frac{l+1/2+m}{2l+1}}Y_{l}^{m-1/2} \\
    \sqrt{\frac{l+1/2-m}{2l+1}}Y_{l}^{m+1/2} \\
  \end{array}
\right),
\end{eqnarray}

for $j=l+1/2$\\
and
\begin{eqnarray}
\phi^{-}_{jm}=
\left(
  \begin{array}{c}
    \sqrt{\frac{l+1/2-m}{2l+1}}Y_{l}^{m-1/2} \\
    -\sqrt{\frac{l+1/2+m}{2l+1}}Y_{l}^{m+1/2} \\
  \end{array}
\right).
\end{eqnarray}

for $j=l-1/2$

In this derivation we have followed
\cite{Chen:2007dja,Cho:2007zi,Cho:2003qe,Cho:2004wj,Jing:2003wq,Jing:2004zb,R:2008ub,Shu:2004fj,Wang:2009hr}.

Defining $\hat{F^{(\pm)}}$ and $\hat{G}^{(\pm)}$ by
\begin{equation}
\left(
  \begin{array}{c}
    \hat{F}^{(\pm)} \\
    \hat{G}^{(\pm)} \\
  \end{array}
\right)
=
\left(
  \begin{array}{cc}
    \sin(\theta_{(\pm)}/2) & \cos(\theta_{(\pm)}/2) \\
    \cos(\theta_{(\pm)}/2) & -\sin(\theta_{(\pm)}/2) \\
  \end{array}
\right)
\left(
  \begin{array}{c}
    F^{(\pm)} \\
    G^{(\pm)} \\
  \end{array}
\right),
\end{equation}
with $\theta_{(\pm)}=\arccot{\left( \kappa_{(\pm)} \left/ mR(r)
\right.\right)}$, $0\leq\theta_{(\pm)}\leq\pi$ and the tortoise
coordinate change $r_{*} = \int{f(r)^{-2}dr}$, we can separate
equations to get, $W_{(\pm)}$ and $V_{(\pm)1,2}$. Eventually, as
shown in the appendix, the wave equations for spinors are ,
\begin{equation}
-\frac{\partial^{2} \hat{F}}{\partial \hat{r}_{*}^{2}}+\left( V_{1}-\omega^{2} \right) \hat{F}=0 \label{eq3.5}.
\end{equation}
\begin{equation}
-\frac{\partial^{2} \hat{G}}{\partial \hat{r}_{*}^{2}}+\left( V_{2}-\omega^{2} \right) \hat{G}=0 \label{eq3.6}.
\end{equation}

The radial potentials are
\begin{equation}
V_{1,2}=W^{2} \pm \frac{\partial W}{\partial \hat{r}_{*}}\label{eq3.6},
\end{equation}
$\hat{r}_{*}$ is the \textit{generalized tortoise coordinate},
\begin{equation}
\hat{r}_{*}= \int \frac{1}{f(r)^{2}}\left(1-\frac{qQ}{\omega r}+\frac{1}{2}f(r)^{2} \frac{\frac{m}{\omega} \kappa}{ \left(\kappa^{2}+m^{2}R(r)^{2}\right) } \frac{\partial R(r)}{\partial r}\right)dr\label{eq3.15},
\end{equation}
and
\begin{equation}
\fl W= f(r) \left(m^{2} + \frac{\kappa^{2}}{R(r)^{2}}\right)^{\frac{1}{2}} \left(1-\frac{qQ}{\omega r}+\frac{1}{2}f(r)^{2} \frac{\frac{m}{\omega} \kappa}{ \left(\kappa^{2}+m^{2}R(r)^{2}\right) } \frac{\partial R(r)}{\partial r}\right)^{-1}\label{eq3.7}.
\end{equation}

The effective potential of scalars $V_{\eta}$ and fermions $V_{1,2}$
and also superpotential $W^{2}$ have the common factor
$(1-\frac{r_{+}}{r})(1-\frac{r_{-}}{r})^{\frac{1-3\alpha^{2}}{1+\alpha^{2}}}\frac{1}{r^{2}}$
which approximately determines their main characteristics; the
location of the  maximum, its height, the behaviour  at infinity and
at the event horizon, for  different charges and coupling constants.
This factor shows that for $\alpha=1/\sqrt{3}$, there is a turning
point so that for this value the potential becomes independent of
the charge. Note that since
$\frac{r_{-}}{r_{+}}=(1+\alpha^{2})\frac{Q^{2}}{r_{+}^{2}}$, when
$r_{+}$ is kept constant $\frac{r_{-}}{r_{+}} $ or effectively
$r_{-}$ represents the square of the charge. So when $\alpha $
changes it passes through the turning point as  a special point
discussed more in the following.

The maximum of effective potential for both scalars and fermions is approximately,
\begin{eqnarray}
\fl \frac{r_{max}}{r_{+}}=1-\frac{1}{4} \left( \frac{3-\alpha^{2}}{1+\alpha^{2}} \epsilon - 2 \frac{1-\alpha^{2}}{1+\alpha^{2}} \right) + \frac{1}{4}\left( \left( \frac{3-\alpha^{2}}{1+\alpha^{2}} \epsilon - 2 \frac{1-\alpha^{2}}{1+\alpha^{2}} \right)^2 + 8\epsilon \right)^{\frac{1}{2}},
\end{eqnarray}
where we have $\epsilon=1-\frac{r_{-}}{r_{+}}$.

For neutral black holes ($\epsilon=1$) the location of maximum is at
$r_{max}=\frac{3}{2} r_{+}$. By gradually increasing the charge, the
position of the maximum changes. The direction of its change depends
on the value of coupling constant. When $0\leq\alpha<1/\sqrt{3}$,
addition of charge, either positive or negative, pushes the position
of the maximum away from the horizon which tends to  $r_{max}
\rightarrow \frac{2 r_{+}}{1+\alpha^{2}}$ at extremal limit when
$\epsilon$ goes to zero. In the case of $\alpha=1/\sqrt{3}$ the
location of maximum  doesn't change by change of black hole charge
and is always in constant location $r_{max}=\frac{3}{2} r_{+}$. In
the case $1/\sqrt{3}<\alpha<1$, by increase  of the charge, the
location of maximum decreases and approaching  extremality it tends
to $r_{max} \rightarrow \frac{2 r_{+}}{1+\alpha^{2}}$. In the case
of $\alpha\geq 1$, when $r_{-}$  approaches  the extremal limit
($\epsilon \rightarrow 0$), the position of the maximum moves toward
the event horizon ($r_{max} \rightarrow r_{+}$). In the case of
$\alpha \rightarrow \infty$ the maximum point approaches to
$r_{max}=\left(1+\frac{\epsilon}{2}\right)r_{+}$. Hence the maximum
is always in following ranges
\begin{equation}
\left\{
  \begin{array}{cc}
    \frac{3}{2}r_{+}\leqslant r_{max} \leqslant \frac{2}{1+\alpha^{2}} r_{+} & 0\leq\alpha<1/\sqrt{3}, \\
    \frac{2}{1+\alpha^{2}} r_{+} \leqslant r_{max} \leqslant \frac{3}{2}r_{+} & 1/\sqrt{3}\leq\alpha<1, \\
    r_{+}\leqslant r_{max} \leqslant \frac{3}{2} r_{+} & \alpha\geq1. \\
  \end{array}
\right.\label{eq3.19}
\end{equation}
Since $V_{1}$ and $V_{2}$ are supersymmetric partner potentials,
they must have same spectra and maximum height\cite{Cho:2004wj}.

Suppose $\hat{r}_{*max}$ and $\hat{r}_{*max 1,2}$ are maximum of $W$
and $V_{1,2}$ in generalized tortoise coordinate respectively. We
will conclude $\hat{r}_{*max 1}<\hat{r}_{*max}<\hat{r}_{*max 2}$.
The zeros  of the first derivative of effective potentials
($\frac{\partial V_{1,2}}{\partial \hat{r}_{*}}= 2W \frac{\partial
W}{\partial \hat{r}_{*}} \pm \frac{\partial^{2} W}{\partial
\hat{r}_{*}^{2}}=0$) gives $\hat{r}_{*max 1,2}$. In order to obtain
$\hat{r}_{*max 1,2}=\hat{r}_{*max}+\Delta \hat{r}_{*max 1,2}$ we
expand $\frac{\partial V_{1,2}}{\partial \hat{r}_{*}}=0$ around
$\hat{r}_{*max}$. This gives,
\begin{equation}
\left\{
  \begin{array}{cc}
   \Delta \hat{r}_{*max 1}=-{1}\left/\left[{2W + \frac{\partial}{\partial \hat{r}_{*}}\ln \left(\frac{\partial^{2} W}{\partial \hat{r}_{*}^{2}}\right)}\right]_{\hat{r}_{*max}}\right., \\
    \Delta\hat{r}_{*max 2}={1}\left/\left[{2W - \frac{\partial}{\partial \hat{r}_{*}}\ln \left(\frac{\partial^{2} W}{\partial \hat{r}_{*}^{2}}\right)}\right]_{\hat{r}_{*max}}\right. . \\
  \end{array}
\right.
\end{equation}

To conclude that $\hat{r}_{*max 1}< \hat{r}_{*max} < \hat{r}_{*max
2}$ we shall show that the denominators of the above equations are
allways positive. With this aim we calculate $\frac{\partial^{2}
V_{1,2}}{\partial \hat{r}_{*}^{2}}$ at $\hat{r}_{*}=\hat{r}_{*max}$.
\begin{equation}
\frac{\left. \frac{\partial^{2} V_{1,2}}{\partial \hat{r}_{*}^{2}} \right|_{\hat{r}_{*max}}}{\left. \frac{\partial^{2} W}{\partial \hat{r}_{*}^{2}} \right|_{\hat{r}_{*max}}}= \left[ 2W \pm \frac{\partial}{\partial \hat{r}_{*}}\ln \left(\frac{\partial^{2} W}{\partial \hat{r}_{*}^{2}}\right)\right]_{\hat{r}_{*max}}
\end{equation}

\begin{figure*}[htdp]
\centering
\includegraphics[width=0.86\textwidth]{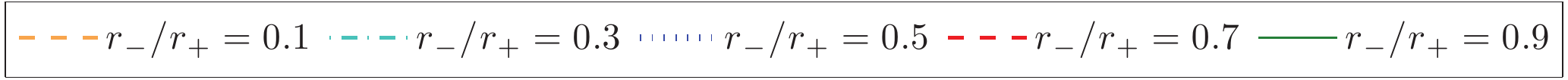}\\
\begin{minipage}[t]{0.43\textwidth}
\centering
\includegraphics[width=1\textwidth]{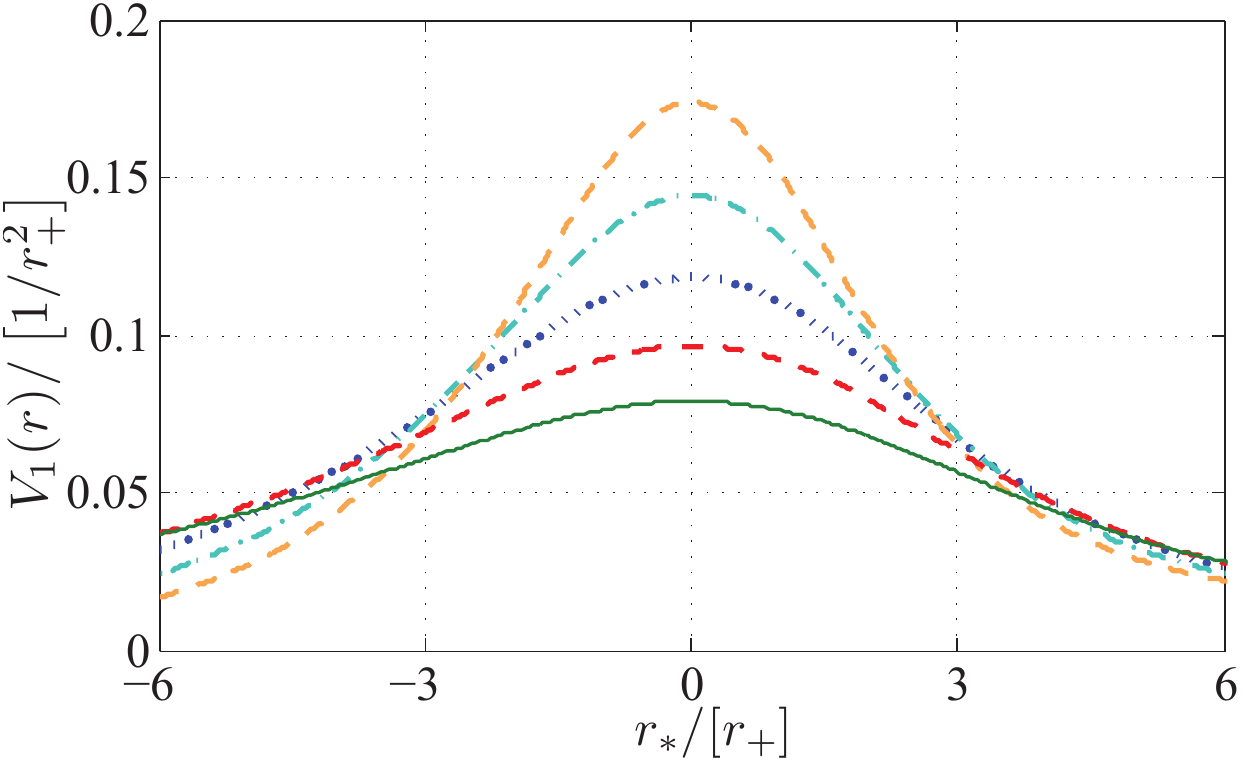}
\subfloat{(a) $\alpha=0$; in terms of $r_{*}/[r_{+}]$.}
\end{minipage}%
~ 
\begin{minipage}[t]{0.43\textwidth}
\centering
\includegraphics[width=1\textwidth]{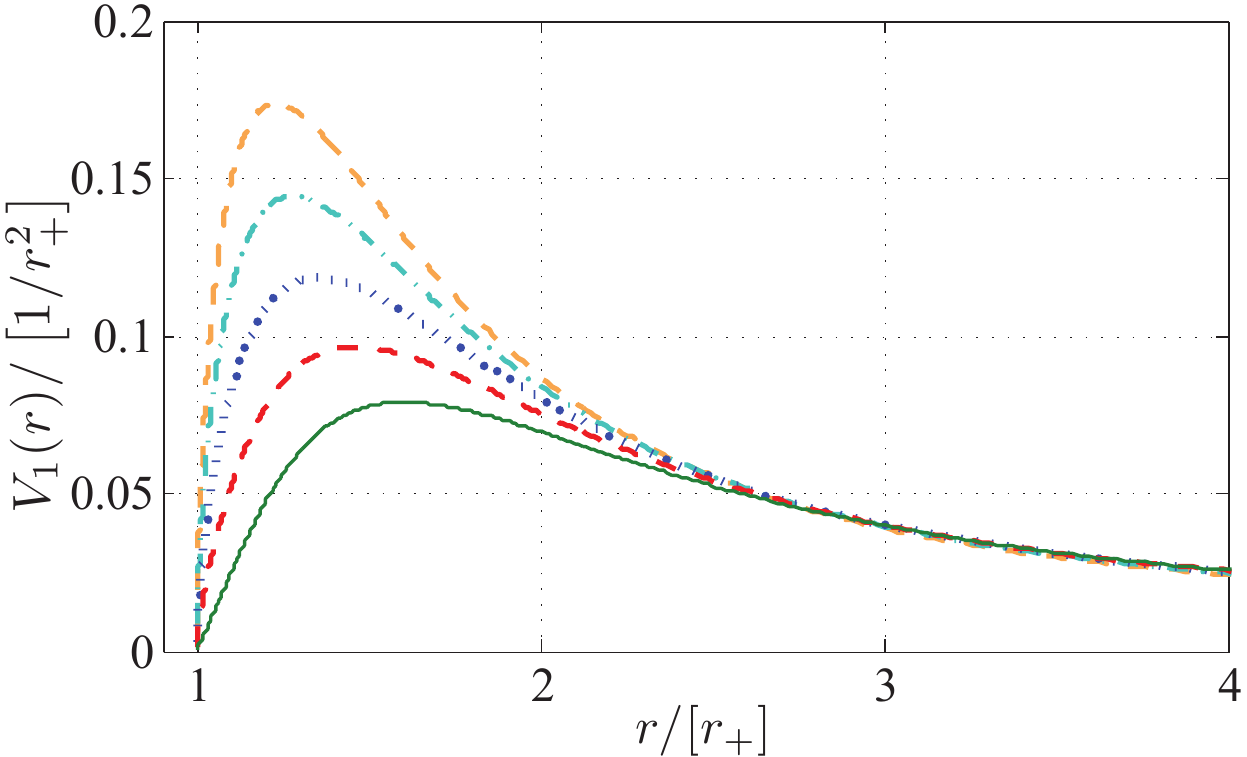}
\subfloat{(b) $\alpha=0$; in terms of $r/[r_{+}]$.}
\end{minipage}\\
\centering
\begin{minipage}[t]{0.43\textwidth}
\centering
\includegraphics[width=1\textwidth]{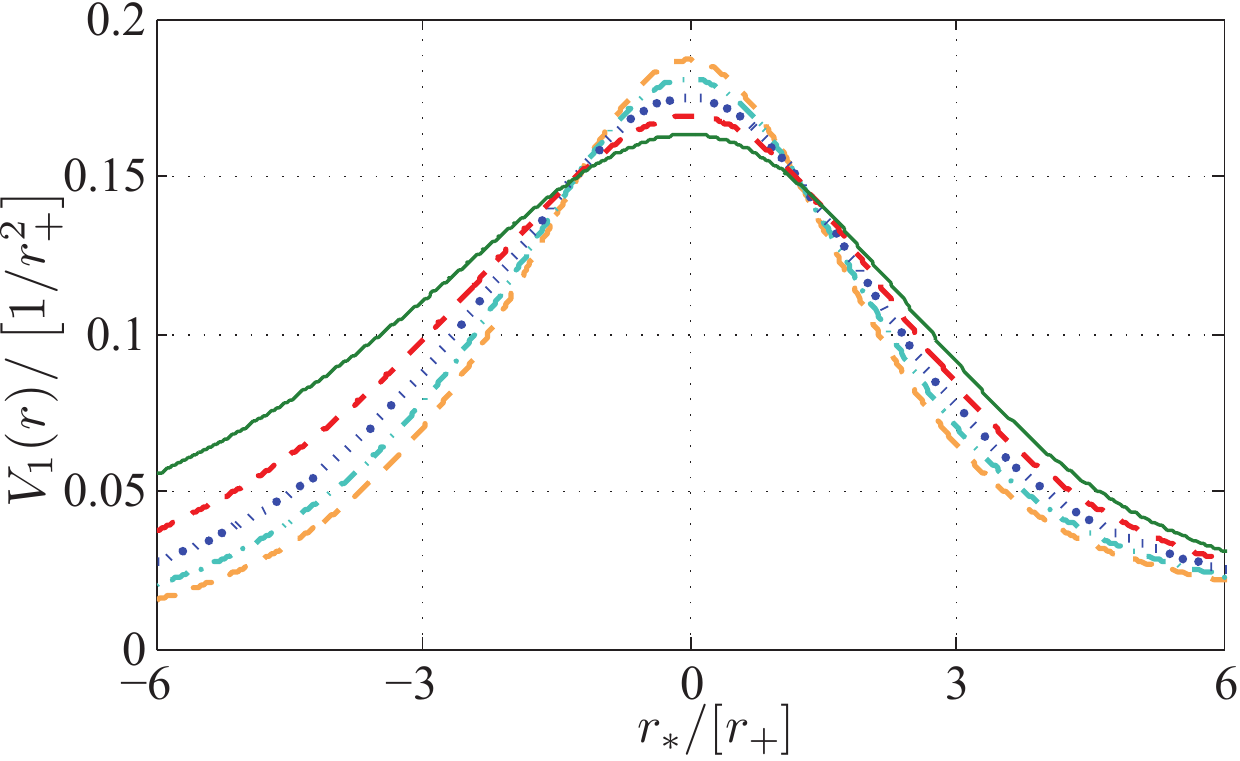}
\subfloat{(c) $\alpha=1/\sqrt{3}$; in terms of $r_{*}/[r_{+}]$.}
\end{minipage}%
~ 
\begin{minipage}[t]{0.43\textwidth}
\centering
\includegraphics[width=1\textwidth]{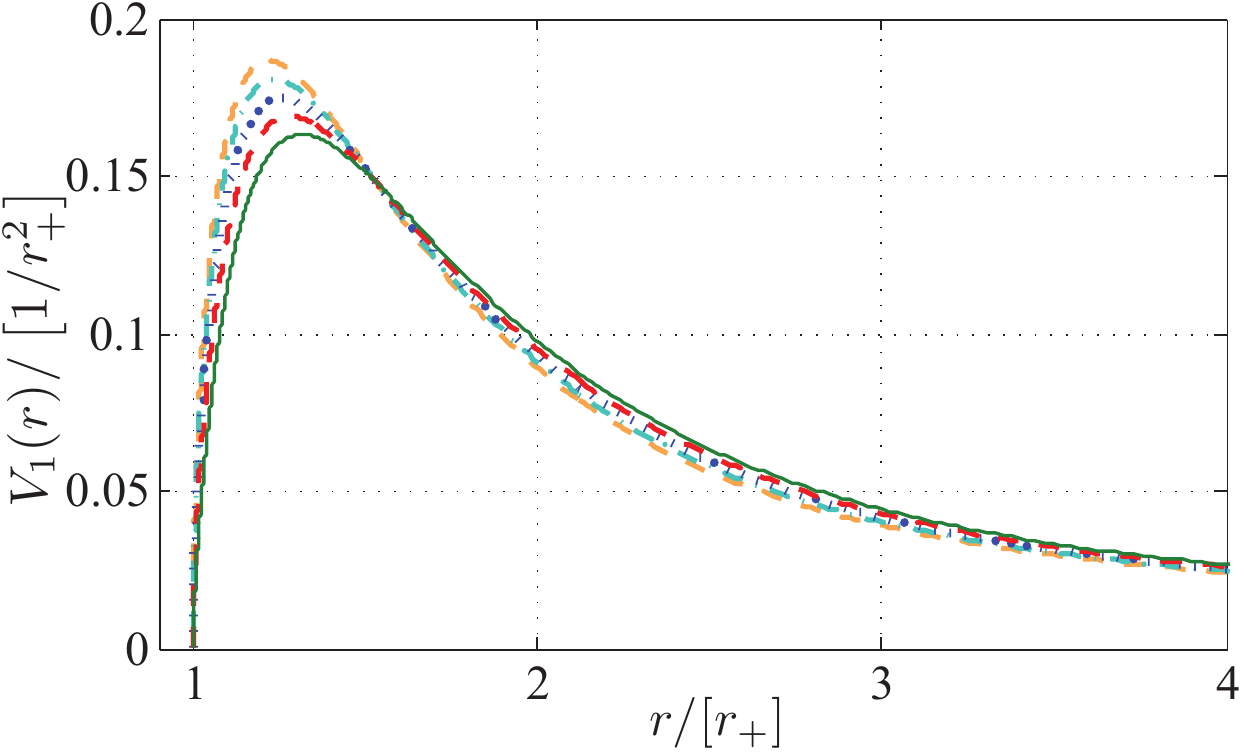}
\subfloat{(d) $\alpha=1/\sqrt{3}$; in terms of $r/[r_{+}]$.}
\end{minipage}
\centering
\begin{minipage}[t]{0.43\textwidth}
\centering
\includegraphics[width=1\textwidth]{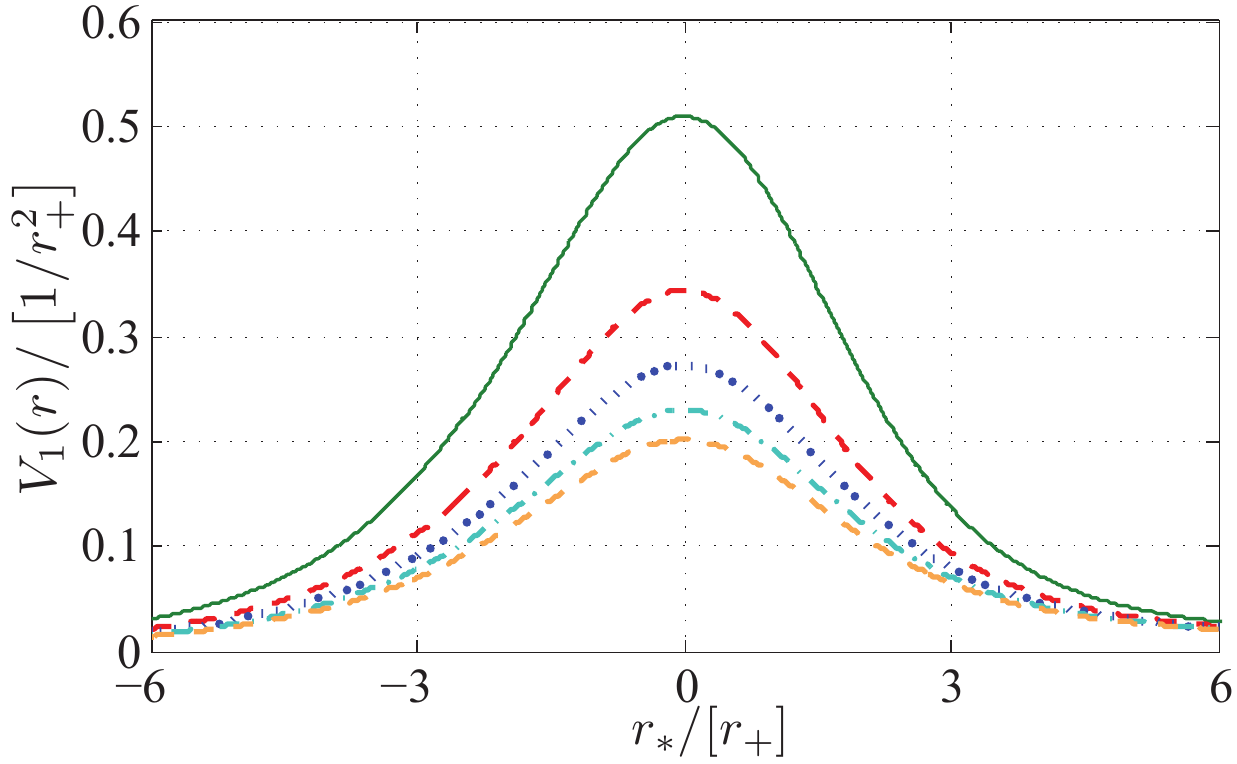}
\subfloat{(e) $\alpha=1$; in terms of $r_{*}/[r_{+}]$.}
\end{minipage}%
~ 
\begin{minipage}[t]{0.43\textwidth}
\centering
\includegraphics[width=1\textwidth]{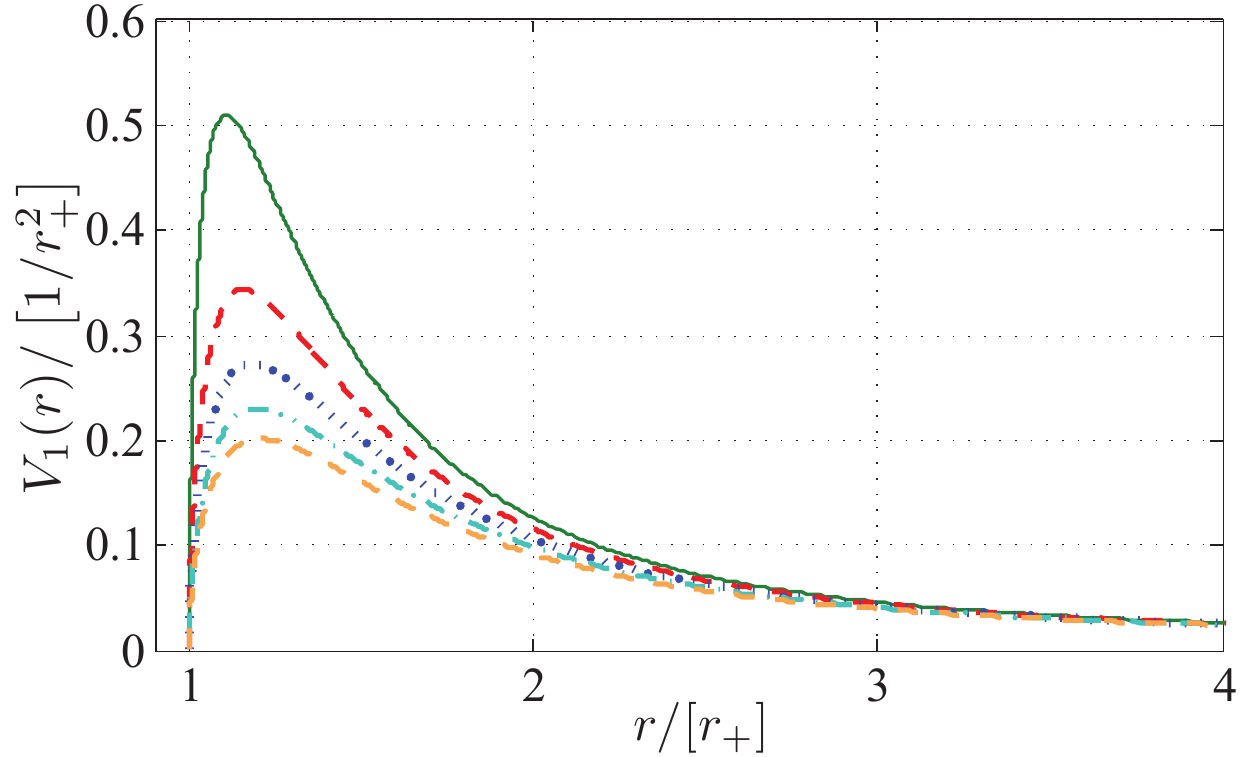}
\subfloat{(f) $\alpha=1$; in terms of $r/[r_{+}]$.}
\end{minipage}
\centering
\begin{minipage}[t]{0.43\textwidth}
\centering
\includegraphics[width=1\textwidth]{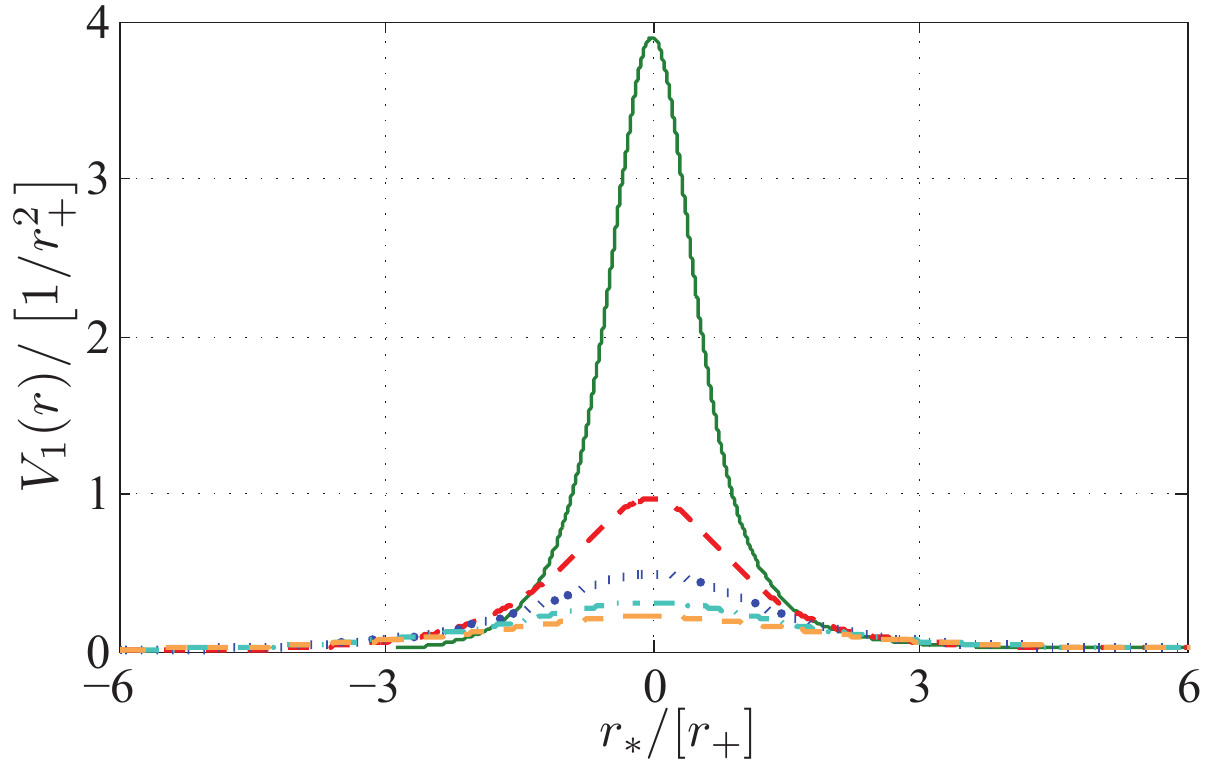}
\subfloat{(g) $\alpha=2$; in terms of $r_{*}/[r_{+}]$. \label{fig_1g}}
\end{minipage}%
~ 
\begin{minipage}[t]{0.43\textwidth}
\centering
\includegraphics[width=1\textwidth]{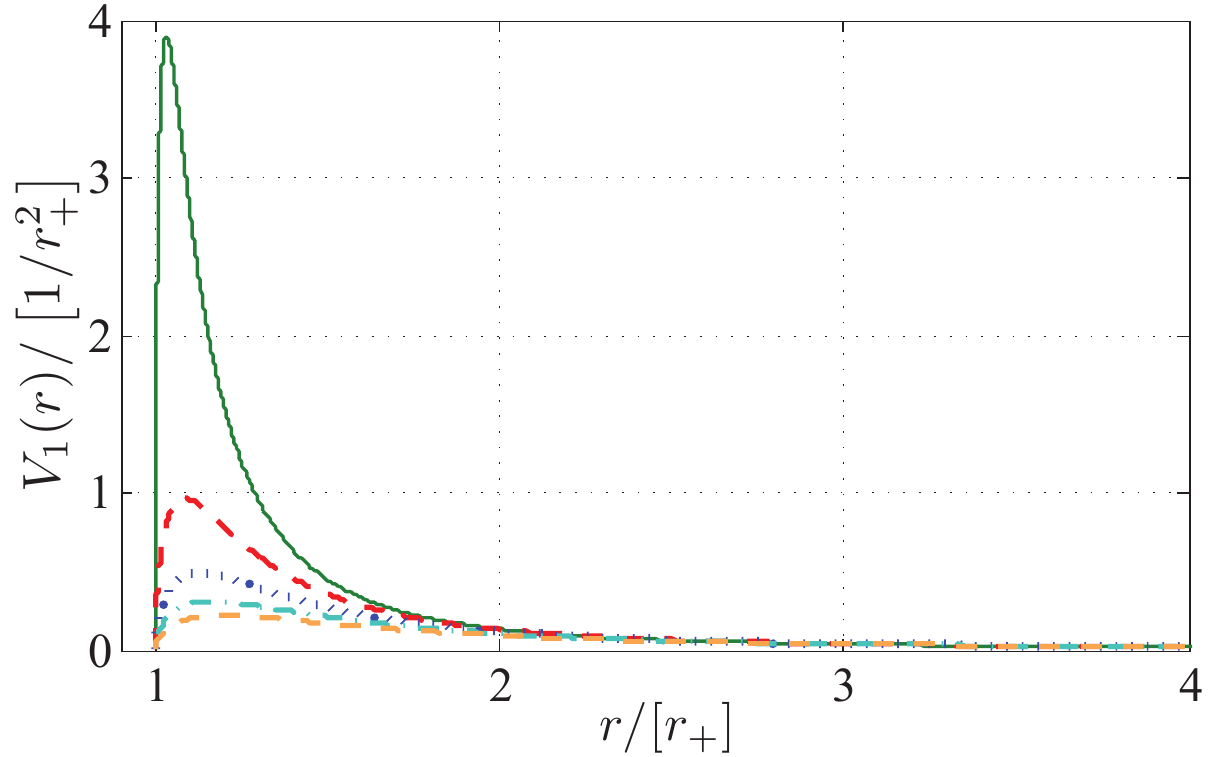}
\subfloat{(h) $\alpha=2$; in terms of $r/[r_{+}]$.}
\end{minipage}
\caption{\label{fig_1}Plots of potential for black holes with
$\alpha=0,1/\sqrt{3},1,2$ and different values of charge
($r_{-}/r_{+}=(1+\alpha^{2})\frac{Q^{2}}{r_{+}}=0.1,\ldots,0.9$) in
natural units and numerical values
$G=\hbar=c=4\pi\varepsilon_{0}=1$, $r_{+}=100$. Spin $\frac{1}{2}$
particles with $\kappa=1$.}
\end{figure*}

Owing to similar behavior of supersymmetric partner potentials
$V_{1,2}$ (Figure \ref{fig_1}) and their superpotential $W$, sign of
their concavity (which is negative) must be equal. Consequently,
above expression is always positive. Furthermore, as we have $\Delta
\hat{r}_{*max 1,2} \simeq \frac{\Delta r_{max 1,2}}{f(r_{max})^{2}}$
from (\ref{eq3.15}), it leads to $r_{max 1}< r_{max} < r_{max 2}$.

As shown before the peak of $V_{1}$ is closer to the horizon, so for
better approximation we analyse  $V_{1}$. Suppose we are  near
extremal condition, we distinguish three different behaviour for the
$r_{max}$.

For $\alpha<1$ we have
$r_{max}\rightarrow\frac{2}{1+\alpha^{2}}r_{+}$,
 and for $\alpha=1$, $r_{max}\rightarrow(1+\sqrt{\frac{\epsilon}{2}})r_{+}$
 and in the case of $\alpha>1$ we obtain
   $r_{max}\rightarrow\left(1+\frac{1}{2}\frac{\alpha^{2}+1}{\alpha^{2}-1}\epsilon\right)r_{+}$.

For fermions the value of the maximum approximately is
\begin{equation}
\left(V_{1,2}\right)_{max} \simeq \left( \frac{1+\alpha^{2}}{2r_{+}} \right)^{2} \frac{ \kappa \left(\kappa + \frac{1-\alpha^{2}}{2}\right)}{ \left(1-\frac{qQ}{\omega r_{+}}\right)^{2} \left(\frac{1-\alpha^{2}}{2} \right)^{\frac{2\alpha^{2}-2}{\alpha^{2}+1}} } ,\ \ \ \ \ \ \ \ \alpha<1\label{eq3.19},
\end{equation}
and
\begin{equation}
\left(V_{1,2}\right)_{max} \simeq \frac{ \kappa^{2} }{ r_{+}^{2} \left(1-\frac{qQ}{\omega r_{+}}\right)^{2}  }, \ \ \ \ \ \ \ \ \ \ \ \ \ \ \ \ \ \ \ \ \ \ \ \ \ \ \ \ \ \ \ \ \ \ \ \alpha=1 \label{eq3.20},
\end{equation}
and
\begin{equation}
\left(V_{1,2}\right)_{max} \simeq \frac{ \kappa \left(\kappa + \frac{1-\alpha^{2}}{1+\alpha^{2}}\right)}{ r_{+}^{2} \left(1-\frac{qQ}{\omega r_{+}}\right)^{2} \left[\frac{1}{2}\frac{\alpha^{2}-1}{\alpha^{2}+1} \left(1-\frac{r_{-}}{r_{+}}\right)\right]^{\frac{2\alpha^{2}-2}{\alpha^{2}+1}} },\ \ \ \ \  \alpha>1\label{eq3.9},
\end{equation}

For scalars the height of the maximum can be approximately obtained
from (\ref{eq2.15}) as follows
\begin{equation}
\left(V_{\eta}\right)_{max} \simeq \left( \frac{1+\alpha^{2}}{2r_{+}} \right)^{2} \frac{\left( (l+\frac{1}{2})^{2}+\frac{1}{4}(1-\alpha^{2}) \right)}{\left( \frac{1-\alpha^{2}}{2} \right)^{\frac{2\alpha^{2}-2}{1+\alpha^{2}}}},\ \ \ \ \ \ \ \ \ \ \alpha<1,
\end{equation}

and
\begin{equation}
\left(V_{\eta}\right)_{max} \simeq \frac{ (l+\frac{1}{2})^{2} }{ r_{+}^{2} },  \ \ \ \ \ \ \ \ \ \ \ \ \ \ \ \ \ \ \ \ \ \ \ \ \ \ \ \ \ \ \ \ \ \ \ \  \ \ \ \ \ \ \ \  \alpha=1, \label{eq3.24}
\end{equation}

and
\begin{equation}
\left(V_{\eta}\right)_{max} \simeq \frac{ (l+\frac{1}{2})^{2} - \frac{1}{4}\left(\frac{1-\alpha^{2}}{1+\alpha^{2}}\right)^{2} }{ r_{+}^{2} \left[\frac{1}{2}\frac{\alpha^{2}-1}{\alpha^{2}+1} \left(1-\frac{r_{-}}{r_{+}}\right)\right]^{\frac{2\alpha^{2}-2}{\alpha^{2}+1}}}, \ \ \ \ \ \ \ \ \ \ \ \ \ \ \ \ \ \ \ \ \ \ \alpha>1 \label{eq3.17}.
\end{equation}

In the case of neutral black holes
($r_{max}\rightarrow\frac{3}{2}r_{+}$) the maximums for fermions and
scalars are
\begin{equation}
\left(V_{1,2}\right)_{max} \simeq \frac{4}{27}\frac{\kappa^{2}}{ r_{+}^{2} },
\end{equation}

\begin{equation}
\left(V_{\eta}\right)_{max} \simeq \frac{4}{27}\frac{ l(l+1)+\frac{2}{3} }{ r_{+}^{2} }.
\end{equation}

Figure \ref{fig_1} shows plots of potential for fermions for
different values of $\alpha$ and black hole charge, from which one
can observe the behavior discussed above.

\begin{figure*}[htdp]
\begin{minipage}[b]{0.5\textwidth}
\centering
\includegraphics[width=1\textwidth]{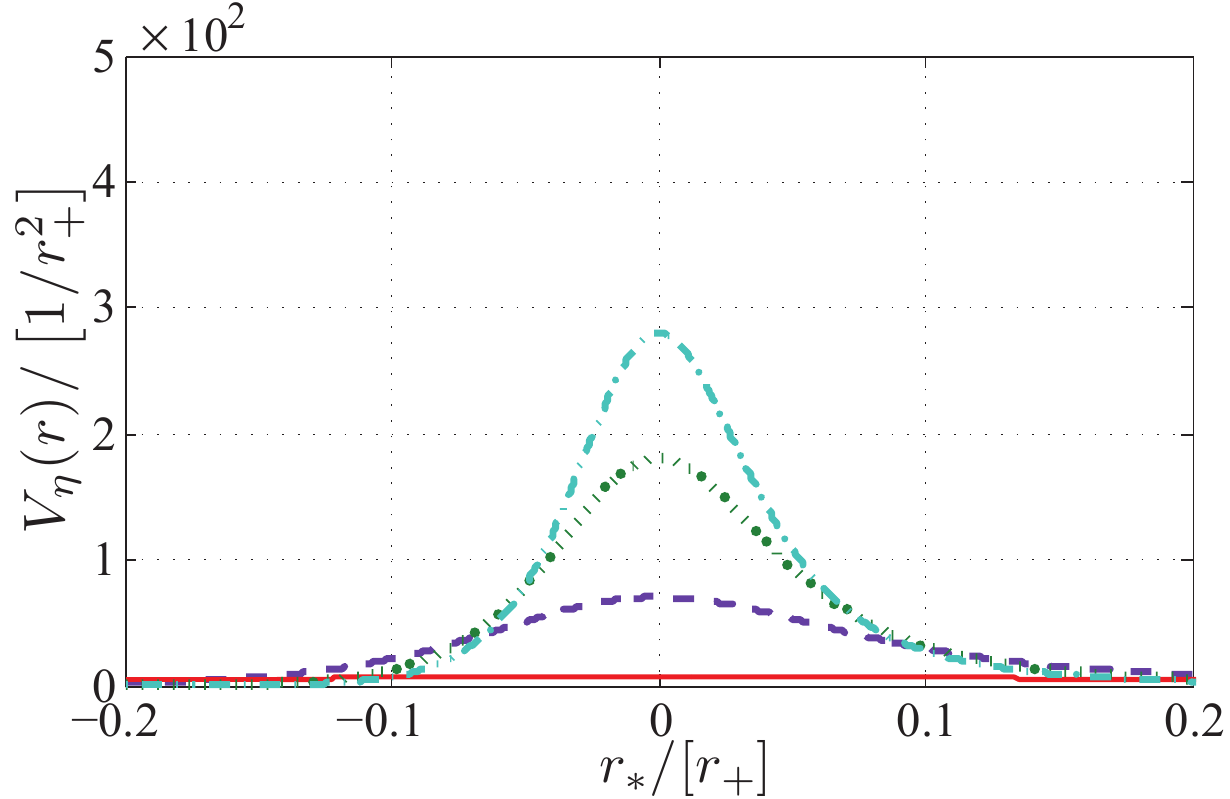}
\subfloat{(a) Effective potentials for scalars.\label{fig_2a}}
\end{minipage}
\begin{minipage}[b]{0.5\textwidth}
\centering
\includegraphics[width=1\textwidth]{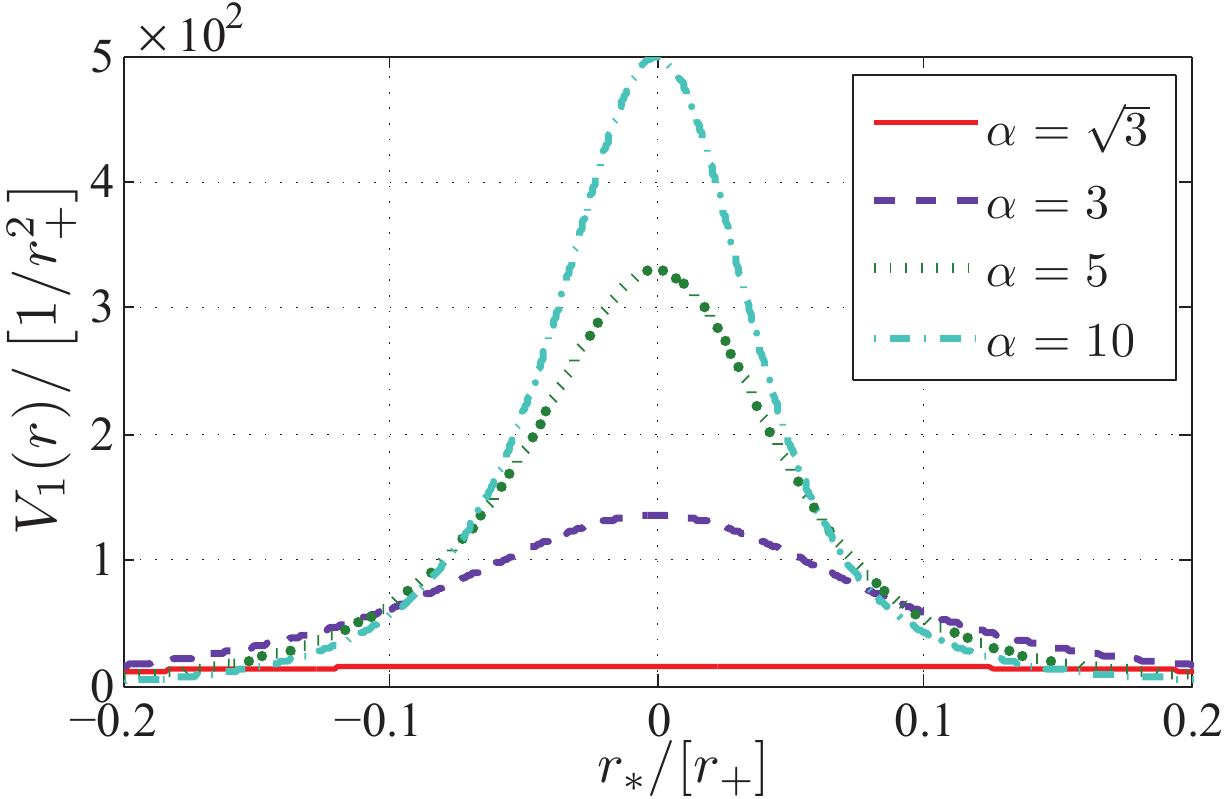}
\subfloat{(b) Effective potentials for fermions.\label{fig_2b}}
\end{minipage}
\begin{minipage}[b]{0.5\textwidth}
\centering
\includegraphics[width=1\textwidth]{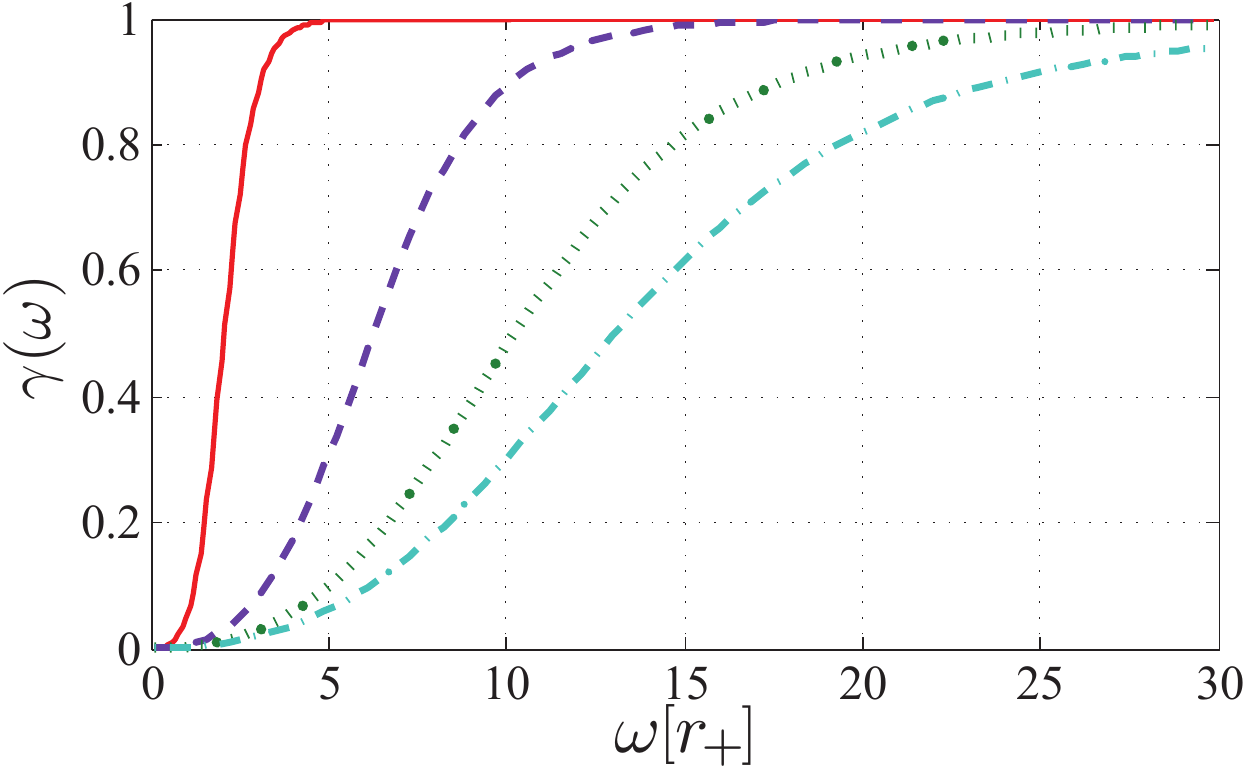}
\subfloat{(c) Greybody factors for scalars.\label{fig_2c}}
\end{minipage}
\begin{minipage}[b]{0.5\textwidth}
\centering
\includegraphics[width=1\textwidth]{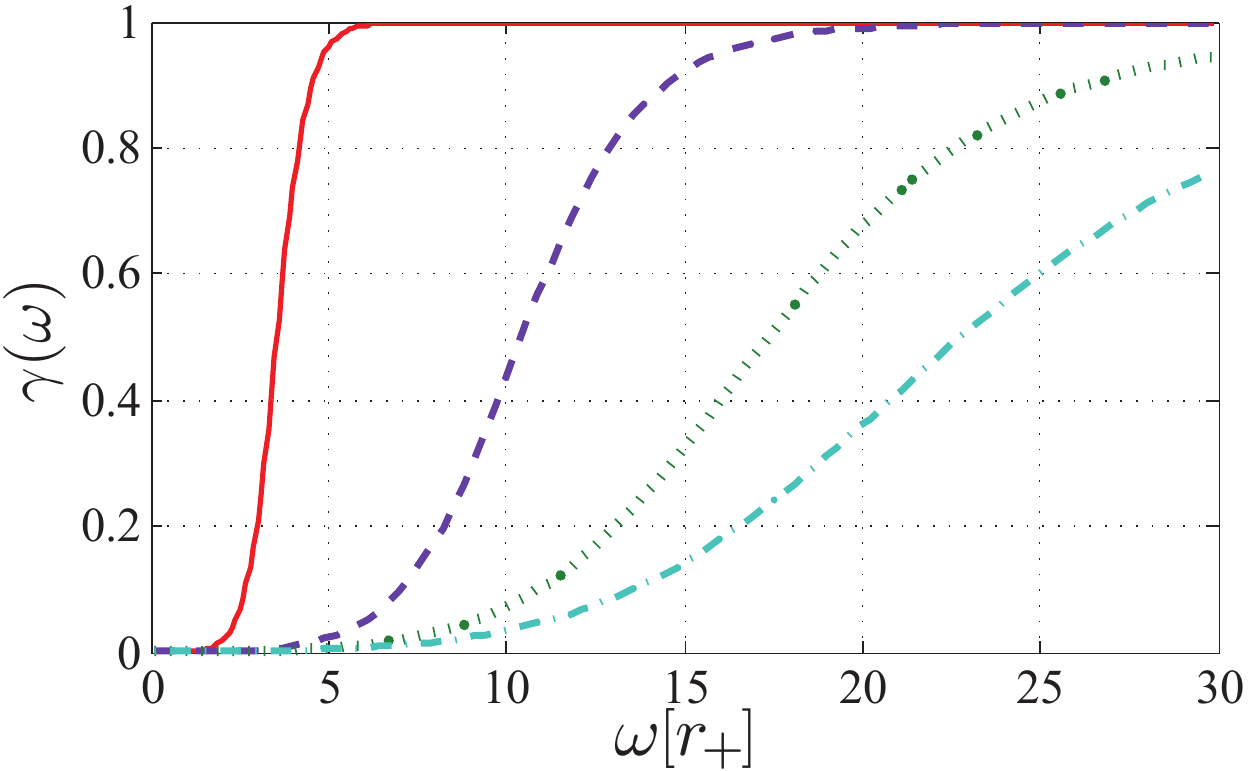}
\subfloat{(d) Greybody factors for fermions.\label{fig_2d}}
\end{minipage}
\begin{minipage}[b]{0.5\textwidth}
\centering
\includegraphics[width=1\textwidth]{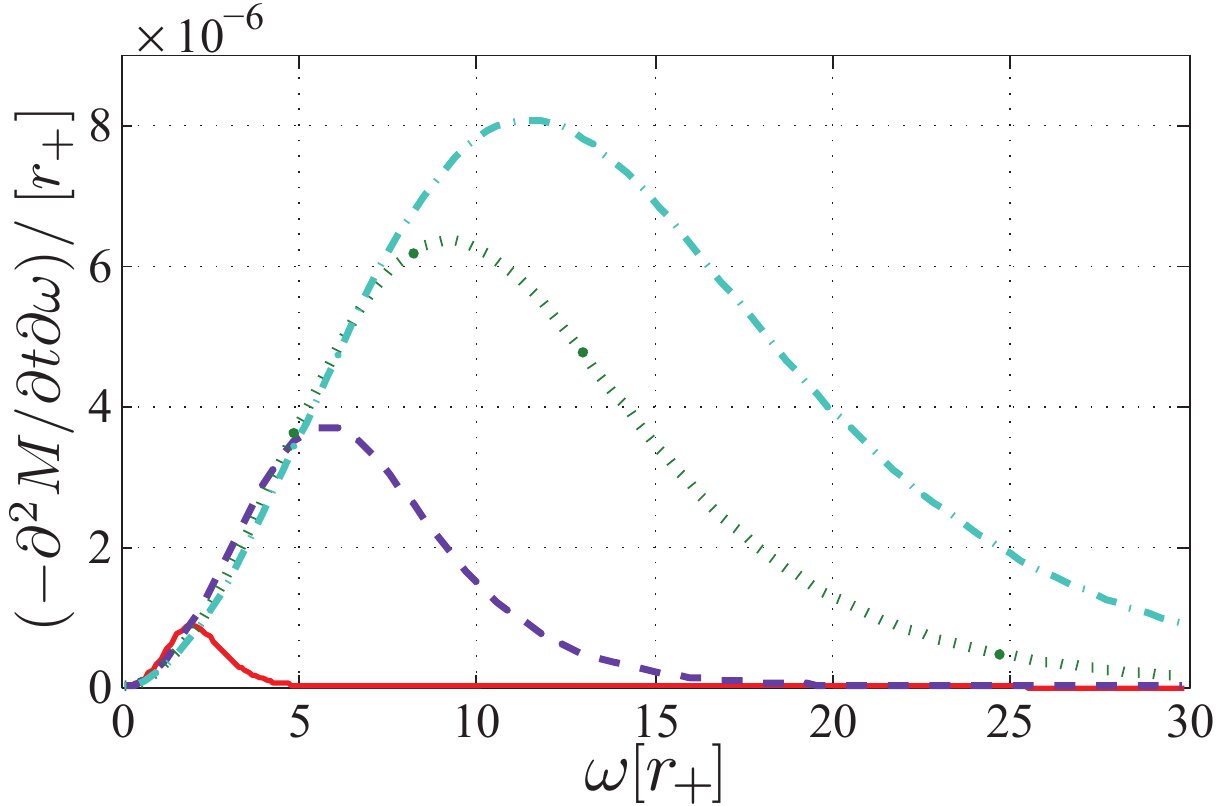}
\subfloat{(e) Energy evaporation rates for scalars.\label{fig_2e}}
\end{minipage}
\begin{minipage}[b]{0.5\textwidth}
\centering
\includegraphics[width=1\textwidth]{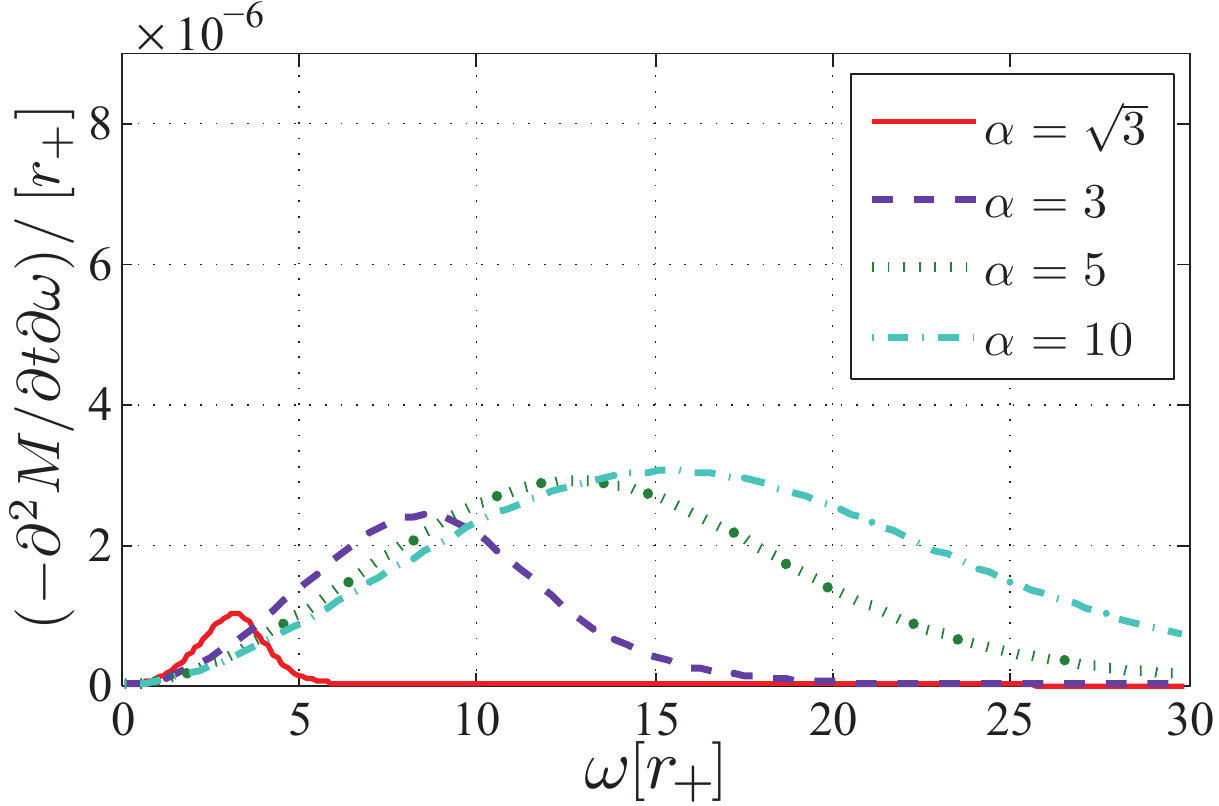}
\subfloat{(f) Energy evaporation rates for fermions.\label{fig_2f}}
\end{minipage}\\
\caption{\label{fig_2}Comparison of dominant mode ($l=0, \kappa=1$)
of potentials, greybody factors and energy evaporation rates of
scalars with fermions for different values of $\alpha$ in natural
units and numerical values $G=\hbar=c=4\pi\varepsilon_{0}=1$,
$r_{-}/r_{+}=0.98$, $r_{+}=100$.}
\end{figure*}

Comparing the maximum of the potentials for scalars and fermions
from previous equations and figures \ref{fig_2a} and \ref{fig_2b} we
see that the value for the  scalars is always less than that of
fermions. Consequently, the greybody factors for scalars (figure
\ref{fig_2c}) would be greater than that of fermions (figure
\ref{fig_2d}). Besides, as scalars obey Bose-Einstein statistics in
thermal emission they would provide a much larger share of the black
hole energy emission with respect to fermions (figure \ref{fig_2e}
and \ref{fig_2f}). Note that the charge of the emitted particle
appears as the product $qQ$ in the denominators of $W$ and in the
maximum of the potential (\ref{eq3.9}). The maximum is higher for
the case when the emitted charge has the same sign as the black
hole. Surprisingly emission of the opposite charge is easier. We
shall see this effect quantitatively when we calculate the greybody
factors. When $q=0$, we obtain the result for the emission of
uncharged fermions such as neutrinos. In this case only the
gravitational force acts on the particle. Also, it can be seen that
the metric factor  at the horizon is zero $f(r_{+})^{2}=0$, where
the potential also vanishes. At  infinity $W^{2}$ and the potential
will be equal to $m^{2}$. The expression for  superpotential $W^{2}$
shows there is an angular  term $\frac{\kappa^{2}}{R(r)^{2}}$ which
vanishes at the extremal limit.  The same factor decreases as
$\alpha$ increases. Therefore, the height of potential barrier
increases as the value of coupling constant increases indicating
that  the coupling constant can have significant effect on greybody
factors and evaporation rates of the black hole.

The potential barrier grows as the second power of the angular
variable $\kappa$ as a result of which we expect lower emission of
higher angular momenta. The angular momentum term
$\frac{\kappa^{2}}{R(r)^{2}}$ is very distinct for the dilaton black
hole and approaches $\frac{\kappa^{2}}{r^{2}}$ for the normal black
hole. At the extremal limit where $r_{-}\rightarrow r_{+}$ , this
term $\frac{\kappa^{2}}{R(r)^{2}}$ diverges. But the maximum value
of the potential (\ref{eq3.19}) for $\alpha<1$ remains finite
although small, and for $\alpha>1$, (\ref{eq3.9}) becomes very large
and divergent.

At $\alpha=1$ the maximum (\ref{eq3.20}) remains finite.

\section{\label{Greybody factors}Greybody factors and dilaton black hole evolution}
In this section we find the greybody factors which are essential in
the evolution of the black hole. The backreaction which also heavily
modifies both  elements is taken into account. The greybody factors
are calculated  with analytical approximation and also numerical
method. Although our conclusions rely upon the numerical
copmutations, the analytical approximation gives insight to the
results obtained.
\subsection{Inclusion of backreaction correction}
As stated  previously the temperature of dilaton black holes plays a
significant role on their behavior. The temperature at the extremal
limit  vanishes for $\alpha<1$, tends to $\frac{1}{4\pi r_{+}}$ for
$\alpha=1$ and diverges for $\alpha>1$. For $\alpha<1$ as
approaching the extremal limit the temperature tends to zero and the
black hole cools . While for $\alpha>1$ the divergence of the
temperature seems to lead to  the eruption of the black hole. On the
other hand, in this case the inclusion of backreaction will show
that the geometry outside the event horizon prevents the decay of
the black hole and cools off  its radiation completely at the
extremal limit. Thanks to this correction the solution provides a
more dynamical picture. In this part first we will explain the
inclusion of this correction into the solution and then try to show
the effect of this correction by analytical approach, for a better
clarification of  the situation. Due to similarity between equations
of scalars and fermions analytical solution is provided only for
fermions. Finally the inclusion of this correction which is more
evident in high frequency are presented in figures \ref{fig_4b},
\ref{fig_7a}, \ref{fig_7d}, \ref{fig_10} with numerical solution.
The $\alpha=1$ case needs higher order corrections which is
delegated to the future works.

Indeed, emission of a quanta of energy $\omega$ and charge $q$
changes the mass $M$ and charge $Q$ of the black hole. Hence, to the
first order of backreaction correction one can subtract the lost
energy and charge from the black hole and solve the equations in the
new background geometry. With this aim we insert
$\acute{M}=M-\omega$ and $\acute{Q}=Q-q$ into all the equations. In
this way we resort to an adiabatic approximation and use the
formulae (\ref{eq18}) and (\ref{eq19}) at every step of the
evolution of the black hole.

According to (\ref{eq3.9}) for fermions and (\ref{eq3.17}) for
scalars in $\alpha>1$ the  maximum of the effective potentials grows
as the black hole approaches  the extremal limit, where it
diverges. Consequently, the emitted particle carrying an amount of
energy and charge of the black hole  pushes it toward  the extremal
limit or $\frac{\acute{r_{-}}}{\acute{r_{+}}}>\frac{r_{-}}{r_{+}}$.
This causes the potential barrier to grow and prevent the process.

As was obtained previously the place of the maximum of the effective
potential,  for both fermions and scalars, approaches  the event
horizon as the black hole approaches  the extremal limit, $r_{+}
\rightarrow \left( 1+\frac{1}{2} \frac{\alpha^{2} +1}{\alpha^{2}-1}
\right)$. In the analytical solution of wave equation at the
extremal limit we need the solution of generalized tortoise
coordinate (\ref{eq3.15}) near the event horizon. In this limit
given $\frac{qQ}{\omega r_{+}} \simeq \frac{q}{\omega
\sqrt{1+\alpha^{2}}}$ assuming $Q>0$ we have,
\begin{eqnarray}
\fl \left. \hat{r}_{*} \right|_{r_{+}\simeq r_{-},\  r \simeq r_{+} } \simeq \left. \int_{r} \frac{\left( 1- \frac{q}{\omega \sqrt{1+\alpha^{2}}} \right) }{\left(1-\frac{r_{+}}{\acute{r}}\right)^{\frac{2}{1+\alpha^{2}}}} d\acute{r} \right|_{r \simeq r_{+}} \simeq \frac{\alpha^{2}+1}{\alpha^{2}-1} r_{+}^{\frac{2}{\alpha^{2}+1}} \left( 1- \frac{q}{\omega \sqrt{1+\alpha^{2}}} \right) (r-r_{+})^{\frac{\alpha^{2}-1}{\alpha^{2}+1}}, \nonumber\\
\label{eq4.1}
\end{eqnarray}

Contrary to the expectations,  this tortoise coordinate at the
extremal limit and for $\alpha>1$ at the limit $r \rightarrow r_{+}$
is finite, while in other cases it tends to $-\infty$. This can be
observed in figure \ref{fig_1g} where the width of the potential
barrier decreases as the black hole approaches  the extremal limit.

In order to obtain the magnitude of the influence of the
backreaction on the high frequency results, we first need to
calculate the change in the  potential barrier  under the emission
of quantum of energy $\omega=-\delta M$ and charge $q=-\delta Q$,
respectively the  changes in   mass and charge of the of black hole.
We have used  the maximum height of the potential barrier obtained
in (\ref{eq3.9}) on the grounds that it can be approximated with
potential barrier near the event horizon. Then, the effect reflected
on the in inner ($\acute{r}_{-}=r_{-}+\delta r_{-}$) and outer
($\acute{r}_{+}=r_{+}+\delta r_{+}$) horizon assuming near extremal
limit ($Q=\sqrt{1+\alpha^{2}} M$, $r_{+} \simeq r_{-} \simeq
(1+\alpha^{2}) M$) are calculated,
\begin{equation}
\left\{
  \begin{array}{cc}
   \delta r_{+}=\left(1+\frac{1}{\alpha^{2}} \right) \delta M + \frac{(\alpha^{2}-1)(\alpha^{2}+1)^{1/2}}{\alpha^{2}} \delta Q, \\
    \delta r_{-}=- \left(1+\frac{1}{\alpha^{2}} \right) \delta M + \frac{(\alpha^{2}+1)^{3/2}}{\alpha^{2}} \delta Q. \\
  \end{array}
\right.
\end{equation}

For the case $ \alpha >1$ the denominator of (\ref{eq3.9}), where
the term $1-\frac{r_{-}}{r_{+}}$ causes the effective potential to
diverge at the extremal limit, plays a significant role on the black
hole behaviour when backreaction correction is applied. The  change
of this term under the change of black hole mass $\delta M$ and
charge $\delta Q$ is ,
\begin{equation}
1-\frac{\acute{r_{-}}}{\acute{r_{+}}}=\epsilon + \frac{2}{\alpha^{2}}\left( \frac{\delta M}{M} - \frac{\delta Q}{Q} \right).
\end{equation}
where we have $\epsilon=1-\frac{r_{-}}{r_{+}}$.

Suppose under the emission of the quantum of energy $\omega=-\delta
M$ and charge $q=-\delta Q$ this term vanishes. Consequently, the
effective potential diverges and as a result the passage of  the
particle  through the potential barrier is impeded. Hence,
$1-\frac{\acute{r_{-}}}{\acute{r_{+}}}=0$ puts an upper limit on the
high cutoff frequency ($\omega_{HCF}$).
\begin{equation}
\omega_{HCF}=\frac{\alpha^{2}}{2} M \epsilon + \frac{q}{\sqrt{1+\alpha^{2}}}\label{eq4.4}.
\end{equation}
Of course as we will discuss below the true high frequency cut off
is much smaller.

As the black hole approaches the extremal limit ($\epsilon
\rightarrow 0$), this high cutoff frequency tending to $\omega_{HCF}
\rightarrow \frac{q}{\sqrt{1+\alpha^{2}}}$ decreases. Obviously,
because of the emission of other neutral particles  with zero second
term  ($\frac{q}{\sqrt{1+\alpha^{2}}}=0$) and opposite sign charged
particles, this high cutoff frequency would be smaller than the
obtained value. In addition, as can be observed from figure
\ref{fig_10a}, as the black hole approaches the extremal limit the
low cutoff frequency $\omega_{LCF}$, specified by the greybody
factor,  increases as a function of temperature of black hole
($\omega_{LCF} \propto T_{H}$)  divergent in  this limit.  Hence in
this limit the Hawking radiation of the black hole would be strongly
suppressed.

After the emission of particle from event horizon, the maximum of
the potential (\ref{eq3.9}) changes,  $(\acute{V_{1,2}})_{max} =
(V_{1,2})_{max} + \delta (V_{1,2})_{max}$. In this limit given
$\frac{qQ}{\omega r_{+}} \simeq \frac{q}{\omega
\sqrt{1+\alpha^{2}}}$ we have,
\begin{equation}
(\acute{V}_{1,2})_{max} \simeq \frac{ \kappa \left(\kappa + \frac{1-\alpha^{2}}{1+\alpha^{2}}\right)}{ r_{+}^{2} \left(1-\frac{q}{\omega \sqrt{1+\alpha^{2}}} \right)^{2} \left[\frac{1}{2}\frac{\alpha^{2}-1}{\alpha^{2}+1} \left( \epsilon - \frac{2}{\alpha^{2}}\left( \frac{\omega}{M} - \frac{q}{Q} \right) \right)\right]^{\frac{2\alpha^{2}-2}{\alpha^{2}+1}}},\label{eq 415}
\end{equation}

This equation shows as the black hole emits particle with
$\omega>\frac{M}{Q} q$, the height of potential barrier grows and it
reduces the greybody factors.

In order to obtain the change in the greybody factors
$\acute{\gamma} (\omega)$ after inclusion of backreaction
correction,  we approximate the change of the maximum of the
potential barrier $\delta (V_{1,2})_{max}$  to the first order under
this correction,
\begin{equation}
\delta (V_{1,2})_{max} \simeq \frac{4}{\alpha^{2} \epsilon}\frac{\alpha^{2}-1}{\alpha^{2}+1} \left( \frac{\omega}{M}-\frac{q}{Q} \right) \left(V_{1,2}\right)_{max}.
\end{equation}

To obtain the  greybody factors  WKB approximation is used. This
approximation gives the transmission probability from a potential
well \cite{Cho:2004wj}. However, because the change of the greybody
factors is needed only and  the peak of the potential barrier is
near the horizon,  we separate the $\delta (V_{1,2})_{max}$ part and
integrate over this near the event horizon and obtain,
\begin{eqnarray}
\acute{\gamma}(\omega) \simeq \exp\left(-2\int \sqrt{V_{1,2}(r)+\delta V_{1,2}(r) - \omega^{2}}d \hat{r}_{*}\right) \nonumber \\
\ \ \ \ \ \ \ \simeq \gamma(\omega) \exp\left(-2\int \frac{1}{2} \frac{\delta V_{1,2}(r)}{V_{1,2}(r)-\omega^2} d \hat{r}_{*}\right) \nonumber \\
\ \ \ \ \ \ \ \simeq \gamma(\omega) \exp\left(-\int^{\left(1+\frac{\alpha^{2}+1}{\alpha^{2}-1}\epsilon\right)r_{+}}_{r \simeq r_{+}} \frac{\delta (V_{1,2})_{max}}{(V_{1,2}-\omega^2)_{max}} d \hat{r}_{*}\right).
\end{eqnarray}
where $\gamma(\omega)$ is greybody factors without the backreaction
correction. On the grounds that maximum height of potential at the
extremal limit  is near the event horizon ($r_{max} \simeq
\left(1+\frac{1}{2}\frac{\alpha^{2}+1}{\alpha^{2}-1}\epsilon\right)r_{+}$),
the integral is taken over $r \simeq \left[r_{+} \ ,
\left(1+\frac{\alpha^{2}+1}{\alpha^{2}-1}\epsilon\right)r_{+}
\right]$ region. In this range of integration  we can assume $\delta
(V_{1,2})_{max}$ be a constant. Finally, the greybody factors under
the inclusion of backreaction is given by,
\begin{eqnarray}
\fl \acute{\gamma}(\omega) \simeq \gamma(\omega) \exp\left(- \frac{4 \omega }{\epsilon} \left( 1+\frac{1}{\alpha^{2}} \right) \left( 1 - \frac{q}{\omega \sqrt{1+\alpha^{2}}} \right)^{2} \left( \frac{\alpha^{2}+1}{\alpha^{2}-1}\epsilon\right)^{\frac{\alpha^{2}-1}{\alpha^{2}+1}}   \right).\label{eq4.8}
\end{eqnarray}
Note that here we have assumed that the charge of black hole is
positive ($\frac{qM}{\omega
Q}=\frac{q}{\omega\sqrt{1+\alpha^{2}}}$).

This expression shows that the backreaction correction always
reduces the greybody factors. Moreover the term $\left( 1 -
\frac{q}{\omega \sqrt{1+\alpha^{2}}} \right)^{2}$ shows that the
greybody factors at high frequency make the black hole to lose
charge, as can be observed from figure \ref{fig_7d} in high
frequency. Besides, we obtain the important result in equation
(\ref{eq4.8}) which is also shown in figure \ref{fig_10a} that as
the black hole approaches the extremal limit $\epsilon \rightarrow
0$ the greybody factors $\acute{\gamma}(\omega)$ vanishes.

One may question the contributions that change of temperature due to
backreaction will  have on the emission rates and other quantities
of interest. We draw the attention of the reader that this effect
is already taken into account. We have used the formulae (\ref{eq18})
and (\ref{eq19}) for decay rates that are given in terms of $\Delta S$
which include all changes as well as the change in the temperature.
What is implicit in our approach is the  use of $r_{+}$, $r_{-}$ ,
the radii of the outer and  inner horizons and their changes to
compute other quantities. The change in temperature can also be
expressed in terms of $\delta r_{\pm}$, reflected in $\Delta S$ used
in (\ref{eq2.18}) and the effective potentials (\ref{eq 415}) discussed.
 Hence one need not consider it separately.
\subsection{Analytical approximation and numerical solution}
To calculate the greybody
factors\cite{Cvetic:1997ap,Kanti:2002ge,Creek:2007tw,Das:1999pt,Gubser:1997cm,alBinni:2009cu,Sampaio:2009ra,Casals:2006xp,Sampaio:2009tp,Maldacena:1996ix}
we need to solve (\ref{eq3.5}) and (\ref{eq3.6}). Analytical
solutions to such equations are formidable. We choose two methods of
approximations to solve these equations. First we approximate the
potential by a solvable potential, the Rosen-Morse
potential\cite{Eckart:1930zza,Rosen:1932,Boonserm:2011} and then we
solve the equations numerically. Using  this potential quasi-normal
modes of several black holes have been
obtained\cite{Shu:2004fj,Chen:2005rm,Cho:2011sf,Cardoso:2003sw,Konoplya:2011qq,Nagar:2006eu}.
The numerical solution helps us to estimate the errors of our
analytical approach. As we shall see later the approximation is
quite reliable.

We also include corrections due to the backreaction. This correction
becomes important when the black hole approaches the extremal limit.

We use adiabatic approximation to include the effect of the emission
of particles. Like previous works we assume the emission rate is not
too large and hence at any moment the black hole configuration
remains unchanged as (\ref{eq2.6}) except that the mass and charge
have changed, $M\rightarrow M-\delta m$ and $Q\rightarrow Q-\delta
q$. This enables us to use the rates calculated by taking $M$ and
$Q$ constant. Obviously approaching the extremal limit without
applying the correction the results are not as reliable.

Rosen-Morse potential, originally devised to investigate diatomic
molecules has an attractive core approximating the vibrational
states, but approaches a constant to allow dissociated states. In
our case it is parameterized as \cite{Boonserm:2011}
\begin{eqnarray}
V(\hat{r}_{*}) = \frac{V_{+\infty} + V_{-\infty}}{2}+\frac{V_{+\infty} - V_{-\infty}}{2} \tanh \left( \frac{\hat{r}_{*}}{\lambda} \right) + \frac{V_{0}}{\cosh^{2}( \hat{r}_{*}/\lambda )}. \label{eq4.5}
\end{eqnarray}

\begin{figure}[t]
\centering
\includegraphics[width=0.7\textwidth]{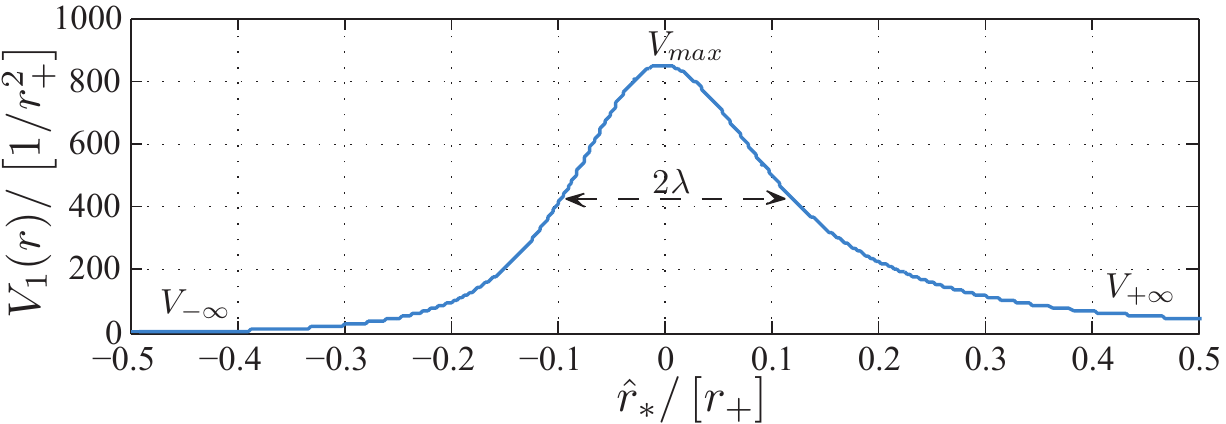}
\caption{\label{fig_3}Plot of potential in terms of $\hat{r}_{*}/[r_{+}]$. $V_{+\infty}$ and $V_{-\infty}$ are black hole potential asymptotic values and $V_{max}$ is its maximum value.}
\end{figure}

To fix the parameters we match it with the black hole potential
(\ref{eq3.6}) at the boundaries $r^* \rightarrow \pm\infty$
\begin{equation}
\left\{
\begin{array}{c}
  V_{+\infty} = \lim_{\hat{r}_{*} \rightarrow +\infty } V_{1,2} = m^{2}, \\
  V_{-\infty} = \lim_{\hat{r}_{*} \rightarrow -\infty } V_{1,2} = 0,
\end{array}
\right.
\end{equation}
and its value at the maximum,
\begin{equation}
V_{0} = \frac{1}{2} V_{max} - \frac{1}{4} m^{2} + \frac{1}{2} \sqrt{ V_{max}^{2} - m^{2} V_{max} } \simeq V_{max}.
\end{equation}
where $V_{max}$ is maximum value of black hole potential. When
$\frac{m^{2}}{V_{max}}<<1$ it will become equal to $V_{0}$. In
(\ref{eq4.5}), $\lambda$ indicates the width of black hole
potential. An approximation in obtaining $\lambda$ is finding the
distance $\hat{r}_{*\lambda}$ of maximum of black hole potential
location from its half in tortoise coordinate,
\begin{equation}
\frac{V_{max}}{2} \simeq \frac{V_{max}}{\cosh^{2}( \hat{r}_{*\lambda}/\lambda )},
\end{equation}
In the above equation for simplicity the mass term is neglected.
Hence, $\lambda$ as a function of $\hat{r}_{*\lambda}$ is given by,
\begin{equation}
\lambda \approx \frac{\hat{r}_{*\lambda}}{\ln(\sqrt{2}\pm1)}.
\end{equation}
These parameters are schematically  shown in figure \ref{fig_3}.

Solving the equations with  Rosen-Morse potential replaced for the
true black hole potential, one obtains  the reflection coefficient
and the greybody factors\cite{Boonserm:2011},
\begin{equation}
\gamma(\omega,\kappa) = \frac{ \sinh \left( \pi \lambda \sqrt{\omega^{2}-m^{2}} \right) \sinh \left( \pi \lambda \omega \right) }{ \sinh^{2} \left( \left. \pi \lambda \left( \omega + \sqrt{\omega^{2}-m^{2}} \right) \right/2 \right) + \cosh^{2} \left( \pi \sqrt{\lambda^{2} V_{0} - 1/4} \right) }\label{eq4.10}.
\end{equation}

\begin{figure*}[t]
\begin{minipage}[b]{1\textwidth}
\centering
\includegraphics[width=0.7\textwidth]{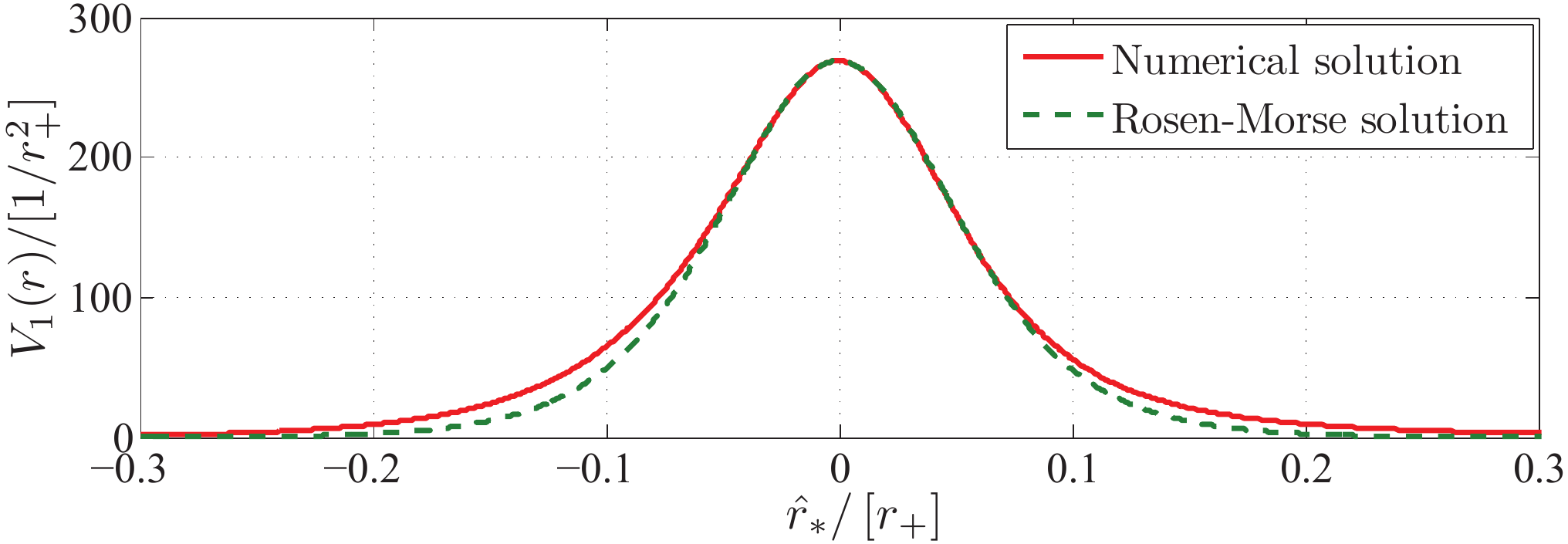}\\
(a) Rosen-Morse potential and exact potential in terms of $\hat{r}_{*}/[r_{+}]$.
\subfloat{\label{fig_4a}}
\end{minipage}
\begin{minipage}[b]{1\textwidth}
\centering
\includegraphics[width=0.7\textwidth]{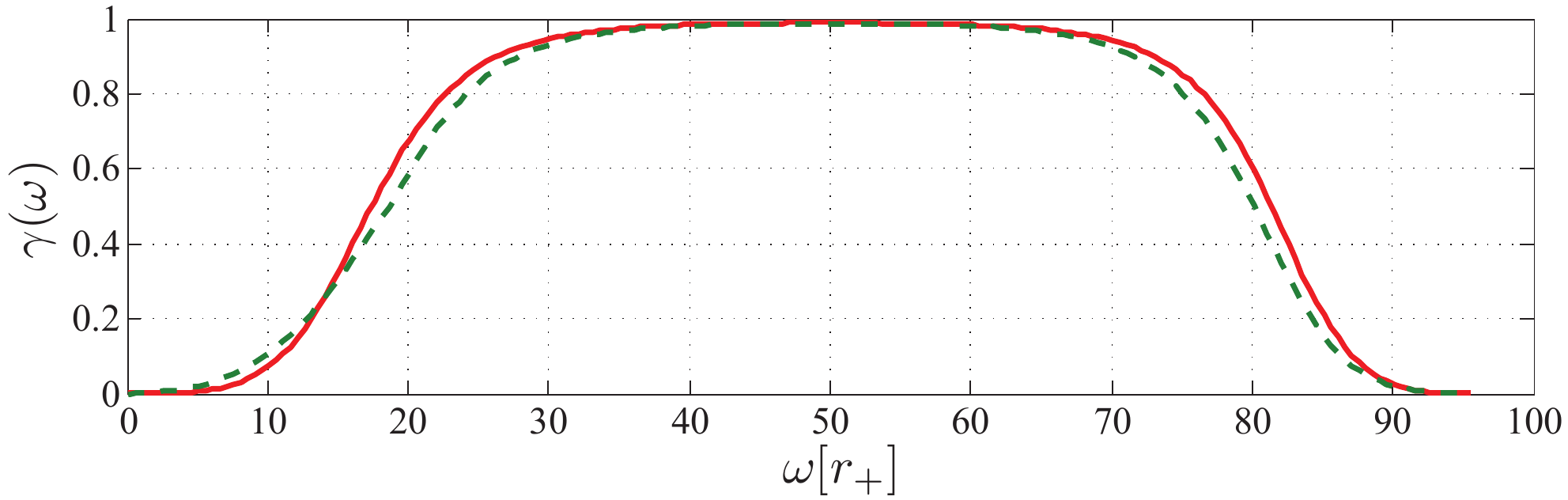}\\
(b) Greybody factors obtained by Rosen-Morse and numerical calculation in terms of frequency $\omega [r_{+}]$.
\subfloat{\label{fig_4b}}
\end{minipage}
\caption{\label{fig_4}Black hole with parameters $r_{-}/r_{+}=0.98$,
$\alpha=5$ and particles with angular momentum $\kappa=1$. Relative
errors are almost $2\%$. Natural units and numerical values are
$G=\hbar=c=4\pi\varepsilon_{0}=1$ and $r_{+}=100$.}
\end{figure*}

The Rosen-Morse potential for our case is compared with numerical
plot of potential (\ref{eq3.6}) in figure \ref{fig_4a}. As it can be
seen the difference looks negligible. In figure \ref{fig_4b} we will
also compare the results of the greybody factors (obtained by the
two methods). The relative error is less than 2$\%$. The errors at
the low frequency limit is larger than the errors in the high
frequency limit. This make the low frequency errors more serious
since they have stronger effect on the evaporation rates.

Although the numerical results are more reliable but the analytic
results obtained by above approximation allows us to explore the
behaviour of the greybody factors more intuitively.


\subsection{\label{Evolution and fate of dilaton black hole}Evolution and fate of the dilaton black hole }
\begin{figure}[t]
\centering
\includegraphics[width=1\textwidth]{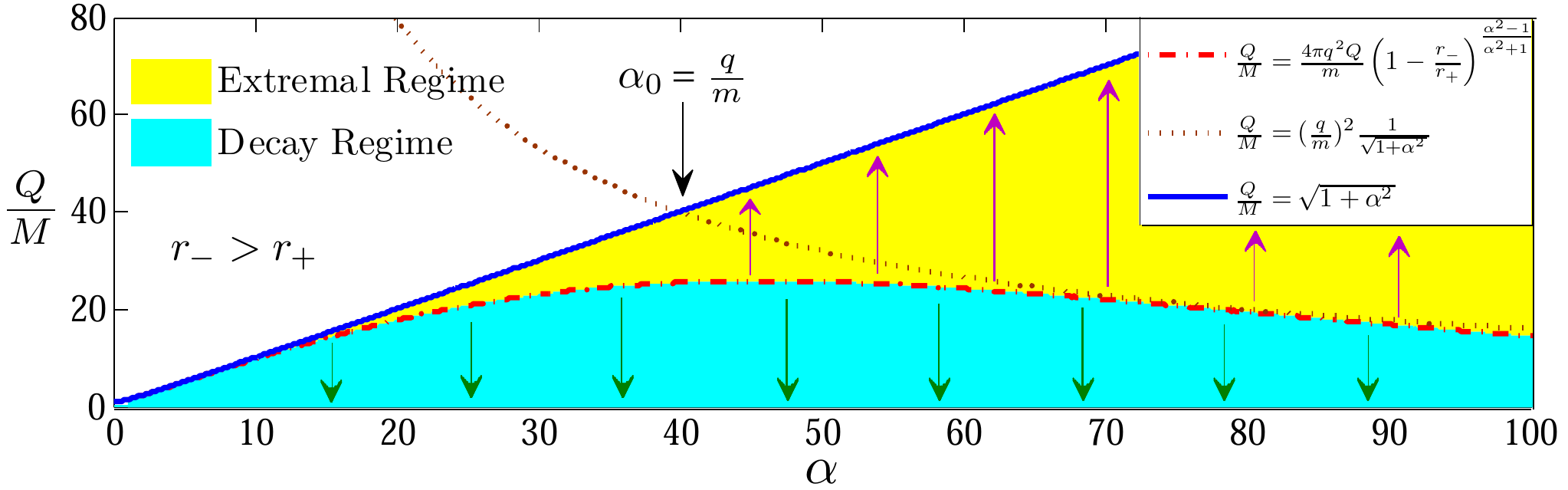}
\caption{\label{fig_5}Plot of transition line, extremal line, extremal regime, decay regime, and direction of evolution and fate of dilaton black hole. Upward arrows in the extreml regime, show the direction of black hole evolution towards extreml limit (flat geometry for $\alpha>>1$), downward arrows in the decay regime show its direction of evolution towards neutral black hole (Schwarzschild black hole) which finally leads to complete evaporation. Numerical values are $\alpha_{0}=40$, $mM=1/2\pi$}
\end{figure}

In this part  we  explore the evolution and fate of the dilaton
black hole. We shall pay particular attention to the conditions
under which  the black hole evolves to two possible final states,
spontaneously evaporating towards extremal limit, or complete
evaporation. We call the boundary of the separation of these two
conditions in the  $(Q/M, \alpha)$ plane (figure \ref{fig_5}), the
transition line. So we distinguish two regimes in this plane, a
region of parameter where the final fate is an extremal black hole
which we call "extremal regime"  and a region in which the final
condition is total evaporation called "decay regime". We have
avoided to call these conditions "phase" since that will cause a
misunderstanding with thermodynamical phases.

The existence of this transition line is easily shown by going to
certain limits. First we prove that there is certain regions that
the final fate of the black hole is inevitably extremal case. We
show this for at least two regions of the parameters, for very large
$\alpha$ and the other when $Q/M > q/m $. First we discuss the case
for large $\alpha$.

If we go to large $\alpha$, equation (\ref{eq17}) shows that the
positive and the negative charges are emitted with equal rate. To
see this more carefully, note that in the charge flux both greybody
factors and Boltzmann factor in Hawking radiation, depend on
$\frac{|qQ|}{\omega r_{+}}$. Also note that when both sign of charge
are emitted the charge evaporation reduces and it may hinder the
motion of the black hole in the ($Q/M, \alpha$) plane away from the
extremal line ($\frac{Q}{M}=\sqrt{1+\alpha^{2}}$). Since
$\frac{qQ}{\omega r_{+}}<\frac{q}{\omega \sqrt{1+\alpha^{2}}}$, we
see that for large $\alpha$, $\frac{qQ}{\omega r_{+}} \rightarrow
0$, and this causes the net charge emission rate to vanish. But mean
while the energy keeps being radiated. Since for a fixed charge
there is a lower limit for the mass of the black hole
($M>Q/\sqrt{\alpha^{2}+1}$), finally the energy radiation must also
come to halt resulting in a extremal state.

Before discussing the other case let us explore an illuminating
property of the large $\alpha$ case. When $\alpha$ is much larger
than  $1$, the geometry (\ref{eq2.6}) becomes flat in the extremal
black hole case. So any black hole geometry in this regions
($\alpha$ large) and extremal regime (figure \ref{fig_5}) taking
$r'=r-r_{-}$ will end up in an extremal state with a flat background
geometry.
\begin{equation}
ds^{2}=dt^{2}-dr'^{2}-r'^{2}d\Omega ^{2}.\label{eq55}
\end{equation}
Therefore it will look like an elementary particle
\cite{Holzhey:1991bx} in a flat space.

Now let us turn to other example which will also help us to find a
mathematical discipline of the line separating two regimes
mentioned.

Another case where similar phenomena  happens is where in the
process of emission of charge and mass, the black hole moves close
to the extremal limit. Since it emits both positive and negative
charges $\pm q$, the effective charge emitted is a positive fraction
of the same sign charge $\xi q$ ($0 \leq \xi < 1$). Then the
effective charge $q'=-\delta Q$ emitted per one quanta of mass
$m=-\delta M$, is less than $q$ the quantum of charge. So we define
$q'/q=\xi$ a ratio to be found. This parameter can be obtained from
$\frac{q}{m}\xi=\frac{dQ}{dM}$, where $dQ = \acute{Q}- Q $ and $dM =
\acute{M}- M $ are the change of black hole charge and mass during
an infinitesimal radiation process and is a function of $(\alpha, Q,
M, \frac{q}{m})$. Then if $\acute{Q}/\acute{M}$ is larger than
$Q/M$, the black hole has moved closer to the extremal limit, and
when this process continues finally it will end up with extremal
parameters and the radiation stops. Such condition is satisfied if
$\frac{Q-\xi q}{M-m}>\frac{Q}{M}$ or equivalently considering the
condition $\frac{Q}{M}<\sqrt{\alpha^{2}+1}$;
\begin{equation}
\frac{\xi q}{m}<\frac{Q}{M}<\sqrt{\alpha^{2}+1}.\label{eq60}
\end{equation}

This completes the proof for the non-emptiness of the extremal
regime. It covers a large area in the ($Q/M, \alpha$) plane as shown
in figure \ref{fig_5}.

To see that there are also black holes that evaporate completely we
give two examples;

One is the case where $\alpha=0$ i.e. the RN black hole which is
well known to lose charge very quickly and then totally evaporate
\cite{Gibbons:1975,Page:1977}. The other example when a black hole
with a given $Q/M$ moves away from the extremal condition after
emission of charge $\xi q$ and mass $m$. This means when
$\frac{Q}{M}<\xi \frac{q}{m}$.

So we have proven the existence of the two regimes which must be
separated with a transition line mathematically specified by
$\frac{Q}{M}=\xi \frac{q}{m}$ from the equation (\ref{eq60}) and is
plotted in figure \ref{fig_5}.

Having the definition for the transition line ($\frac{Q}{M}=\xi
\frac{q}{m}$) we can proceed to calculate it.

Before performing any detailed calculation let us estimate the
transition line for large $\alpha$. As discussed previously as
$\alpha$ increases charge flux decreases and tends to zero as
$\frac{q}{m}\frac{1}{\sqrt{\alpha^{2}+1}} \rightarrow 0$, while the
energy flux remains finite. Hence, at this limit $\xi$ which is the
ratio of charge flux to energy flux ($\frac{q}{m}\xi=\frac{dQ}{dM}$)
tends to zero. As we will show bellow
$\xi<\frac{q}{m}\frac{1}{\alpha^{2}+1}<1$ (\ref{eq4.25}), for small
$\frac{q}{m}\frac{1}{\sqrt{\alpha^{2}+1}}$.

Also, the line
$\frac{Q}{M}=\frac{(q/m)^{2}}{\sqrt{\alpha^{2}+1}}=\frac{\alpha_{0}^{2}}{\sqrt{\alpha^{2}+1}}$
gives an upper limit approximation for the transition line which is
shown in figure \ref{fig_5}. This holds for
$\alpha>\alpha_{0}(=\frac{q}{m})$ and $T_{H}>m$.

This line intersects the boundary line of $r_{-} \leq r_{+}$ given
by $\frac{Q}{M}=\sqrt{\alpha^{2}+1}$ at
$\alpha=\sqrt{\alpha_{0}^{2}-1}$. In physical case say electron
emission where $\alpha_{0}=\frac{1}{\sqrt{4 \pi
\varepsilon_{0}G}}\frac{e}{m_{e}}=2 \times 10^{21}$ is very large,
it is equal to $\alpha_{0}$. It breaks down for $\alpha<\alpha_{0}$.
For this range the transition line can be well approximated with
$\frac{Q}{M}=\sqrt{\alpha^{2}+1}$ same as the line specifying the
extremal condition for black hole with mass above to solar mass.
More accurate calculation bellow shows that the error in the above
estimate is extremely small which becomes order of $10^{-17}$
relatively near $\alpha=\alpha_{0}$ for physical cases.

In order to evaluate $\xi$ one can see from $\frac{q}{m}
\xi=\frac{dQ}{dM}$, (\ref{eq16}), and (\ref{eq17}),
\begin{equation}
\fl \xi = \frac{m}{q} \frac{dQ}{dM}=\frac{ \int_{m}^{\infty}\frac{d\omega}{2\pi}\sum_{mods\ n}\frac{ \gamma_{n} (\omega - \frac{qQ}{r_{+}} ) }{\exp(\left(\omega-\frac{qQ}{r_{+}}\right)/T_{H}) + 1} - \frac{\gamma_{n} (\omega + \frac{qQ}{r_{+}} )}{\exp(\left(\omega+\frac{qQ}{r_{+}}\right)/T_{H}) +1} } {  \frac{1}{m} \frac{\partial M_{neutral}}{\partial t} + \int_{m}^{\infty} \frac{\omega}{m} \frac{d\omega}{2\pi}\sum_{mods\ n}\frac{ \gamma_{n} (\omega - \frac{qQ}{r_{+}} ) }{\exp(\left(\omega-\frac{qQ}{r_{+}}\right)/T_{H}) + 1} + \frac{ \gamma_{n} (\omega + \frac{qQ}{r_{+}} ) }{\exp(\left(\omega+\frac{qQ}{r_{+}}\right)/T_{H}) + 1} },\label{eq4.20}
\end{equation}
here the term $\frac{\partial M_{neutral}}{\partial t}$ presents
contribution of other neutral particles which reduces $\xi$.
Obviously it can be seen that the above expression is always less
than 1 ($\xi<1$).

For any given value of mass $M$ and charge $Q$ of the black hole and
$\alpha$ the value for $\xi$ determines fate of the black hole or
determines weather the black hole is in extremal regime or in decay
regime. Indeed, $\xi$ presents competition of charge and energy
emission of the black hole.

We can estimate the integral in (\ref{eq4.20}) as function of
$\frac{\omega}{T_{H}}(1 \pm \frac{qQ}{\omega r_{+}})$ by
approximation, taking the peak value of the integrands at
$\omega=\omega_{max}$ and multiplying them by its effective width.
Let us take $\lambda=\frac{\omega_{max}}{T_{H}}$ and
$\eta=\frac{qQ}{\omega_{max} r_{+}}$. According to figure
\ref{fig_9} and \cite{Maldacena:1996ix}, $\omega_{max}$ increases as
the temperature of the black hole increases ($\omega_{max}\approx
T_{H}$). $\eta$ is small for the near extremal black holes for
$\alpha>1$, since $T_{H}$ becomes very large. Then taking $\eta$
small and neglecting the effect of neutral particles we have;
\begin{eqnarray}
\xi \lesssim \frac{ \Delta \omega \sum_{mods\ n}\frac{ \gamma_{n} (\lambda(1 - \eta) ) }{\exp(\lambda(1 - \eta)) + 1} - \frac{\gamma_{n} ( \lambda (1 + \eta) )}{\exp(\lambda(1 + \eta)) +1} } { \Delta \omega \frac{\omega}{m} \sum_{mods\ n}\frac{ \gamma_{n} ( \lambda(1 - \eta) ) }{\exp(\lambda(1 - \eta)) + 1} + \frac{ \gamma_{n} (\lambda(1 + \eta)) }{\exp( \lambda(1 + \eta) ) + 1} }   \nonumber \\
\ \ \  \simeq -\lambda \eta \frac{\sum_{mods\ n} \gamma^{\prime}_{n} (\lambda)} {\sum_{mods\ n} \gamma_{n} (\lambda)} -\lambda \eta \frac{1}{1+e^{\lambda}} +\lambda \eta .\label{eq4.24}
\end{eqnarray}
where $\gamma^{\prime}_{n} (\lambda)=\frac{\partial \gamma_{n} (\lambda)}{\partial \lambda}$.

Hence the upper limit for $\xi$ is given by $\xi<\lambda \eta
(\xi<\frac{qQ}{T_{H}r_{+}})$. As $ \omega_{max} \approx T_{H}$, and
$\omega \geq m$, for small black holes (black holes with high
temperature or $8\pi mM<1$) we have $T_{H}>m$. Consequently these
approximations leads to the following inequalities,
\begin{eqnarray}
\xi < \frac{qQ}{T_{H} r_{+}} < \frac{qQ}{m r_{+}} < \frac{q}{m}\frac{1}{\sqrt{1+\alpha^{2}}} .\label{eq4.25}
\end{eqnarray}

In order to calculate the transition line more precisely we assume
the upper bound $\xi < \frac{qQ}{T_{H} r_{+}}$ and insert it in
(\ref{eq60}).
\begin{equation}
\frac{4 \pi q^{2}Q}{m} \left( 1-\frac{r_{-}}{r_{+}} \right)^{\frac{\alpha^{2}-1}{\alpha^{2}+1}}<\frac{Q}{M}<\sqrt{1+\alpha^{2}}. \label{eq4.28}
\end{equation}

Or in another form,
\begin{equation}
\frac{4 \pi q^{2} \sqrt{r_{-} r_{+}}}{m} \frac{1}{\sqrt{1+\alpha^{2}}} \left( 1-\frac{r_{-}}{r_{+}} \right)^{\frac{\alpha^{2}-1}{\alpha^{2}+1}}<\frac{Q}{M}<\sqrt{1+\alpha^{2}}.
\end{equation}

In order to obtain the transition line we put $\frac{\xi
q}{m}=\frac{Q}{M}$.
\begin{equation}
\frac{4 \pi q^{2}Q}{m} \left( 1-\frac{r_{-}}{r_{+}} \right)^{\frac{\alpha^{2}-1}{\alpha^{2}+1}}=\frac{Q}{M}.
\end{equation}

Numerical solution of this equation is given in figure \ref{fig_5}.
Solution of this equation gives us the transition line as a function
of $\alpha$. Assuming $\alpha>>1$, the solution of the above
equation gives,
\begin{equation}
\left. \frac{Q}{M}\right|_{Transition} = \frac{8 \pi m M \alpha_{0}^{2}/\alpha}{1+ 8 \pi m M \alpha_{0}^{2}/\alpha^{2}}
\end{equation}

We see that this line decreases as a function of
$\alpha_{0}^{2}/\alpha$ at large $\alpha$ as like as former
approximation to transition line. Besides, contrary to the former
approximation, this transition line does not intersect the boundary
of $r_{-}\leq r_{+}$ at $\alpha=\alpha_{0}$. For small
$\alpha<<\alpha_{0}\sqrt{8 \pi mM}$ this transition line is very
close to the extremal boundary $Q/M=\sqrt{\alpha^{2}+1}$ which is
shown in figure \ref{fig_5}. The relative difference to the extremal
line for $\alpha>>1$ is given by,
\begin{equation}
\frac{\sqrt{\alpha^{2}+1} - \left. \frac{Q}{M}\right|_{Transition}}{\sqrt{\alpha^{2}+1}} \simeq \frac{1}{1+8 \pi m M \alpha_{0}^{2}/\alpha^{2}}
\end{equation}
In the system of standard units we assume that $M_{\odot}=2 \times
10^{30}$ as solar mass and for electron this distance becomes
$\frac{1}{1+4 \times 10^{59} \frac{M}{M_{\odot}}
\frac{1}{\alpha^{2}}}$. One can check that for small couplings this
relative distance reduces to $2.5 \times 10^{-60}
\frac{M_{\odot}}{M} \alpha^{2}$. Also at $\alpha=\alpha_{0}$ this
ratio becomes $10^{-17}$. This relative distance for a solar mass
black hole and small $\alpha$ shows that how much the black hole
needs to be near the extremal limit to be in extremal regime. While
for $\alpha>>\alpha_{0}\sqrt{8 \pi mM}$ it reduces to 1 and covers
all the $0<\frac{Q}{M}<\sqrt{\alpha^{2}+1}$ region.

One can compare this line with former approximation ($\frac{Q}{M}=
\alpha_{0}^{2}/\sqrt{\alpha^{2}+1}$). Both have similar behaviour as
$\alpha_{0}^{2}/\alpha$. Hence, the ratio of their differences to
the width of the region ($\sqrt{\alpha^{2}+1}$) becomes,
\begin{equation}
\frac{ \frac{\alpha_{0}^{2}}{\sqrt{\alpha^{2}+1}} - \frac{8 \pi m M \alpha_{0}^{2}/\alpha}{1+8 \pi m M \alpha_{0}^{2}/\alpha^{2}}}{\sqrt{\alpha^{2}+1}}  \simeq  \frac{1 + 8 \pi mM \left(\frac{\alpha_{0}^{2}}{\alpha^{2}}-1\right) }{1+8 \pi m M \alpha_{0}^{2}/\alpha^{2}} \frac{\alpha_{0}^{2}}{\alpha^{2}}
\end{equation}

At $\alpha>\alpha_{0}$, as $\alpha$ increases this ratio decreases
and tends to zero. However, for large black holes where $8 \pi mM$
is greater than 1, it is possible that the above expression became
negative (as explained before); the black hole becomes cooler and
the inequality $T_{H}>m$ no longer holds.

From our discussion in the first part of this section and numerical
calculation of next section, we see that for $\alpha>1$ the
backreaction and high potential barrier impedes the radiation near
the extremal limit. Then the radiation vanishes; as a result, for
the black holes in the extremal regime the final stage can be
stable. Furthermore, at the limit $\alpha>>1$ the background
geometry tends to the flat one.

As the black hole tends to the extremal limit, its temperature
increases and diverges, while its area vanishes. However, at this
limit the potential barrier outside the event horizon of the black
hole, as mentioned previously impedes the radiation of this hot
body. Hence this potential barrier acting as an isolator outside the
event horizon prevents the black hole to become in thermal
equilibrium with outside world. Consequently, the black hole stays
stable.

\section{\label{Results and discussion}Results and discussion}
In this section we analyse and discuss effects of various parameters
on the potential and greybody factors obtained in previous sections.
Modification of the Hawking radiation on the light of the behaviour
of the greybody factors is also discussed.

The discussion is based on numerical results in solving the basic
equations (\ref{eq3.1}). Since there are several parameters the
discussion becomes complicated. So we discuss different factors
separately.

We shall consider the effects of charge, mass and angular momentum
of the emitted particle, dilaton coupling constant $\alpha$, the
near extremal condition and finally the difference between scalars
and fermions.

\subsection{Effects of the charges of the emitted particle and of the black hole}

\begin{figure*}[t]
\centering
\begin{minipage}[b]{0.31\textwidth}
\centering
\includegraphics[width=\textwidth]{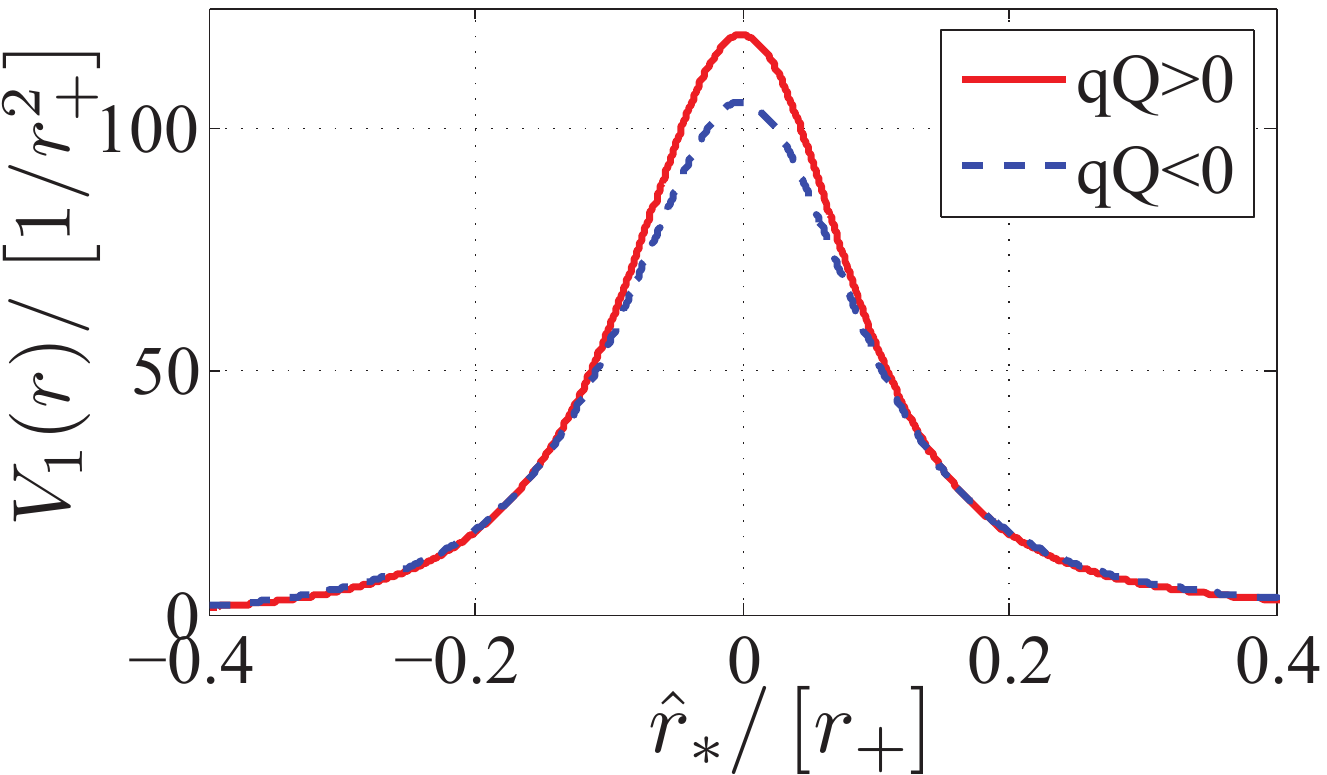}
(a) Potential for $qQ>0$ and $qQ<0$ in terms of $\hat{r}_{*}/[r_{+}]$.
\subfloat{\label{fig_6a}}
\end{minipage}
\quad
\begin{minipage}[b]{0.31\textwidth}
\centering
\includegraphics[width=\textwidth]{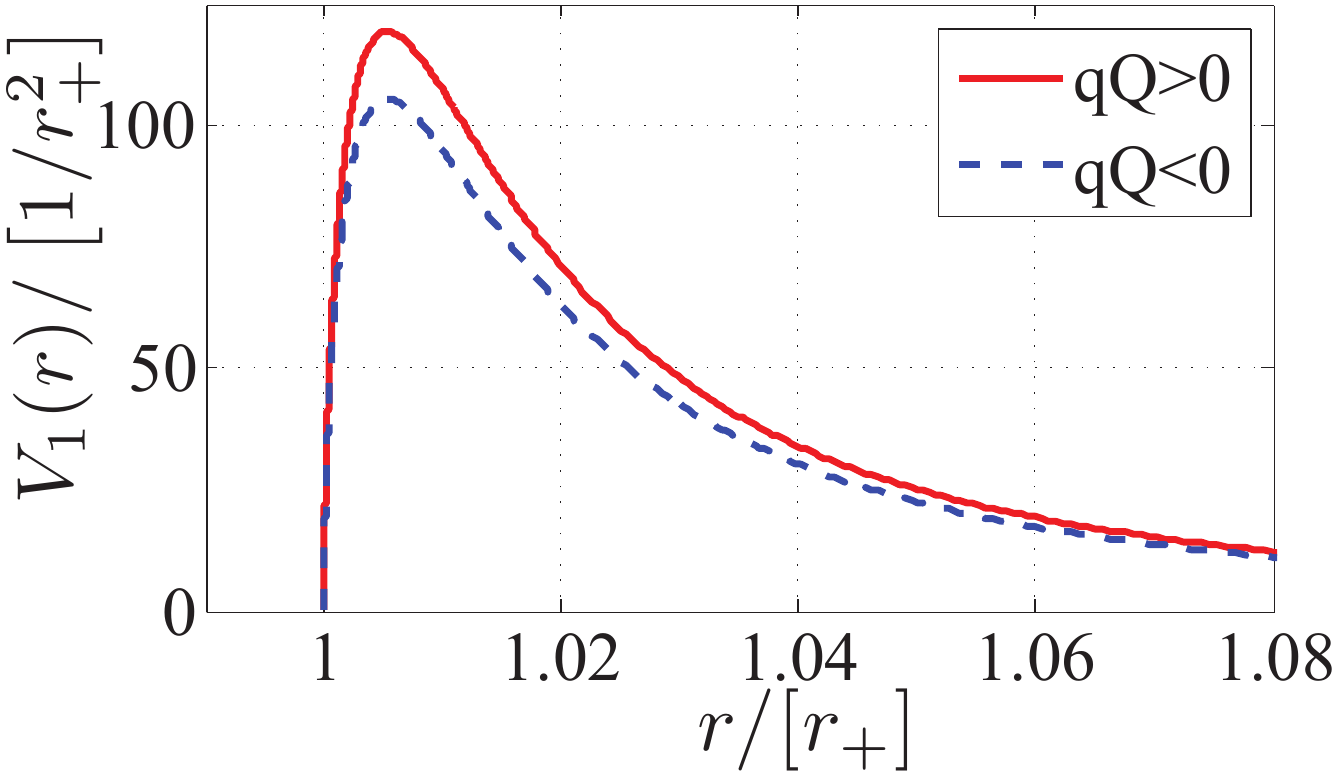}
(b) Potential for $qQ>0$ and $qQ<0$ in terms of $r/[r_{+}]$.
\subfloat{\label{fig_6b}}
\end{minipage}
\quad
\begin{minipage}[b]{0.31\textwidth}
\centering
\includegraphics[width=\textwidth]{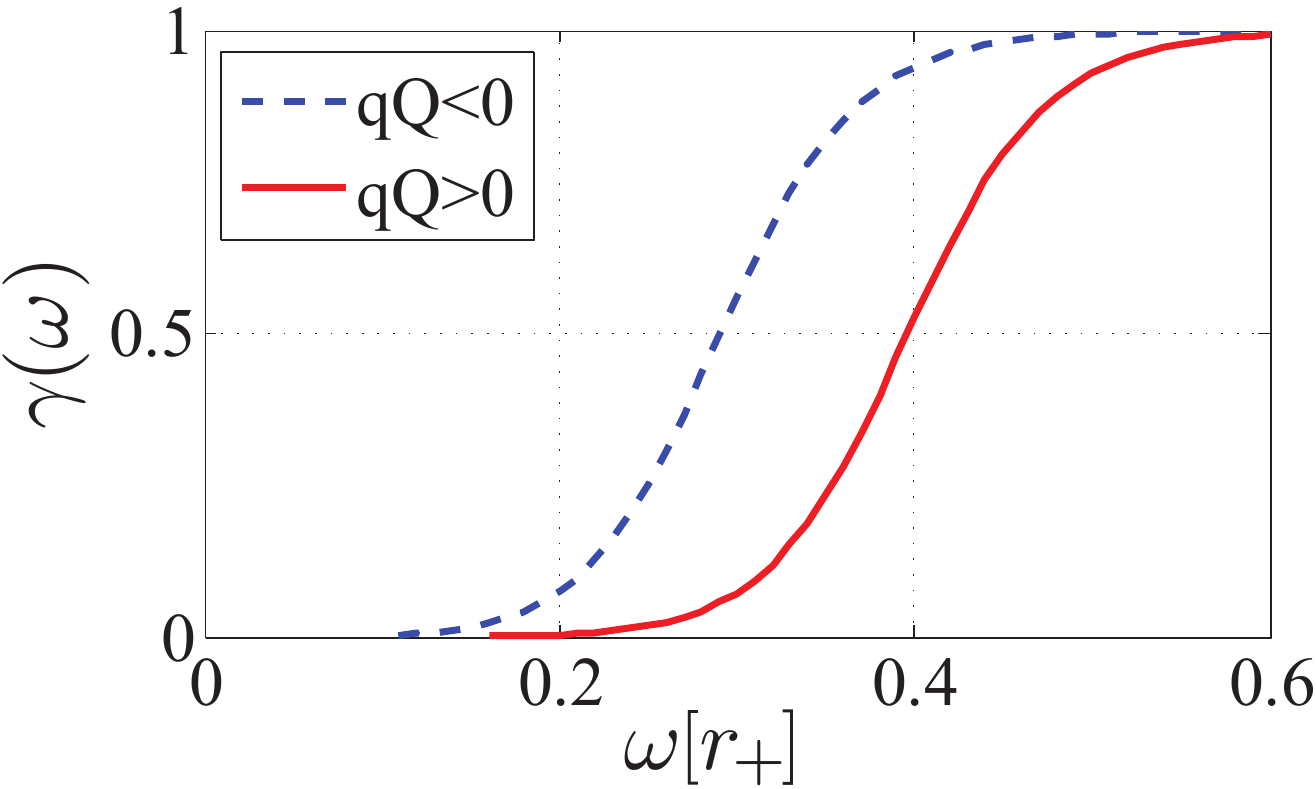}
(c) Greybody factor for $qQ>0$ and $qQ<0$ in terms of $\omega [r_{+}]$.
\subfloat{\label{fig_6c}}
\end{minipage}
\caption{\label{fig_6}Dilaton black hole with parameters
$r_{-}/r_{+}=0.98$ and $\alpha=0.7$. Charged spin $\frac{1}{2}$
particles with $\kappa=1$. Natural units and numerical values are
$G=\hbar=c=4\pi\varepsilon_{0}=1$, $r_{+}=100$ and $q[r_{+}]=0.1$.}
\end{figure*}

The following analysis is based on computations leading to Figures
\ref{fig_6a} and \ref{fig_6b} that show the potential barrier for
both q having the same sign  and opposite to the black hole.

\begin{figure*}[htdp]
\begin{minipage}[b]{1\textwidth}
\centering
\includegraphics[width=0.85\textwidth]{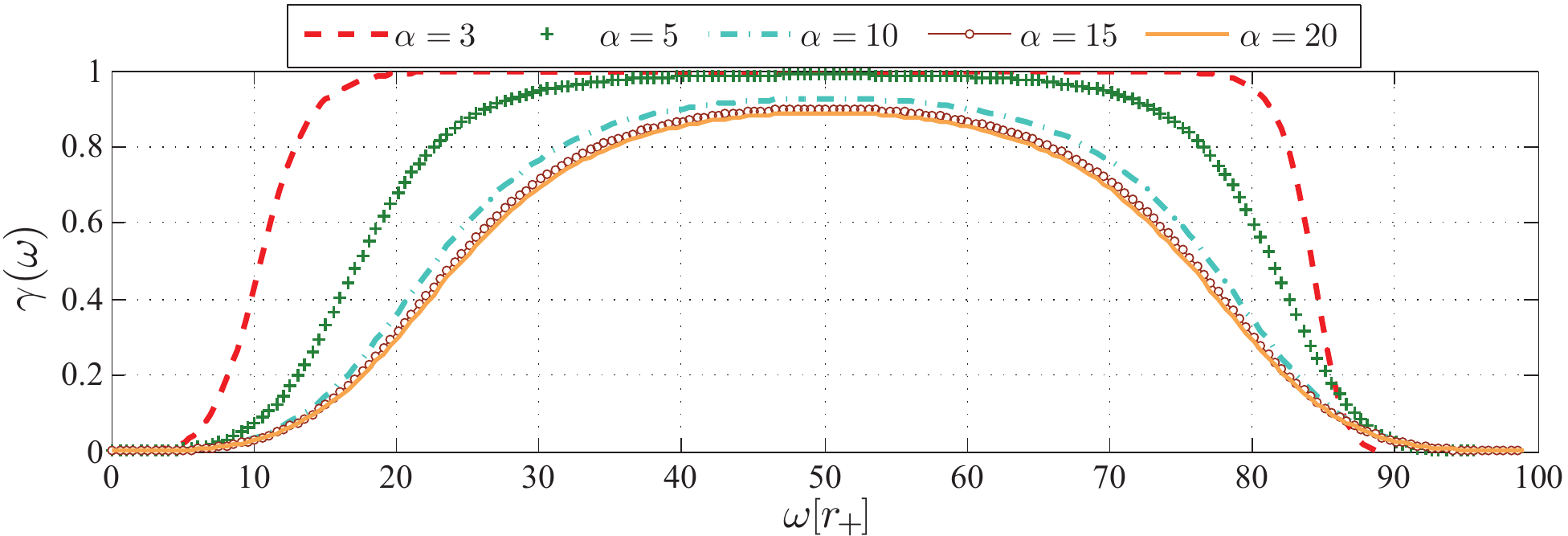}
\subfloat{(a) Spectrum of greybody factors in terms of $\omega [r_{+}]$.\label{fig_7a}}
\end{minipage}
\begin{minipage}[b]{1\textwidth}
\centering
\includegraphics[width=0.85\textwidth]{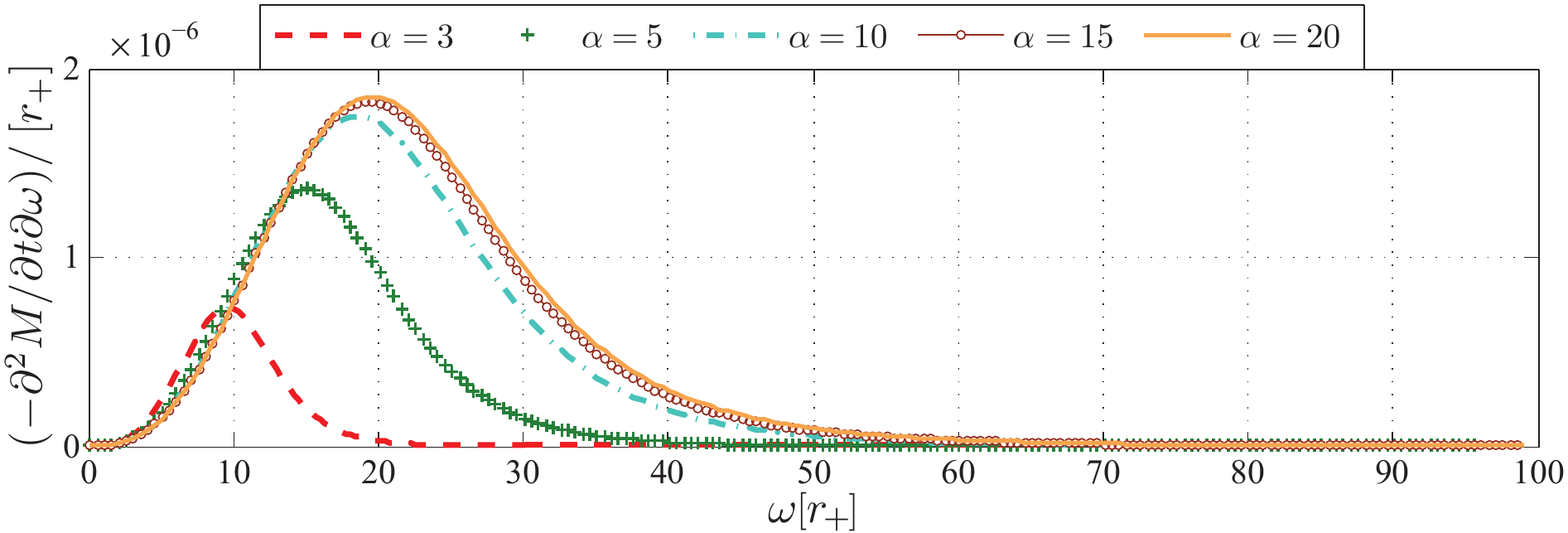}
\subfloat{(b) Spectrum of energy flux:
$F_{E}/[r_{+}]=-[r_{+}^{-1}]\partial M /\partial t$ in terms of
$\omega [r_{+}]$.\label{fig_7c}}
\end{minipage}
\begin{minipage}[b]{1\textwidth}
\centering
\includegraphics[width=0.85\textwidth]{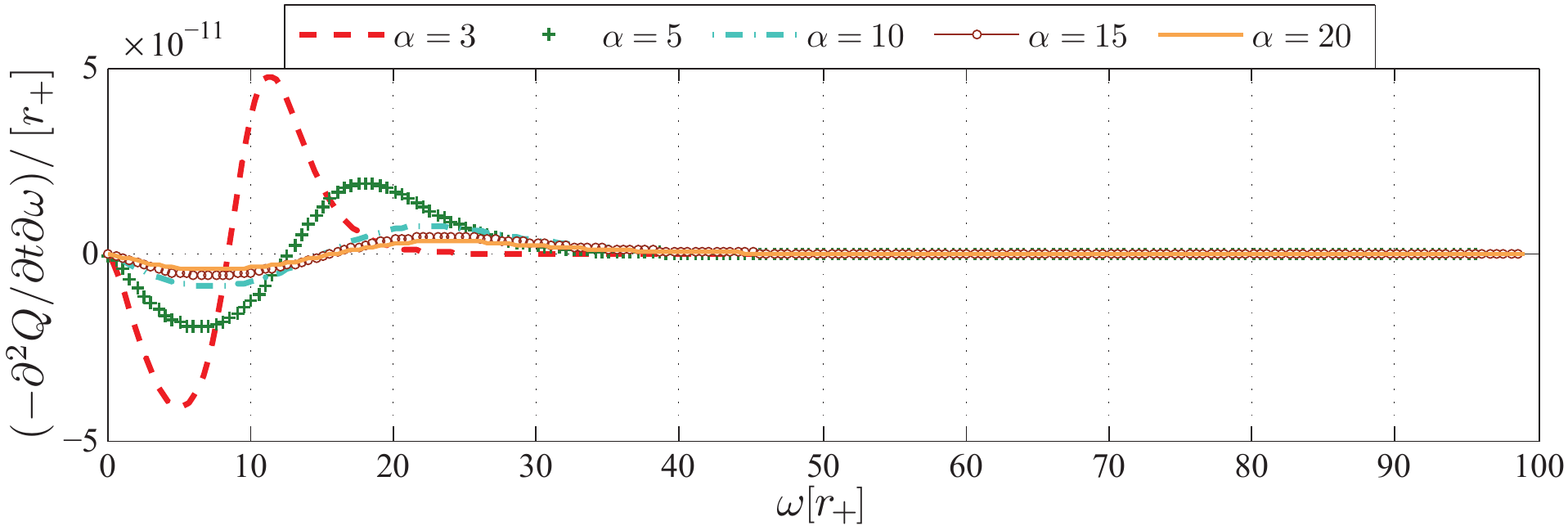}
\subfloat{(c) Spectrum of charge flux:
$F_{Q}/[r_{+}]=-[r_{+}^{-1}]\partial Q /\partial t$ in terms of
$\omega [r_{+}]$.\label{fig_7b}}
\end{minipage}
\begin{minipage}[b]{1\textwidth}
\centering
\includegraphics[width=0.85\textwidth]{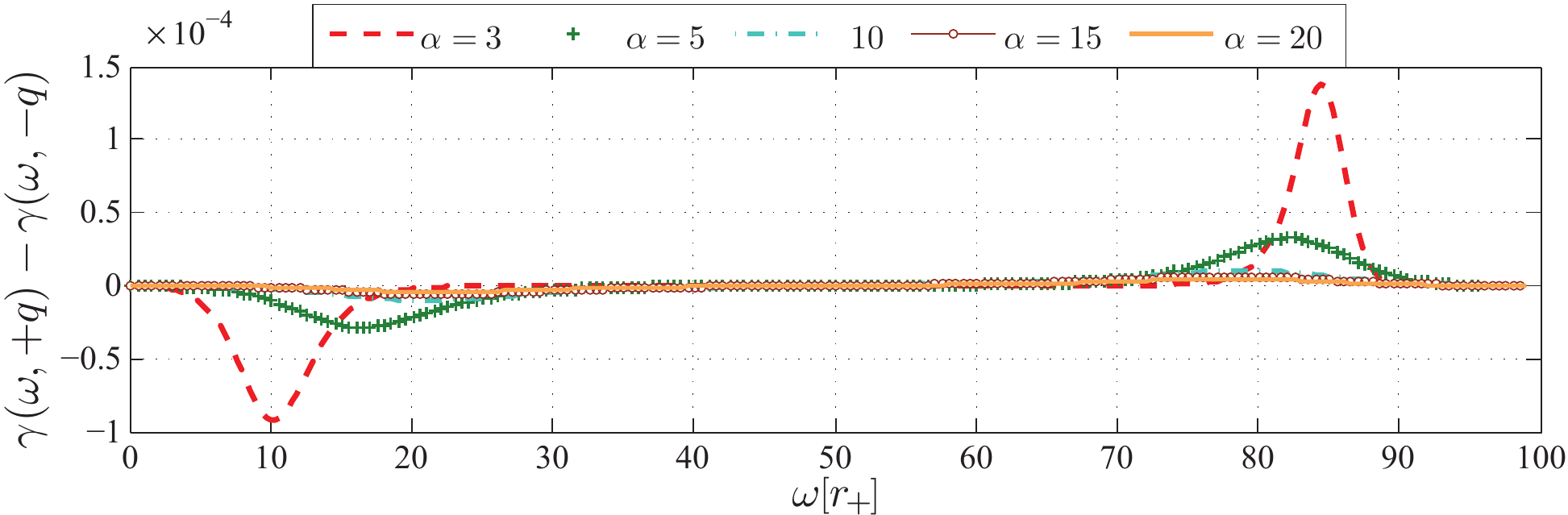}
\subfloat{(d) Subtraction of greybody factors for $qQ>0$ and $qQ<0$ in terms of $\omega [r_{+}]$.\label{fig_7d}}
\end{minipage}
\caption{\label{fig_7}Parameters are: $r_{-}/r_{+}=0.98$, $\alpha=3,5,10,15,20$, $\kappa=1$. Natural units and numerical values are $G=\hbar=c=4\pi\varepsilon_{0}=1$, $r_{+}=100$ and $q[r_{+}]=0.1$.}
\end{figure*}

It shows that the potential for the particles with the same signs as
the black hole ($qQ>0$) is higher than the potential when the two
signs are opposite ($qQ<0$). The result on the greybody factors can
be seen in  figure \ref{fig_6c} which shows that for low frequencies
($\omega r_{+}<<1$ or more precisely $\frac{\omega}{T_{H}}<<1$) the
greybody factors for the same sign particles are always lower and
hence less particles with the same sign are emitted at low
frequencies i.e. they do not have sufficient energy to escape the
barrier. The effect changes at high frequencies by including
backreaction correction which can be seen from the figure
\ref{fig_7d} that shows the difference of the  two greybody factors.
When a high energy particle with opposite sign  ($qQ<0$) is emitted,
the black hole gets closer to the  extremal limit and the height of
potential barrier for $\alpha>1$ sharply increases. This higher
potential barrier itself forces the emitted particle back toward the
black hole.

Despite the fact that potential barrier prevents more of charged
particles with same sign as the black hole than opposite sign
particles to pass through, creation rate of charged particles with
same sign to black hole due to thermal  radiation is more than
opposite sign particles. Hence, a competition arises between the
thermal  radiation and the greybody factors. At low energies
particles cannot easily pass through the potential barrier hence the
greybody factors will be dominant. On the other hand because high
energy particles can pass through the potential barrier more easily,
the thermal  Hawking radiation will be dominant at high frequencies.
As a result, at low frequencies sign of charge flux is opposite to
the black hole charge, while at high frequencies it is equal. Exact
calculations show that the total flux coming out of the black hole
is always with the same sign of the black hole.

Figures \ref{fig_1}, \ref{fig_8} and \ref{fig_9} show that as the
charge of black hole increases, the radiation  becomes more
sensitive to the change of the coupling constant. On the other hand,
at very low charges, the black hole behaves  similar to the
Reissner-Nordstr\"{o}m black hole. In the range
$0\leq\alpha<1/\sqrt{3}$ as the charge of black hole increases the
height of potential barrier decreases and the maximum  recedes from
the black hole resulting in the increase of   the greybody factors
and  the location of the peak of the power spectrum $\omega_{max}$
decreases. In the case $\alpha=1/\sqrt{3}$ the potential and
greybody factors and $\omega_{max}$ do not  change with the change
of the charge of the black hole.  When  $1/\sqrt{3}<\alpha\leq1$ the
potential grows and its maximum  moves toward the event horizon, the
greybody factors decrease and $\omega_{max}$ increases. At
$\alpha>1$ and approaching  the extremal limit, (\ref{eq3.9}) shows
that, the height of the potential grows indefinitely  and the
location of its maximum  $r_{max}$ approaches the event horizon, the
greybody factors decrease and $\omega_{max}$ increases. For a better
understanding these results are shown on table \ref{table_1}.
\begin{figure*}[htdp]
\includegraphics[width=1\textwidth]{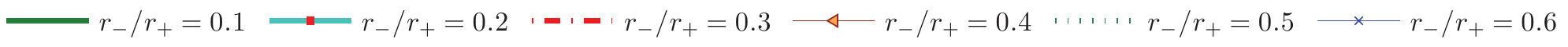}\\
\includegraphics[width=1\textwidth]{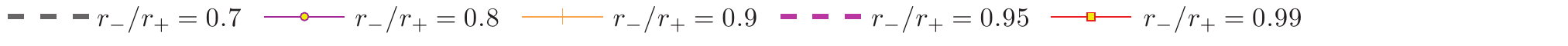}\\
\begin{minipage}[b]{1\textwidth}
\centering
\includegraphics[width=1\textwidth]{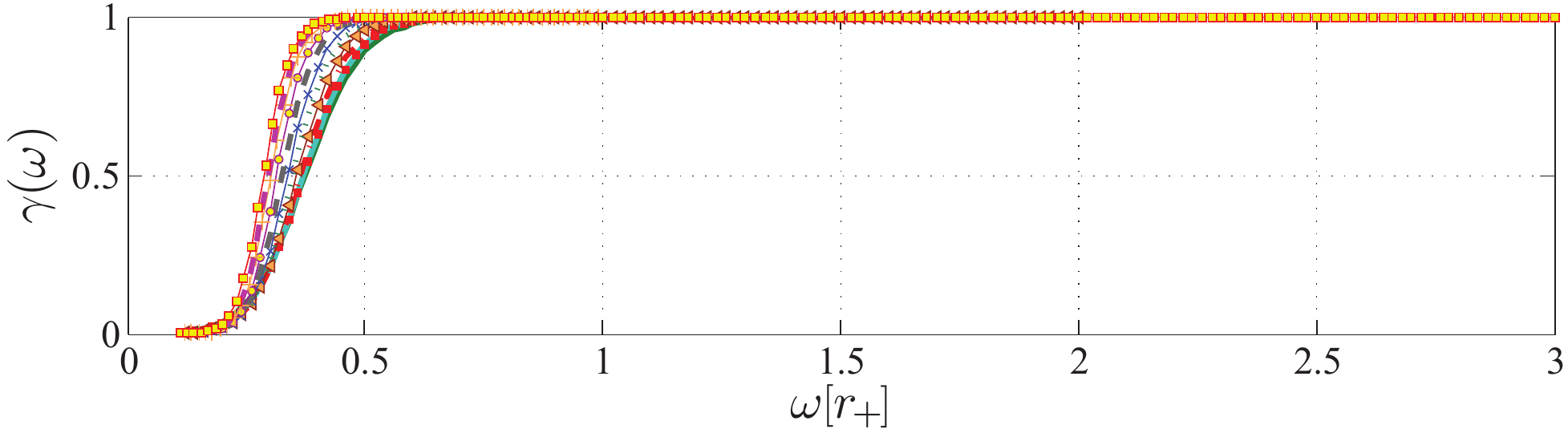}
\subfloat{(a) Greybody factors of black hole with $\alpha=0$.\label{fig_8a}}
\end{minipage}
\begin{minipage}[b]{1\textwidth}
\centering
\includegraphics[width=1\textwidth]{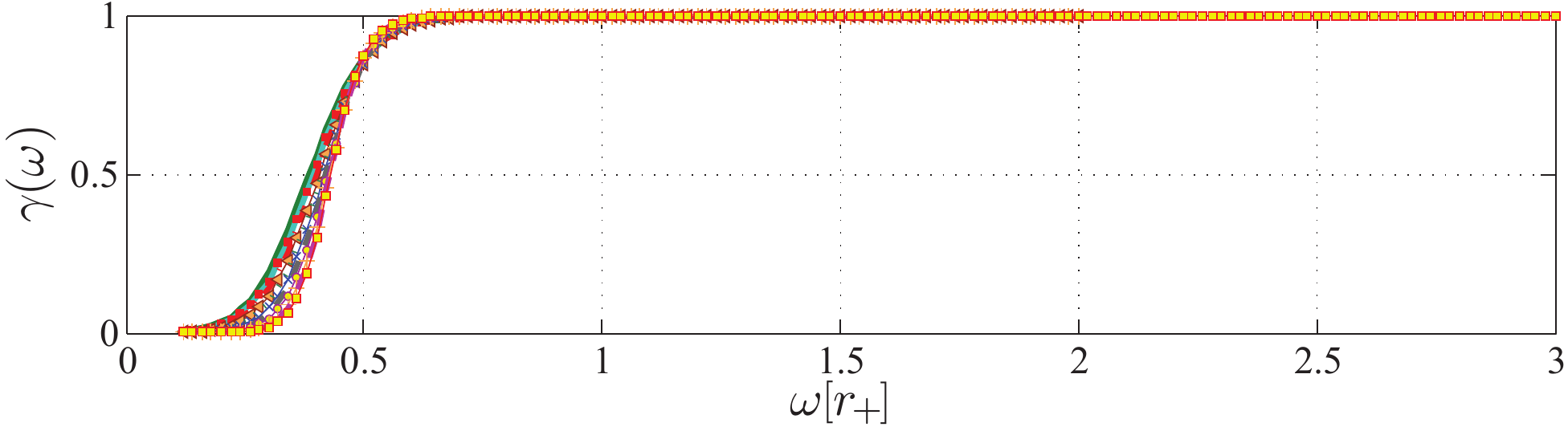}
\subfloat{(b) Greybody factors of black hole with $\alpha=1/\sqrt{3}$.\label{fig_8b}}
\end{minipage}
\begin{minipage}[b]{1\textwidth}
\centering
\includegraphics[width=1\textwidth]{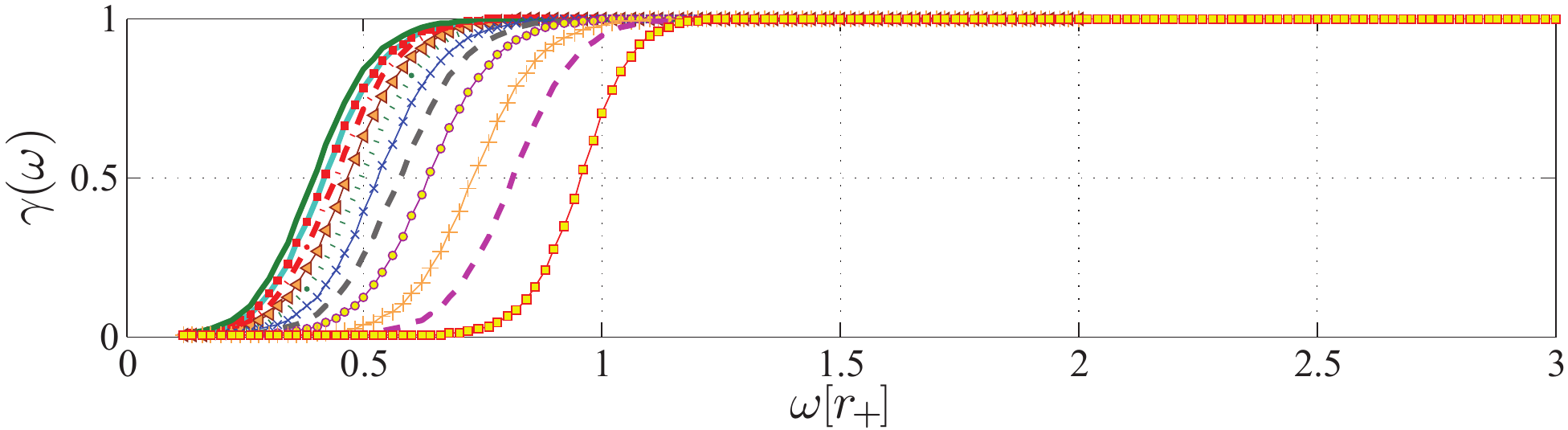}
\subfloat{(c) Greybody factors of black hole with $\alpha=1$.\label{fig_8c}}
\end{minipage}
\begin{minipage}[b]{1\textwidth}
\centering
\includegraphics[width=1\textwidth]{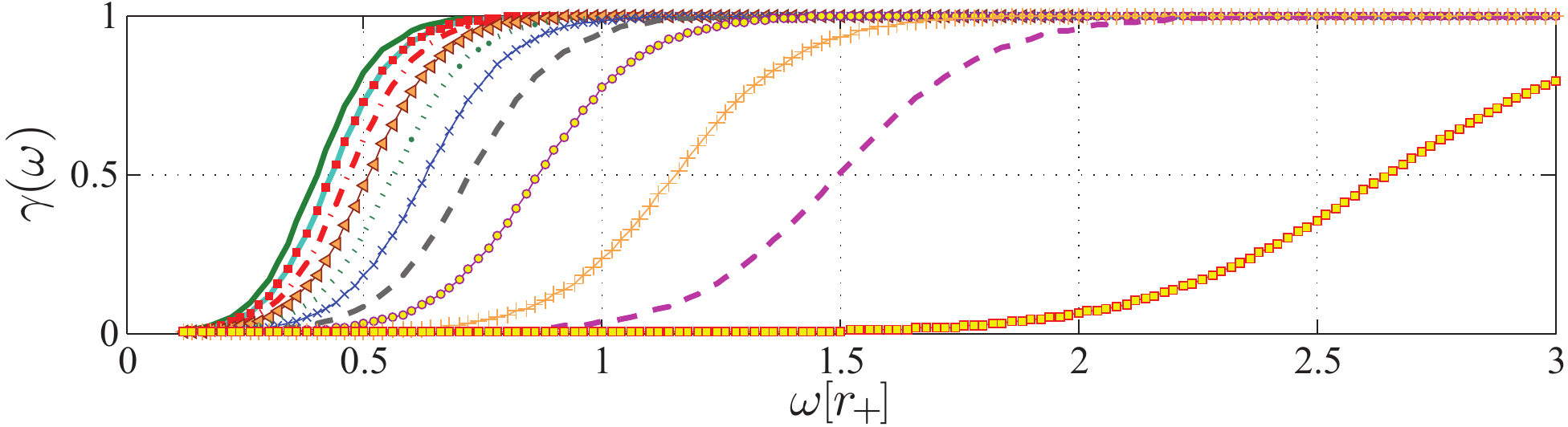}
\subfloat{(d) Greybody factors of black hole with $\alpha=1.4$.\label{fig_8d}}
\end{minipage}\\ 
\caption{}
\end{figure*}
\begin{figure*}[htdp]
\ContinuedFloat
\includegraphics[width=1\textwidth]{fig_8_legend1}\\
\includegraphics[width=1\textwidth]{fig_8_legend2}\\
\begin{minipage}[b]{1\textwidth}
\centering
\includegraphics[width=1\textwidth]{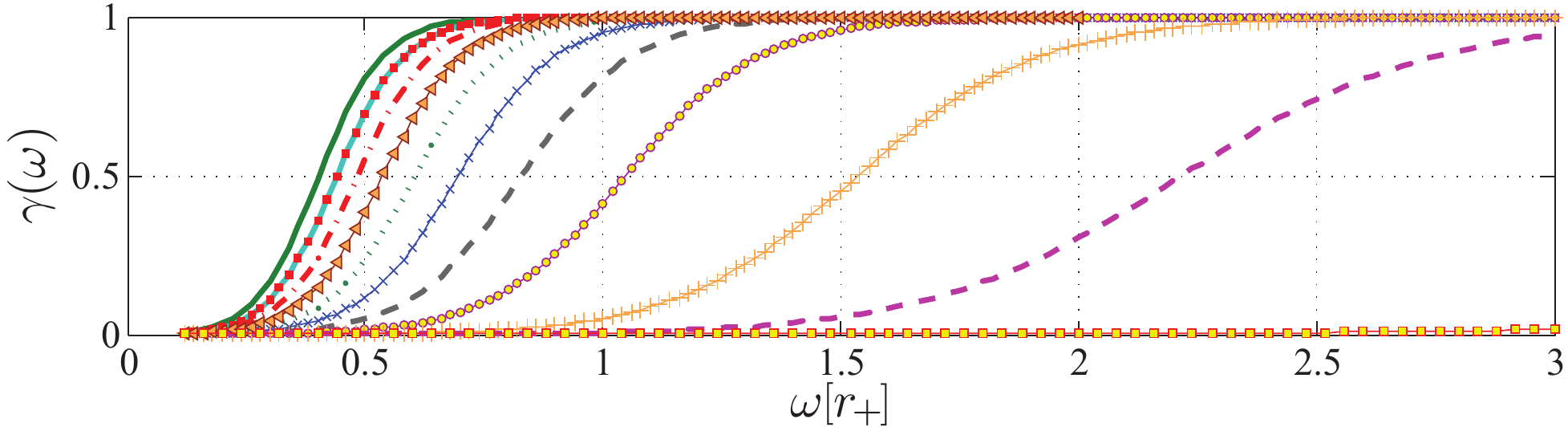}
\subfloat{(e) Greybody factors of black hole with $\alpha=\sqrt{3}$.\label{fig_8e}}
\end{minipage}
\caption{\label{fig_8}Greybody factors of dilaton black hole with
different values of $\alpha$ and charge
$r_{-}/r_{+}=(1+\alpha^{2})\frac{Q^{2}}{r_{+}^{2}}$ in terms of
$\omega [r_{+}]$. Spin $1/2$ particles with $\kappa=1$. Natural
units and numerical values are $G=\hbar=c=4\pi\varepsilon_{0}=1$ and
$r_{+}=100$.}
\end{figure*}
\begin{figure*}[htdp]
\includegraphics[width=1\textwidth]{fig_8_legend1}\\
\includegraphics[width=1\textwidth]{fig_8_legend2}\\
\begin{minipage}[b]{1\textwidth}
\centering
\includegraphics[width=1\textwidth]{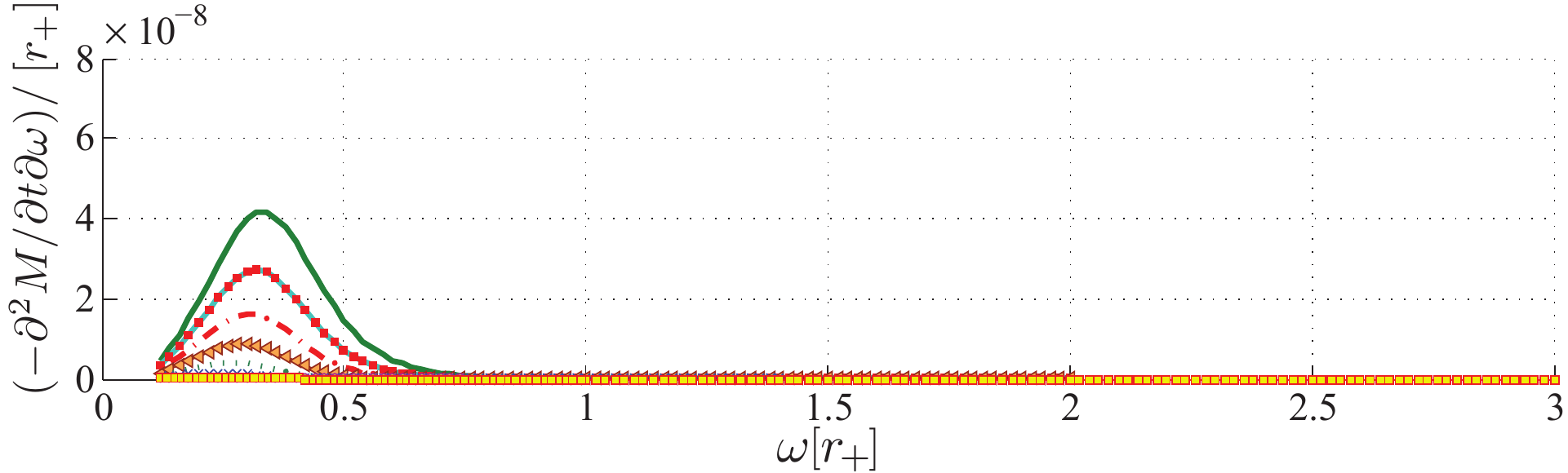}
\subfloat{(a) Energy evaporation rates (power spectrum) of black hole with $\alpha=0$.\label{fig_9a}}
\end{minipage}
\begin{minipage}[b]{1\textwidth}
\centering
\includegraphics[width=1\textwidth]{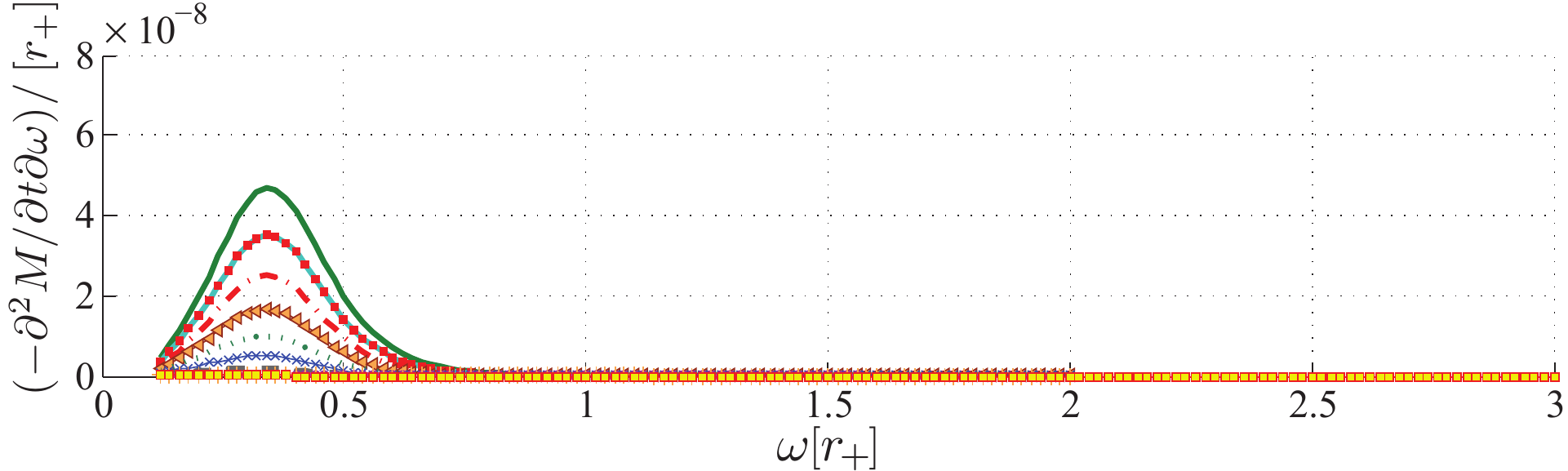}
\subfloat{(b) Energy evaporation rates (power spectrum) of black
hole with $\alpha=1/\sqrt{3}$.\label{fig_9b}}
\end{minipage}\\ \\
\caption{}
\end{figure*}
\begin{figure*}[htdp]
\ContinuedFloat
\includegraphics[width=1\textwidth]{fig_8_legend1}\\
\includegraphics[width=1\textwidth]{fig_8_legend2}\\
\begin{minipage}[b]{1\textwidth}
\centering
\includegraphics[width=1\textwidth]{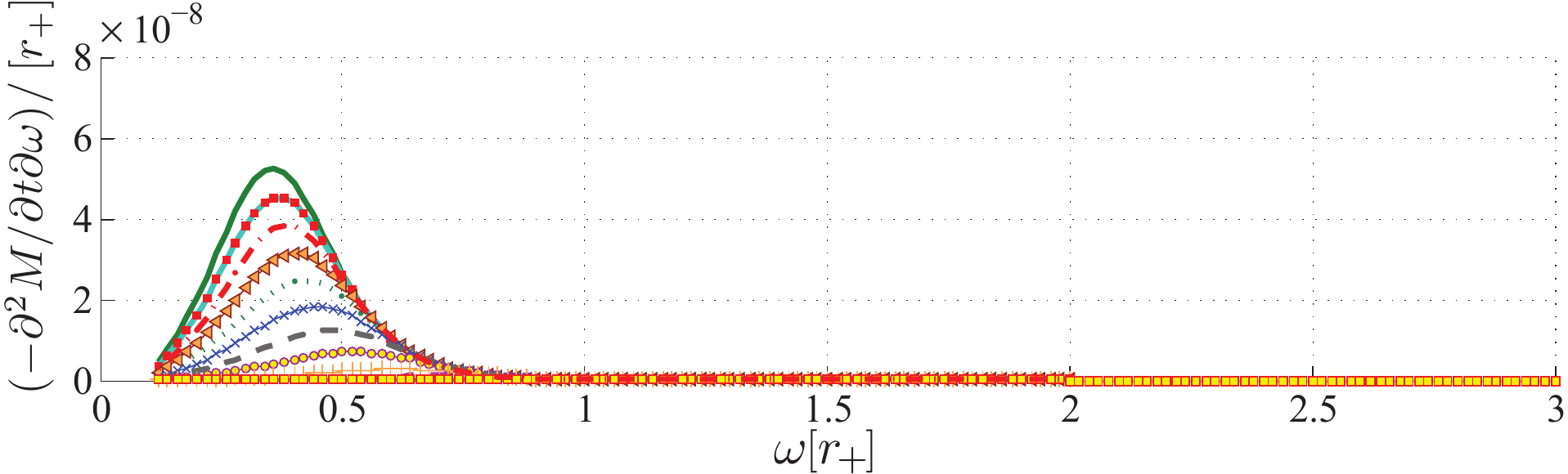}
\subfloat{(c) Energy evaporation rates (power spectrum) of black
hole with $\alpha=1$.\label{fig_9c}}
\end{minipage}
\begin{minipage}[b]{1\textwidth}
\centering
\includegraphics[width=1\textwidth]{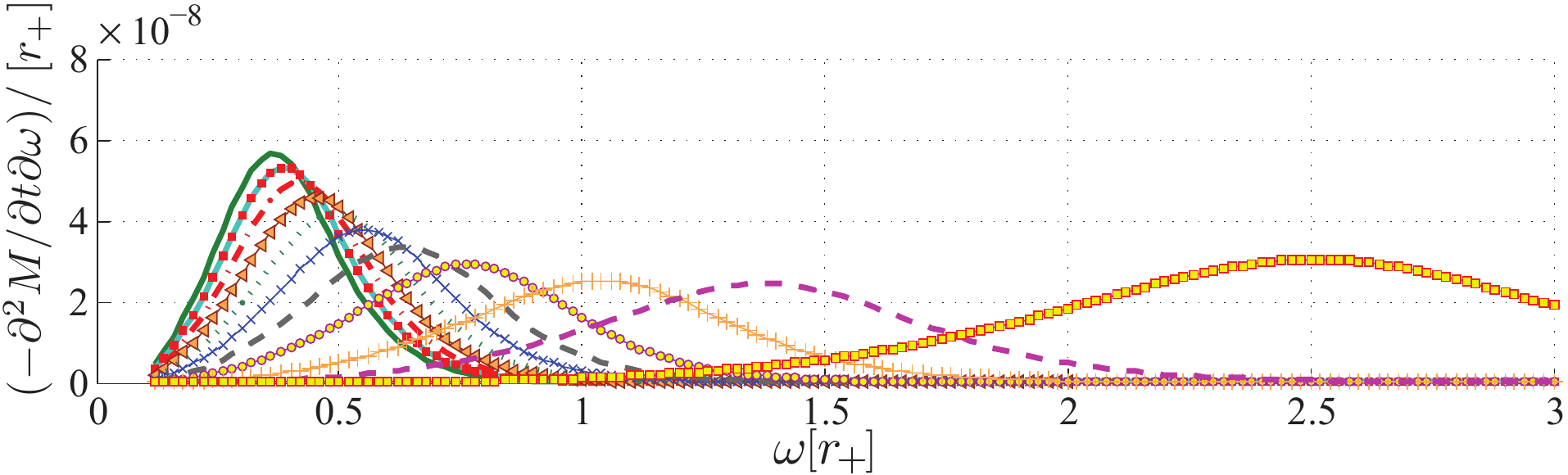}
\subfloat{(d) Energy evaporation rates (power spectrum) of black
hole with $\alpha=1.4$.\label{fig_9d}}
\end{minipage}
\begin{minipage}[b]{1\textwidth}
\centering
\includegraphics[width=1\textwidth]{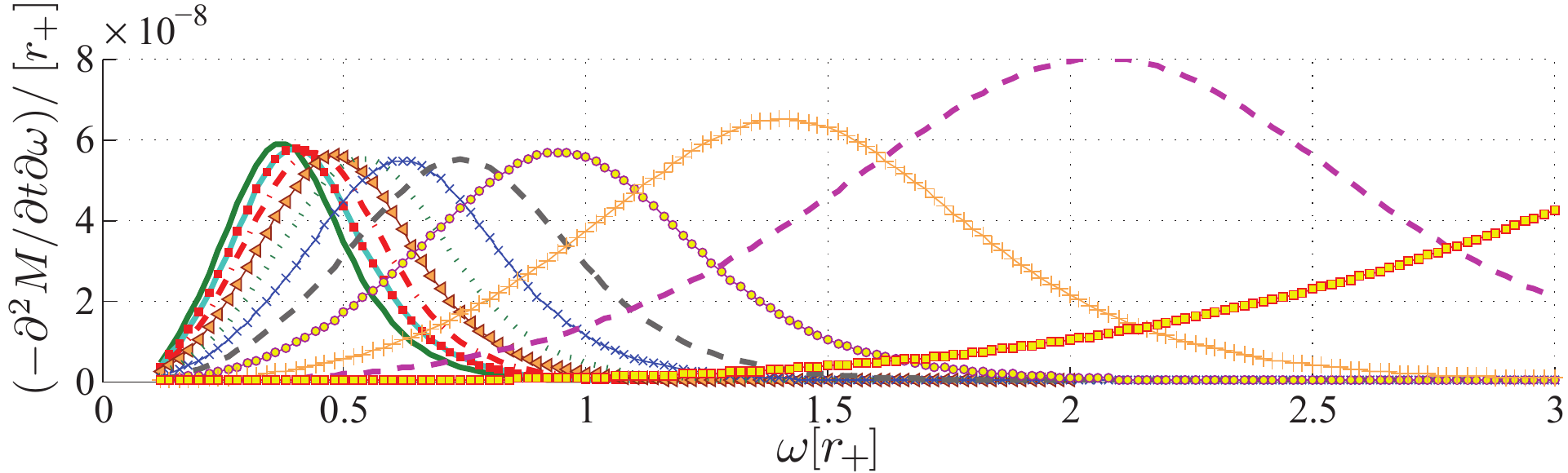}
\subfloat{(e) Energy evaporation rates (power spectrum) of black
hole with $\alpha=\sqrt{3}$.\label{fig_9e}}
\end{minipage}
\caption{\label{fig_9}Energy evaporation rates of dilaton black hole
with different values of $\alpha$ and charge
($r_{-}/r_{+}=(1+\alpha^{2})\frac{Q^{2}}{r_{+}^{2}}=0.1,...,0.99$)
in terms of $\omega [r_{+}]$. Spin $1/2$ particles with $\kappa=1$.
Natural units and numerical values are
$G=\hbar=c=4\pi\varepsilon_{0}=1$ and $r_{+}=100$.}
\end{figure*}
\begin{table}[t]
\caption{behaviour of maximum height $V_{max}$, location of maximum
point $r_{max}$ of effective potential, $\gamma(\omega)$, location
of maximum point of power spectrum ($\omega_{max}$) by increment of
BH charge
($\frac{r_{-}}{r_{+}}=(1+\alpha^{2})\frac{Q^{2}}{r_{+}^{2}}
\nearrow^{\ 1}$).} \label{table_1}
\begin{center}
\begin{tabular}{|c|c|c|c|c|}
\hline
& $0\leq \alpha<\frac{1}{\sqrt{3}}$ & $\alpha=\frac{1}{\sqrt{3}}$  & $\frac{1}{\sqrt{3}}<\alpha \leq 1$ & $\alpha>1$ \\
\hline
$V_{max}$ & $V_{max}\downarrow$ & Does not change & $V_{max}\uparrow$ & $V_{max}\nearrow^{\ \infty}$ \\
\hline
$r_{max}$ & $r_{max}\nearrow^{\frac{2}{1+\alpha^{2}} r_{+}}$ & Does not change & $r_{max}\searrow_{\frac{2}{1+\alpha^{2}} r_{+}}$ & $r_{max}\searrow_{\ r_{+}}$ \\
\hline
$\gamma(\omega)$ & $\gamma(\omega)\uparrow$ & Does not change & $\gamma(\omega)\downarrow$ & $\gamma(\omega)\downarrow$ \\
\hline
$\omega_{max}$ & $\omega_{max}\downarrow$ & Does not change & $\omega_{max}\uparrow$ & $\omega_{max}\uparrow$ \\
\hline
\end{tabular}
\end{center}
\end{table}

\begin{figure*}[htdp]
\begin{minipage}[b]{1\textwidth}
\centering
\includegraphics[width=1\textwidth]{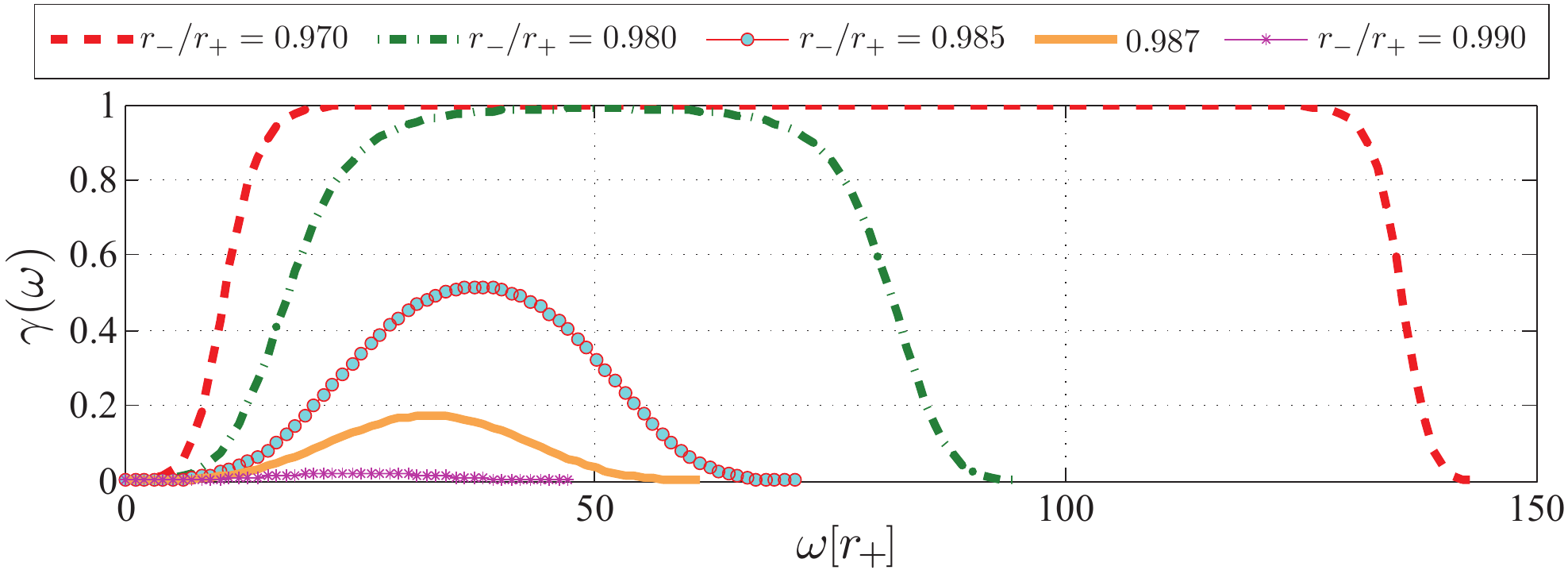}
\subfloat{(a)\label{fig_10a}}
\end{minipage}
\begin{minipage}[b]{1\textwidth}
\centering
\includegraphics[width=1\textwidth]{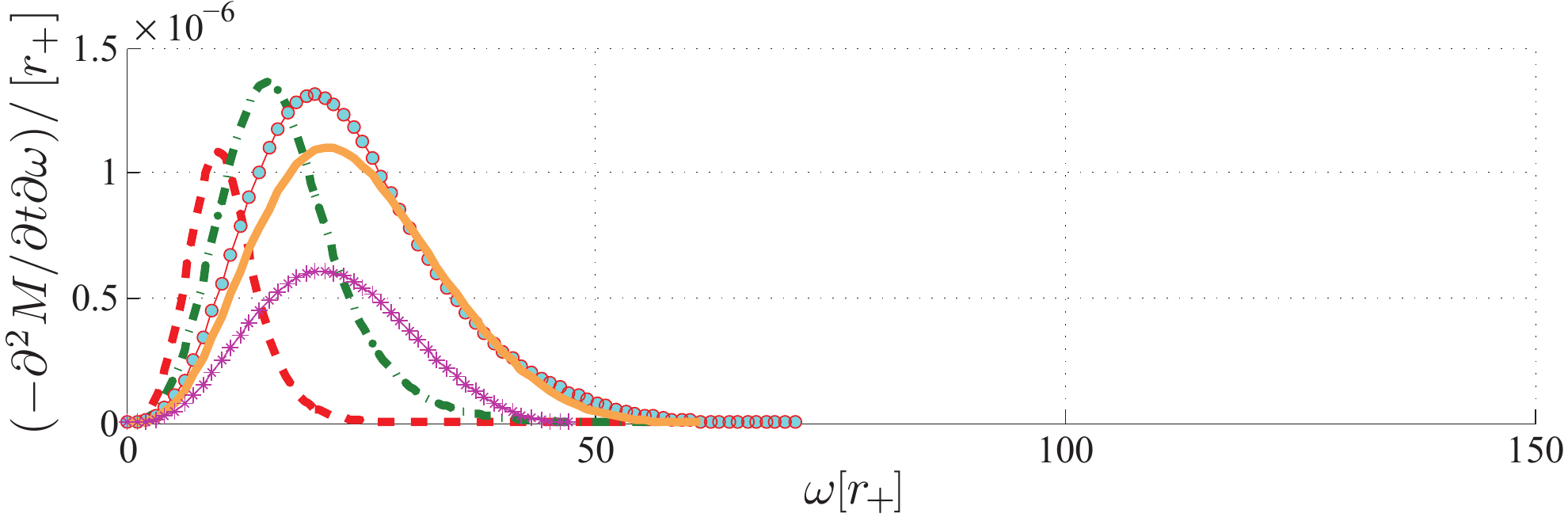}
\subfloat{(b)\label{fig_10b}}
\end{minipage}
\begin{minipage}[b]{1\textwidth}
\centering
\includegraphics[width=1\textwidth]{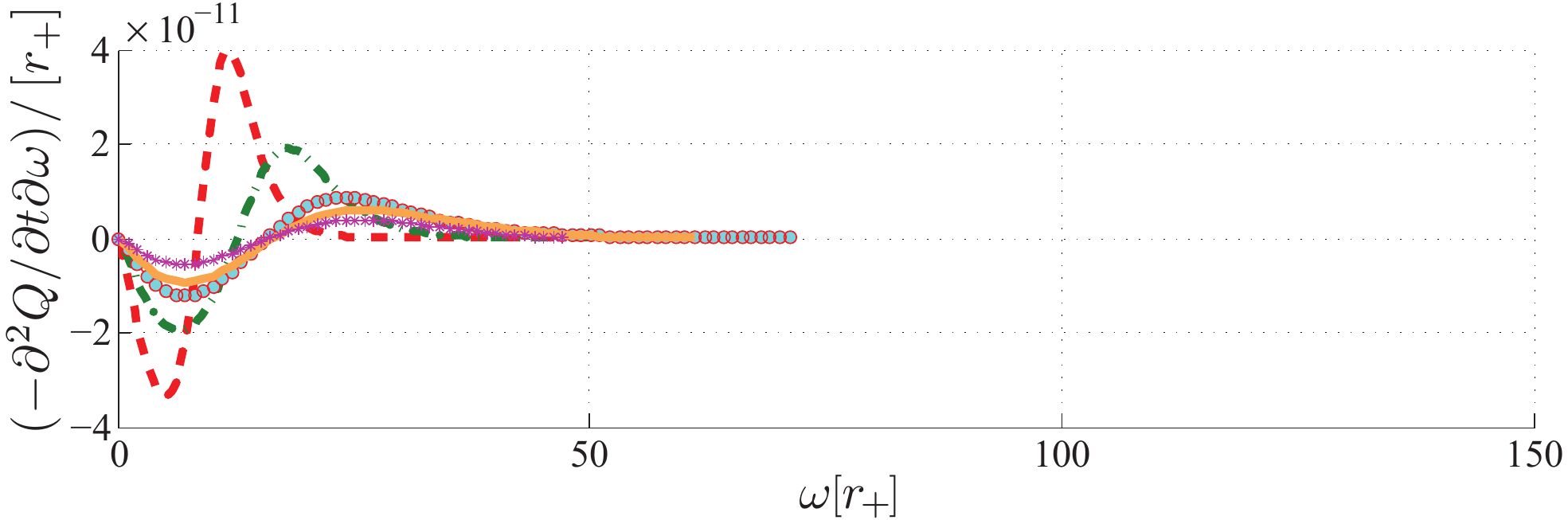}
\subfloat{(c)\label{fig_10c}}
\end{minipage}
\caption{\label{fig_10}Spectrum of greybody factors
$\gamma(\omega,\kappa)$, energy flux
$F_{E}/[r_{+}]=-[r_{+}^{-1}]\partial M /\partial t$ and charge flux
$F_{Q}/[r_{+}]=-[r_{+}^{-1}]\partial M /\partial t$ at the extremal
limit with different values of charge
($r_{-}/r_{+}=(1+\alpha^{2})\frac{Q^{2}}{r_{+}^{2}}=0.97,\ldots,0.99$)
and with $\alpha=5$, in terms of frequency $\omega [r_{+}]$.
Particles with $\kappa=1$. Natural units and numerical values are
$G=\hbar=c=4\pi\varepsilon_{0}=1$, $r_{+}=100$ and $q[r_{+}]=0.1$.}
\end{figure*}

The greybody factors as function of $\omega$, shows two low and high
cutoff frequencies; the low cutoff does not show much sensitivity to
different parameters including the charge. On the other hand the
high cutoff frequency $\omega_{HCF}$ shows strong dependence on
$\frac{r_{-}}{r_{+}}=(1+\alpha^{2})\frac{Q^{2}}{r_{+}^{2}}$, which
can be seen in figure \ref{fig_10a} and equation (\ref{eq4.4}) that
shows it decreases as the charge increases.

The change in the Hawking temperature, as the charge changes depends on the value of $\alpha$.

For $\alpha<1$, as $Q$ increases the Hawking temperature decreases
and approaches zero and the black hole cools, therefore the
radiation is decreased (figures \ref{fig_9a}, \ref{fig_9b} and table
\ref{table_2}).

For $\alpha=1$, as $Q$ increases the temperature $T_{H}$ doesn't
changes, greybody factors decrease, so evaporation rates decrease
(figures \ref{fig_8c}, \ref{fig_9c} and table \ref{table_2}).

For $\alpha>1$ at small $Q$, the behaviour is similar to that of RN
black hole. But as $Q$ increases the Hawking temperature diverges
and the black hole becomes hot, so the evaporation rates grows
significantly with increase in the BW (bandwidth: the range covered
between the two points where the evaporation rate is half of its
maximum). FCFS (frequency of change in flux sign, i.e. the frequency
where the sign of the charge flux is reversed, $\omega_{FCFS}$:
$\left. -\frac{\partial^{2} Q}{\partial t \partial \omega}
\right|_{\omega_{FCFS}}=0$) also grows with the charge $Q$ (figures
\ref{fig_9d}, \ref{fig_9e}, \ref{fig_10c} and table \ref{table_2}).
However near the extremal limit due to presence of high cutoff
frequency $\omega_{HCF}$ the process slows and the emission rates
and bandwidth get suppressed (figure \ref{fig_10} and table
\ref{table_2}).
\begin{table}[t]
\caption{behaviour of  $T_{H}$, $-\frac{\partial M}{\partial t}$,
$-\frac{\partial Q}{\partial t}$, $\omega_{HCF}$, BW of
$-\frac{\partial M}{\partial t}$ and $-\frac{\partial Q}{\partial
t}$, and $\omega_{FCFS}$ by increment of BH charge
($\frac{r_{-}}{r_{+}}=(1+\alpha^{2})\frac{Q^{2}}{r_{+}^{2}}
\nearrow^{\ 1}$).}

\label{table_2}
\begin{center}
\begin{tabular}{|c|c|c|c|c|}
\hline
- & $0\leq \alpha<1$ & $\alpha=1$  & $\alpha>1$ \\
\hline
$T_{H}$ & $T_{H}\searrow_{\ 0}$ & Does not change & $T_{H}\nearrow^{\ \infty}$ \\
\hline
$-\frac{\partial M}{\partial t}$ & $-\frac{\partial M}{\partial t}\searrow_{\ 0}$ & $-\frac{\partial M}{\partial t}\downarrow$ & First(low charge) $-\frac{\partial M}{\partial t}\downarrow$ then $-\frac{\partial M}{\partial t}\uparrow$ \\
\hline
$-\frac{\partial Q}{\partial t}$ & $-\frac{\partial Q}{\partial t}\searrow_{\ 0}$ & $-\frac{\partial Q}{\partial t}\downarrow$ & First(low charge) $-\frac{\partial Q}{\partial t}\downarrow$ then $-\frac{\partial Q}{\partial t}\uparrow$ \\
\hline
$\omega_{HCF}$ & - & - & $\omega_{HCF}\searrow_{\ q/\sqrt{1+\alpha^{2}}}$ \\
\hline
Extremal limit & - & - & $-\frac{\partial Q}{\partial t}\rightarrow 0$, $-\frac{\partial M}{\partial t}\rightarrow 0$ \\
\hline
BW & BW$\searrow_{\ 0}$ & Does not change & BW$\uparrow$ (But at extremal limit BW$\rightarrow 0$) \\
\hline
$\omega_{FCFS}$ & - & - & $\omega_{FCFS} \uparrow$ \\
\hline
\end{tabular}
\end{center}
\end{table}

\subsection{\label{Effect of dilaton coupling constant}Effect of dilaton coupling constant}

In the previous discussion on the effect of charge we have also
discussed some of the effects of change in $\alpha$. In this part we
look at changes in other quantities such as Hawking temperature, the
emission rates, cutoff frequencies, $\ldots$ as the dilaton coupling
changes.

The special point $\alpha=1/\sqrt{3}$ which showed importance for
charge dependence does not show any particular relevance for other
quantities. On the other hand the particular point $\alpha=1$ marks
serious changes in the behaviour of most of the quantities
mentioned.

\begin{figure*}[htdp]
\centering
\includegraphics[width=0.9\textwidth]{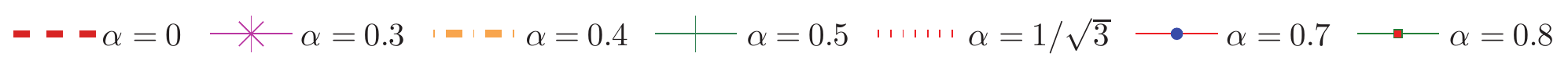}\\
\includegraphics[width=0.9\textwidth]{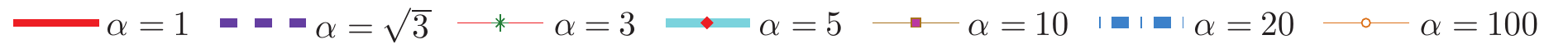}\\
\begin{minipage}[b]{1\textwidth}
\centering
\includegraphics[width=0.9\textwidth]{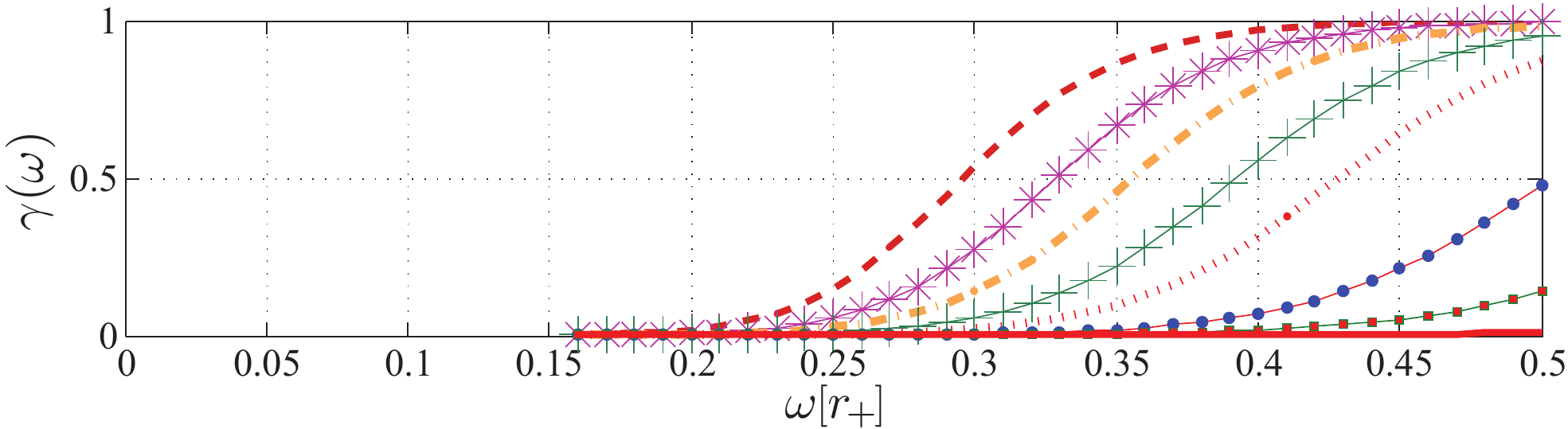}
\subfloat{(a) Greybody factors for $\alpha=0,...,1$.\label{fig_11a}}
\end{minipage}
\begin{minipage}[b]{1\textwidth}
\centering
\includegraphics[width=0.9\textwidth]{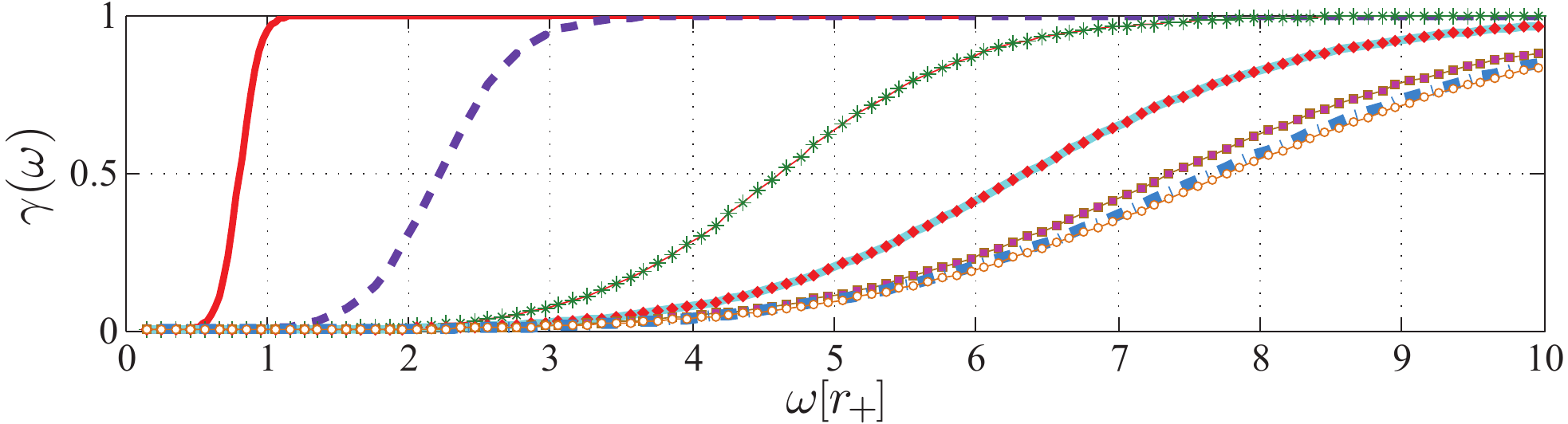}
\subfloat{(b) Greybody factors for $\alpha=1,...,100$.\label{fig_11b}}
\end{minipage}
\begin{minipage}[b]{1\textwidth}
\centering
\includegraphics[width=0.9\textwidth]{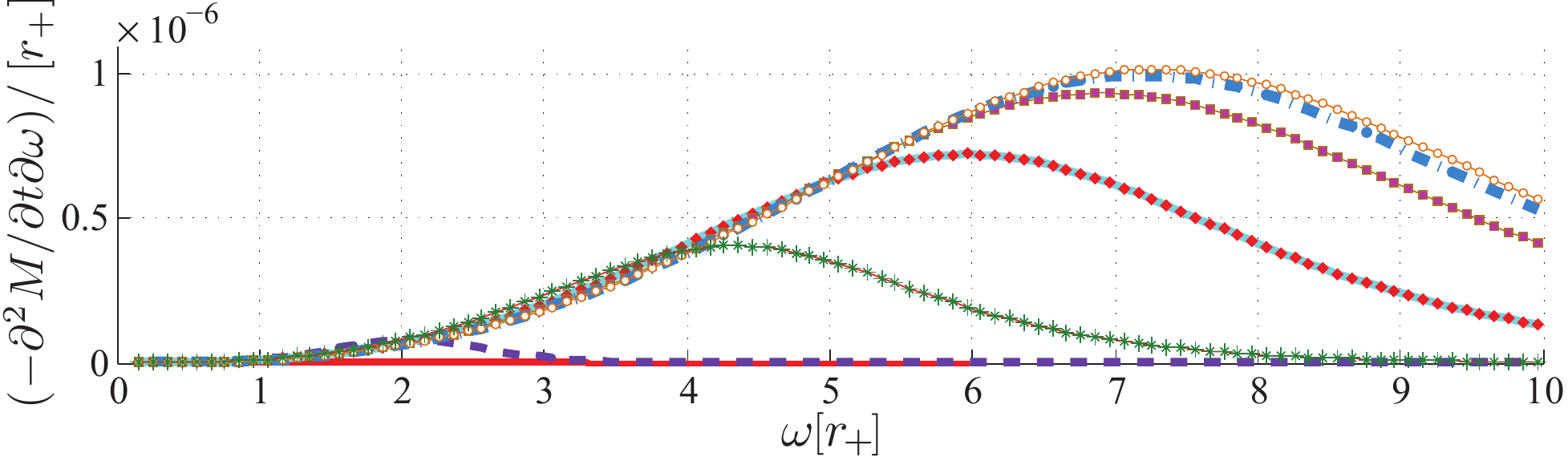}
\subfloat{(c) Energy fluxes $F_{E}/[r_{+}]=-[r_{+}^{-1}]\partial M/\partial t$.\label{fig_11c}}
\end{minipage}
\begin{minipage}[b]{1\textwidth}
\centering
\includegraphics[width=0.9\textwidth]{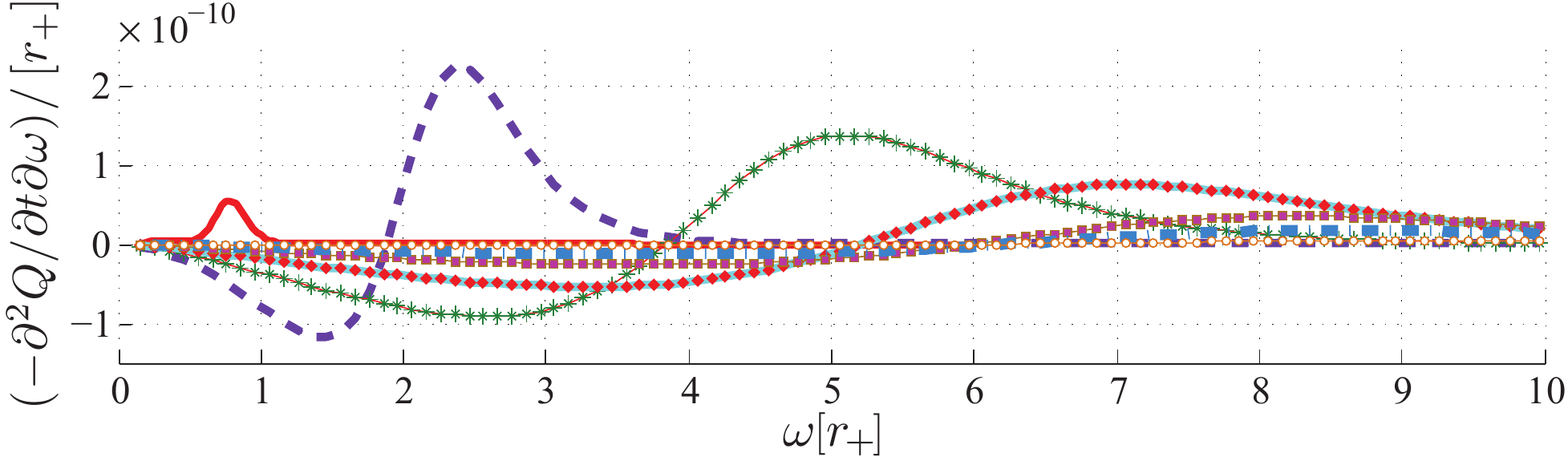}
\subfloat{(d) Charge fluxes $F_{Q}/[r_{+}]=-[r_{+}^{-1}]\partial Q/\partial t$.\label{fig_11d}}
\end{minipage}
\caption{\label{fig_11}Greybody factors (a), (b) and energy (c) and
charge (d) evaporation rates of dilaton black hole with different
values of $\alpha$ with $r_{-}/r_{+}=0.95$ in terms of frequency
$\omega [r_{+}]$. Spin $1/2$ particles with angular momentum
$\kappa=1$. Natural units and numerical values are
$G=\hbar=c=4\pi\varepsilon_{0}=1$, $r_{+}=100$ and $q[r_{+}]=0.1$.}
\end{figure*}

As the coupling constant increases,  the height of the potential
barrier shown in figure \ref{fig_1} increases along with a decrease
in its width, which becomes considerable  near extremal limit for
$\alpha>1$. In this case upon approaching the  extremal limit, the
height of the potential barrier grows indefinitely, and hence the
greybody factors plotted in figures \ref{fig_11a} and \ref{fig_11b}
decrease. In  the limit of very large  $\alpha$ the dependence on
$\alpha$ disappears (figures \ref{fig_7} and \ref{fig_11}). With
increase in $\alpha$, the Hawking temperature of black hole
increases. Hence the energy evaporation rate and its bandwidth shown
in figure \ref{fig_11c} increase and tends to a constant rate. Also
the charge flux in figure \ref{fig_11d} increases with  $\alpha$,
but as discussed in section \ref{Evolution and fate of dilaton black
hole} for $\frac{e}{\omega}\frac{1}{\sqrt{1+\alpha^{2}}}<<1$ it
decreases and tends to zero at the limit
$\frac{e}{\omega}\frac{1}{\sqrt{1+\alpha^{2}}}\rightarrow 0$. The
figure \ref{fig_7c} shows the charge flux for different values of
$\alpha$; the change in the flux sign $\omega_{FCFS}$,  occurs at a
frequency which grows with $\alpha$ due to increase of temperature.
Table \ref{table_3} gives a picture of the behaviour of the
quantities with change in $\alpha$.
\begin{table}[t]
\caption{behaviour of $T_{H}$, $V_{max}$, width of potential,
$\gamma(\omega)$, $-\frac{\partial M}{\partial t}$, $-\frac{\partial
Q}{\partial t}$, bandwidth(BW) of $-\frac{\partial M}{\partial t}$
and $-\frac{\partial Q}{\partial t}$ and frequency of change in flux
sign($\omega_{FCFS}$) by increment of coupling constant $\alpha$.}
\label{table_3}
\begin{center}
\begin{tabular}{@{}|c|c|c|c|c|}
\hline
& $\alpha \uparrow$\\
\hline
$T_{H}$ & $T_{H}\uparrow$\\
\hline
$V_{max}$ & $V_{max}\uparrow$\\
\hline
Width of potential & Decrease\\
\hline
$\gamma(\omega)$ & $\gamma(\omega)\downarrow$\\
\hline
$-\frac{\partial M}{\partial t}$ & $-\frac{\partial M}{\partial t}\uparrow$\\
\hline
$-\frac{\partial Q}{\partial t}$ & $-\frac{\partial Q}{\partial t}\uparrow$ (But at the $\frac{1}{\sqrt{4\pi \varepsilon_{0}G}}\frac{e}{m_{e}} \frac{1}{\sqrt{1+\alpha^{2}}} \rightarrow 0$  limit tends to $-\frac{\partial Q}{\partial t}\rightarrow 0$)\\
\hline
BW & BW$\uparrow$ \\
\hline
$\omega_{FCFS}$ & $\omega_{FCFS}\uparrow$ \\
\hline
\end{tabular}
\end{center}
\end{table}

\subsection{Effects of mass, angular momentum, and the statistic of the emitted particle  }

Again the quantity to study is the height of the potential that has
strong effect on the greybody factors. Equation (\ref{eq3.7}) shows
the height of potential barrier grows as the angular momentum of
particle increases. Increase of the angular momentum $\kappa$,
causes considerable decrease in the energy and charge fluxes. This
is why  it is  rational to consider only the lowest value of the
angular momentum,  in the case of our interest $\kappa=1$.

Equation (\ref{eq4.10}) shows that the mass of the particle affects
the low frequency spectrum of greybody factors. Particles with
energy less than the rest mass $\omega \leq m$ are not real and will
never reach the infinity and hence the relevant greybody factors
vanish and with that, the energy and charge fluxes.

The effective potential obtained for spin 1/2 particles behaves like
the scalar particle potential which is mentioned in section~\ref{A
short review of dilaton black hole, greybody factors and Hawking
radiation} and calculated in \cite{Holzhey:1991bx}. For the class
$\alpha<1$, at the event horizon the potential barrier vanishes.
This barrier has a finite maximum tending to zero by increasing
radius. For the case $\alpha=1$, the height of potential barrier
near the extremal limit remains finite. For $\alpha>1$, in the
extremal limit the height of potential barrier diverges while its
position approaches the event horizon. For non-extremal cases the
height of potential barrier is finite, but its peak (as it tends to
extremal limit) grows as like as
$(r_{+}-r_{-})^{-2(\alpha^{2}-1)/(\alpha^{2}+1)}$.

Figure \ref{fig_2} shows that the height and width of the effective
potentials for scalars are smaller than effective potentials for
fermions and greybody factors and hence evaporation rates for
scalars are always greater than greybody factors and evaporation
rates for fermions.

However, the overall behaviour of dilaton black hole for scalar
particles on increasing or decreasing parameters $\alpha$ and $Q$ is
similar to fermions, the scalar particles dominate over the fermions
on determining the fate of dilaton black hole.

\section{\label{Conclusion}Conclusion}

In this paper we have calculated the greybody factors for charged
spin 1/2 particles in the dilaton black hole background and
investigated its effects. First it was done using  the  Rosen-Morse
potential and WKB approximation. The advantage of these methods are
that they gave us a good general formula that described the
behaviour of greybody factors albeit approximately. In calculation
of evaporation rates by Rosen-Morse method errors become
significant. To avoid these errors  we performed more accurate
numerical computations. Although qualitatively,  and for most of the
time even quantitatively these analytic computations are  reliable.
In the calculation of evaporation  rates semi-classical
approximation was employed. For better accuracy we applied
back-reaction correction.  This correction, adiabatically brought
the dynamics into the solution which shows a considerable effect on
high frequency spectrum of the greybody factors especially  near
extremal limit. Consideration of the backreaction caused highly
non-trivial picture which specially revealed new phenomenon of
evaporating to extremal limit as fate of certain  dilaton black
holes. We obtained a cutoff frequency for greybody factors at high
frequency for $\alpha>1$. Hence greybody factors act like filter
i.e. filter for  low and high frequencies, like a band-pass filter.
With the increase of the charge of the black hole high cutoff
frequency $\omega_{HCF}$ decreases, while low cutoff frequency
$\omega_{LCF}$ increases. As a result in this limit the black hole
stops radiating.

We obtained that for $\alpha>1$ due to the growth and divergence of
the temperature and height of the potential barrier of the black
hole, a competition between these two arises. For low charges the
greybody factors are dominant (as the charge of black hole increases
the evaporation rate decreases) while for large $Q$ the temperature
takes over (as the charge of the black hole increases the
evaporation rates increases). Also increase of the coupling constant
acts in favor of the temperature; as $\alpha$ increases the onset of
the increase of the evaporation rate moves to  lower charges. This
effect does not last forever, at the extremal limit due to the
existence of high cutoff frequency the radiation become suppressed.
Divergence of low cutoff frequency and decrease in high cutoff
frequency at the extremal limit turn off the Hawking radiation of
the black hole. In this situation from viewpoint of an infinite
observer the space-time around the event horizon of the black hole
extinguishes  the Hawking radiation of dilaton black
 hole.

We also obtained another difference that distinguishes dilaton black
hole from other black holes. Because the charge evaporation rate is
always faster than the energy evaporation rate normal black holes
always lose their charge and become neutral, and then, similar to
the Schwarzschild black hole  radiate their energy away and
disappear. But for the dilaton black holes we have found unexpected
result that fate of the black hole depends on different variables,
$Q/M$, $\alpha$, $mM$, and $\alpha_{0}=q/m$ charge over mass ratio
of the emitted particle. A transition line separates the two regions
in the $(Q/M, \alpha )$ plane.

For the first approximation, we obtained this transition line to be
$Q/M=\alpha_{0}^{2}/\sqrt{\alpha^{2}+1}$ for
$\alpha>\sqrt{\alpha_{0}^{2}-1}$. This approximation gave an upper
bound for $\xi$ for small black holes ($8\pi m M<1$). More presice
solution of this transition line shows that there is a small area of
extremal regime for $\alpha < \sqrt{\alpha_{0}^{2}-1}$. Precise
solution shows that for $\alpha < \alpha_{0} \sqrt{8\pi mM}$ a tiny
part of the allowed region  belongs to the extremal regime and
almost all of it belongs to the decay regime. In contrast for $
\alpha >\alpha_{0}\sqrt{8\pi mM}$ a much larger part of the allowed
space belongs to the extremal regime. Therefore for small black
holes, in practice $\alpha_{0}$ marks a critical value for $\alpha$
below which almost all the conditions result in total evaporation
and after which most of the parameter space belongs to the extremal
regime.

In the extremal regime the final state of the black hole  is a
stable situation were it acts like an elementary particle
\cite{Holzhey:1991bx}. Incidently for this region the background
metric become flat as the temperature tends to infinity and its
entropy vanishes; stabilizing the black hole which makes it  more
identifiable with an elementary particle.

During the radiation process the greybody factors always tend to add
charge to the black hole in contrast with Hawking radiation spectrum
that always tends to discharge it. Now the question is why this
competition is always in the interest of the Hawking radiation and
not the greybody factors. The answer to this question is that the
gravity  where  the Hawking radiation originates, near the event
horizon is stronger than the effect  of the  greybody factors which
have their source  outside the event horizon.

The fate of an extremal dilaton black hole may differ from the
Reissner-Nordstr\"{o}m black hole, in the $\alpha=1$ case, since the
role of the two horizons in the charged dilaton black hole are very
different. Evolution of rotating dilaton black holes and
corresponding transition line  may differ from the non rotating case
and needs a separate analysis. It may be possible that at large
$\alpha $ we may be able to consider a class of these black holes as
elementary particles;  being stable and  possesing definite mass,
spin and charge. Throughout  this work we have always considered
$r_{-} < r_{+} $. The case for $r_{-} > r_{+}$ is very different and
also needs its own particular analysis. These questions are under
investigation by the authors.

\ack
We would like to thank Navid Abbasi for fruitful discussions and
reading the manuscript.

\appendix
\setcounter{section}{1}
\section*{\label{Appendix}Appendix}


In this appendix charged massive Dirac equation in the background of
most general spherically symmetric static black hole is changed into
Schr\"{o}dinger like equation by appropriate change of variables.
The methods developed in
\cite{Chen:2007dja,Cho:2003qe,Cho:2004wj,Cho:2007zi,Jing:2003wq,Jing:2004zb,R:2008ub,Shu:2004fj,Wang:2009hr}
are employed.

Consider the most general spherically symmetric static metric
\begin{equation}
ds^{2}=A(r)^{2}dt^{2}-B(r)^{2} dr^{2}-C(r)^{2}d\Omega^{2}.
\end{equation}

Dirac equation with charge $q$ and mass $m$ in the background metric
is given by\cite{Cho:2003qe,Jing:2003wq},
\begin{equation}
\left(i\gamma^{\mu}D_{\mu}-m \right) \Psi=0.
\end{equation}
where
\begin{equation}
D_{\mu}=\partial_{\mu}+\Gamma_{\mu}-iqA_{\mu},
\end{equation}
where $A_{\mu}$ is the Maxwell field and $\Gamma_{\mu}$ the spin
connection defined by
\begin{equation}
\Gamma_{\mu}=\frac{1}{8}\left[\gamma^{a},\gamma^{b}\right] e_{a}^{\nu} e_{b\nu;\mu},
\end{equation}
$e^{a}_{\mu}$ is the tetrad,
\begin{equation}
e^{a}_{\mu}=diag\left(A(r),B(r),C(r),C(r)\sin\theta \right).
\end{equation}

Contraction $\gamma^{\mu} \Gamma_{\mu}$ in Dirac equation gives
\begin{eqnarray}
\gamma^{\mu} \Gamma_{\mu} = \gamma^{r} \frac{1}{4B(r)} \left( \frac{1}{A(r)^{2}} \frac{\partial A(r)^{2}}{\partial r} + \frac{2}{C(r)^{2}} \frac{\partial C(r)^{2}}{\partial r} \right) + \gamma^{\theta} \frac{1}{2 C(r)} \cot{\theta}\label{eqA6}.
\end{eqnarray}

Inserting (\ref{eqA6}) in Dirac equation and taking
$A_{\mu}=(A_{t},0,0,0)$ we obtain
\begin{eqnarray}
\fl \left\{ i\frac{\gamma^{t}}{A(r)} (\partial_{t}-iqA_{t}) + i\frac{\gamma^{r}}{B(r)} \left( \partial_{r} + \frac{1}{4 A(r)^{2}} \frac{\partial A(r)^{2}}{\partial r} + \frac{1}{2 C(r)^{2}} \frac{\partial C(r)^{2}}{\partial r} \right)  \right. \nonumber\\
\left. + i\frac{\gamma^{\theta}}{C(r)} \left( \partial_{\theta} + \frac{1}{2} \cot {\theta} \right) +i\frac{\gamma^{\varphi}}{C(r)\sin{\theta}} \partial_{\varphi} - m \right\} \Psi=0\label{eqA7}.
\end{eqnarray}

With definition of
$\Psi=A(r)^{-\frac{1}{2}}(\sin{\theta})^{-\frac{1}{2}}\Phi$
\cite{Chen:2007dja,R:2008ub} and substitution into (\ref{eqA7})
gives
\begin{eqnarray}
\fl \left\{ i\frac{\gamma^{t}}{A(r)} (\partial_{t}-iqA_{t}) + i\frac{\gamma^{r}}{B(r)} \left( \partial_{r} + \frac{1}{2 C(r)^{2}} \frac{\partial C(r)^{2}}{\partial r} \right) \right.\nonumber\\
\left.+ i\frac{\gamma^{\theta}}{C(r)} \partial_{\theta} + i\frac{\gamma^{\varphi}}{C(r)\sin{\theta}} \partial_{\varphi} - m \right\} \Phi=0.
\end{eqnarray}

In order to remove angular terms let us define the operator $K$
\begin{eqnarray}
K=\gamma^{t} \gamma^{r} \gamma^{\theta} \frac{\partial}{\partial \theta} + \gamma^{t} \gamma^{r} \gamma^{\varphi} \frac{1}{\sin{\theta}} \frac{\partial}{\partial \varphi}.
\end{eqnarray}
Hence we obtain
\begin{eqnarray}
\fl \left\{ i\frac{\gamma^{t}}{A(r)} (\partial_{t}-iqA_{t}) + i\frac{\gamma^{r}}{B(r)} \left( \partial_{r} + \frac{1}{C(r)} \frac{\partial C(r)}{\partial r} \right) - i\frac{\gamma^{r} \gamma^{t}}{C(r)} K - m \right\} \Phi=0.
\end{eqnarray}

Introducing the ansatz
\begin{eqnarray}
\Phi=
\left(
  \begin{array}{c}
    \frac{iG^{(\pm)}(r)}{C(r)} \phi^{(\pm)}_{jm}(\theta,\varphi) \\
    \frac{F^{(\pm)}(r)}{C(r)} \phi^{(\mp)}_{jm}(\theta,\varphi) \\
  \end{array}
\right) e^{-i\omega t},
\end{eqnarray}
with
\begin{eqnarray}
\phi^{+}_{jm}=
\left(
  \begin{array}{c}
    \sqrt{\frac{l+1/2+m}{2l+1}}Y_{l}^{m-1/2} \\
    \sqrt{\frac{l+1/2-m}{2l+1}}Y_{l}^{m+1/2} \\
  \end{array}
\right),
\end{eqnarray}

for $j=l+1/2$\\
and
\begin{eqnarray}
\phi^{-}_{jm}=
\left(
  \begin{array}{c}
    \sqrt{\frac{l+1/2-m}{2l+1}}Y_{l}^{m-1/2} \\
    -\sqrt{\frac{l+1/2+m}{2l+1}}Y_{l}^{m+1/2} \\
  \end{array}
\right).
\end{eqnarray}

for $j=l-1/2$

Eigenvalues of $K$ by applying on ansatz are obtained as follows
\begin{eqnarray}
\kappa_{(\pm)}=
\left\{
  \begin{array}{cc}
    (j+\frac{1}{2}) & j=l+\frac{1}{2}, \\
    -(j+\frac{1}{2}) & j=l-\frac{1}{2}. \\
  \end{array}
\right.
\end{eqnarray}
Here $\kappa_{\pm}$ is positive and negative
integers($\kappa_{\pm}=\pm 1,\pm 2, ...$).

With simplification we get
\begin{eqnarray}
\fl \left\{
  \begin{array}{c}
    \frac{1}{A(r)}(\omega+qA_{t})G^{(\pm)}(r)+\frac{1}{B(r)}\frac{\partial}{\partial r} F^{(\pm)}(r) + \frac{1}{C(r)} \kappa_{(\pm)} F^{(\pm)}(r)-mG^{(\pm)}(r)=0, \\
    -\frac{1}{A(r)}(\omega+qA_{t})F^{(\pm)}(r)+\frac{1}{B(r)}\frac{\partial}{\partial r} G^{(\pm)}(r) - \frac{1}{C(r)} \kappa_{(\pm)} G^{(\pm)}(r)-mF^{(\pm)}(r)=0. \\
  \end{array}
\right.
\end{eqnarray}

Introducing tortoise coordinate $r_{*} = \int{(B(r)/A(r))dr}$, the
radial equations for $F^{(\pm)}$ and $G^{(\pm)}$ become
\begin{eqnarray}
\fl \frac{d}{dr_{*}}
\left(
  \begin{array}{c}
    F^{(\pm)}(r) \\
    G^{(\pm)}(r) \\
  \end{array}
\right)
-A(r)
\left(
  \begin{array}{cc}
    -\frac{\kappa_{(\pm)}}{C(r)} & m \\
    m & \frac{\kappa_{(\pm)}}{C(r)} \\
  \end{array}
\right)
\left(
  \begin{array}{c}
    F^{(\pm)}(r) \\
    G^{(\pm)}(r) \\
  \end{array}
\right) \nonumber\\
=
\left(
  \begin{array}{cc}
    0 & -(\omega+qA_{t}) \\
    (\omega+qA_{t}) & 0 \\
  \end{array}
\right)
\left(
  \begin{array}{c}
    F^{(\pm)}(r) \\
    G^{(\pm)}(r) \\
  \end{array}
\right)\label{eqA16}.
\end{eqnarray}

Defining $\hat{F}^{(\pm)}$ and $\hat{G}^{(\pm)}$ by
\begin{eqnarray}
\left(
  \begin{array}{c}
    \hat{F}^{(\pm)} \\
    \hat{G}^{(\pm)} \\
  \end{array}
\right)
=
\left(
  \begin{array}{cc}
    \sin(\theta_{(\pm)}/2) & \cos(\theta_{(\pm)}/2) \\
    \cos(\theta_{(\pm)}/2) & -\sin(\theta_{(\pm)}/2) \\
  \end{array}
\right)
\left(
  \begin{array}{c}
    F^{(\pm)} \\
    G^{(\pm)} \\
  \end{array}
\right),\label{eqA17}
\end{eqnarray}
with
\begin{eqnarray}
\theta_{(\pm)}=\arccot{\left(\kappa_{(\pm)} \left/ mC(r) \right.\right)}, \ \ \ \ \ \ 0\leq\theta_{(\pm)}\leq\pi.
\end{eqnarray}

Applying (\ref{eqA17}) into (\ref{eqA16}) gives
\begin{eqnarray}
\frac{\partial}{\partial\hat{r}_{*}}
\left(
  \begin{array}{c}
    \hat{F}^{(\pm)} \\
    \hat{G}^{(\pm)} \\
  \end{array}
\right)
+W_{(\pm)}
\left(
  \begin{array}{c}
    -\hat{F}^{(\pm)} \\
    \hat{G}^{(\pm)} \\
  \end{array}
\right)
=
\omega
\left(
  \begin{array}{c}
    \hat{G}^{(\pm)} \\
    -\hat{F}^{(\pm)} \\
  \end{array}
\right)\label{eqA19},
\end{eqnarray}

Equation (\ref{eqA19}) can be easily separated. So with definition
of $V_{(\pm)1,2} = W_{(\pm)}^{2} \pm \left. \partial W_{(\pm)}
\right/ \partial \hat{r}_{*}$ and for simplification removing
$(\pm)$ we get
\begin{eqnarray}
    \frac{\partial^{2}}{\partial \hat{r}_{*}^{2}} \hat{F} + \left(\omega^{2} - V_{1} \right) \hat{F}=0, \\
    \frac{\partial^{2}}{\partial \hat{r}_{*}^{2}} \hat{F} + \left(\omega^{2} - V_{2} \right) \hat{G}=0,\label{eqA23}
\end{eqnarray}
where
\begin{eqnarray}
V_{1,2} = W^{2} \pm \frac{\partial W}{\partial \hat{r}_{*}},
\end{eqnarray}
and
\begin{eqnarray}
\fl W=  A(r) \left(m^{2} + \frac{\kappa^{2}}{C(r)^{2}}\right)^{\frac{1}{2}} \left(1+\frac{q}{\omega}A_{t}+\frac{1}{2}\frac{A(r)}{B(r)} \frac{\frac{m}{\omega} \kappa}{ \left(\kappa^{2}+m^{2}C(r)^{2}\right) } \frac{\partial C(r)}{\partial r}\right)^{-1},
\end{eqnarray}
In (\ref{eqA23}) following change of variable is applied
\begin{eqnarray}
\hat{r}_{*}= \int \frac{B(r)}{A(r)}\left(1+\frac{q}{\omega}A_{t}+\frac{1}{2} \frac{A(r)}{B(r)} \frac{\frac{m}{\omega} \kappa}{ \left(\kappa^{2}+m^{2}C(r)^{2}\right) } \frac{\partial C(r)}{\partial r}\right)dr.
\end{eqnarray}

This new coordinate is obtained from tortoise coordinate
$r_{*}=\int\left(B(r)/A(r)\right)dr$. Hence, we call this coordinate
\textit{generalized tortoise coordinate} change.

In studying the black holes with anisotropic metric factors like
Lifshitz, where we have anisotropic scale transformation,
understanding the effect of these factors is critically important,
so we calculated the effective potential for most general
spherically symmetric static family of black hole.

Considering the superpotential $W$ we see that the factor of metric
$A(r)$ can have crucial effect on potential with respect to the
factor $B(r)$.  Hence for the black holes which factor $A(r)$ has a
distinct behaviour we can expect different results.



\section*{References}
\begin{thebibliography}{99}

\bibitem{Gibbons:1987ps}
  G.~W.~Gibbons and K.~-i.~Maeda,
  Nucl.\ Phys.\ B {\bf 298} (1988) 741.

\bibitem{Garfinkle:1990qj}
  D.~Garfinkle, G.~T.~Horowitz and A.~Strominger,
  Phys.\ Rev.\ D {\bf 43} (1991) 3140
   [Erratum-ibid.\ D {\bf 45} (1992) 3888].

\bibitem{Holzhey:1991bx}
  C.~F.~E.~Holzhey and F.~Wilczek,
  Nucl.\ Phys.\ B {\bf 380} (1992) 447
  [hep-th/9202014].

\bibitem{Koga:1995bs}
  J.~-i.~Koga and K.~-i.~Maeda,
  Phys.\ Rev.\ D {\bf 52} (1995) 7066
  [hep-th/9508029].

\bibitem{Hawking:1974sw}
  S.~W.~Hawking,
  Commun.\ Math.\ Phys.\  {\bf 43} (1975) 199
   [Erratum-ibid.\  {\bf 46} (1976) 206].

\bibitem{Gibbons:1975}
  G.~W.~Gibbons,
  Commun.\ Math.\ Phys.\  {\bf 44} (1975) 245.

\bibitem{Bardeen:1973gs}
  J.~M.~Bardeen, B.~Carter and S.~W.~Hawking,
  Commun.\ Math.\ Phys.\  {\bf 31} (1973) 161.

\bibitem{Page:1977}
  D.~N.~Page,
  Phys.\ Rev.\ D {\bf 16} (1977) 2402.

\bibitem{Cvetic:1997ap}
  M.~Cvetic and F.~Larsen,
  Phys.\ Rev.\ D {\bf 57} (1998) 6297
  [hep-th/9712118].

\bibitem{Kanti:2002ge}
  P.~Kanti and J.~March-Russell,
  Phys.\ Rev.\ D {\bf 67} (2003) 104019
  [hep-ph/0212199].

\bibitem{Creek:2007tw}
  S.~Creek, O.~Efthimiou, P.~Kanti and K.~Tamvakis,
  Phys.\ Rev.\ D {\bf 76} (2007) 104013
  [arXiv:0707.1768 [hep-th]].

\bibitem{Das:1999pt}
  S.~Das and A.~Dasgupta,
  JHEP {\bf 9910} (1999) 025
  [hep-th/9907116].

\bibitem{Gubser:1997cm}
  S.~S.~Gubser,
  Phys.\ Rev.\ D {\bf 56} (1997) 7854
  [hep-th/9706100].

\bibitem{alBinni:2009cu}
  U.~A.~al-Binni and G.~Siopsis,
  Phys.\ Rev.\ D {\bf 79} (2009) 084041
  [arXiv:0902.2194 [hep-th]].

\bibitem{Sampaio:2009ra}
  M.~O.~P.~Sampaio,
  JHEP {\bf 0910} (2009) 008
  [arXiv:0907.5107 [hep-th]].

\bibitem{Casals:2006xp}
  M.~Casals, S.~R.~Dolan, P.~Kanti and E.~Winstanley,
  JHEP {\bf 0703} (2007) 019
  [hep-th/0608193].

\bibitem{Sampaio:2009tp}
  M.~O.~P.~Sampaio,
  JHEP {\bf 1002} (2010) 042
  [arXiv:0911.0688 [hep-th]].

\bibitem{Gibbons:2008rs}
  G.~W.~Gibbons and M.~Rogatko,
  Phys.\ Rev.\ D {\bf 77} (2008) 044034
  [arXiv:0801.3130 [hep-th]].

\bibitem{Nakonieczny:2011bs}
  L.~Nakonieczny and M.~Rogatko,
  Phys.\ Rev.\ D {\bf 84} (2011) 044029
  [arXiv:1108.3892 [hep-th]].

\bibitem{Chen:2007dja}
  S.~Chen, B.~Wang and R.~Su,
  Class.\ Quant.\ Grav.\  {\bf 23} (2006) 7581
  [gr-qc/0701089].

\bibitem{Cho:2003qe}
  H.~T.~Cho,
  Phys.\ Rev.\ D {\bf 68} (2003) 024003
  [gr-qc/0303078].

\bibitem{Cho:2004wj}
  H.~T.~Cho and Y.~-C.~Lin,
  Class.\ Quant.\ Grav.\  {\bf 22} (2005) 775
  [gr-qc/0411090].

\bibitem{Cho:2007zi}
  H.~T.~Cho, A.~S.~Cornell, J.~Doukas and W.~Naylor,
  Phys.\ Rev.\ D {\bf 75} (2007) 104005
  [hep-th/0701193].

\bibitem{Chakrabarti:2008xz}
  S.~K.~Chakrabarti,
  Eur.\ Phys.\ J.\ C {\bf 61} (2009) 477
  [arXiv:0809.1004 [gr-qc]].

\bibitem{Doran:2005vm}
  C.~Doran, A.~Lasenby, S.~Dolan and I.~Hinder,
  Phys.\ Rev.\ D {\bf 71} (2005) 124020
  [gr-qc/0503019].

\bibitem{Dolan:2006vj}
  S.~Dolan, C.~Doran and A.~Lasenby,
  Phys.\ Rev.\ D {\bf 74} (2006) 064005
  [gr-qc/0605031].

\bibitem{Jin:1998rg}
  W.~M.~Jin,
  Class.\ Quant.\ Grav.\  {\bf 15} (1998) 3163
  [gr-qc/0009009].

\bibitem{Jing:2003wq}
  J.~-l.~Jing,
  Phys.\ Rev.\ D {\bf 69} (2004) 084009
  [gr-qc/0312079].

\bibitem{Jing:2004xv}
  J.~-l.~Jing,
  Phys.\ Rev.\ D {\bf 70} (2004) 065004
  [gr-qc/0405122].

\bibitem{Jing:2004zb}
  J.~Jing,
  Phys.\ Rev.\ D {\bf 72} (2005) 027501
  [gr-qc/0408090].

\bibitem{Jung:2004nh}
  E.~Jung, S.~Kim and D.~K.~Park,
  JHEP {\bf 0409} (2004) 005
  [hep-th/0406117].

\bibitem{LopezOrtega:2005ep}
  A.~Lopez-Ortega,
  Gen.\ Rel.\ Grav.\  {\bf 37} (2005) 167.

\bibitem{LopezOrtega:2010uu}
  A.~Lopez-Ortega,
  Rev.\ Mex.\ Fis.\  {\bf 56} (2010) 44
  [arXiv:1006.4906 [gr-qc]].

\bibitem{LopezOrtega:2011sc}
  A.~Lopez-Ortega and I.~Vega-Acevedo,
  Gen.\ Rel.\ Grav.\  {\bf 43} (2011) 2631
  [arXiv:1105.2802 [gr-qc]].

\bibitem{Moderski:2008nq}
  R.~Moderski and M.~Rogatko,
  Phys.\ Rev.\ D {\bf 77} (2008) 124007
  [arXiv:0805.0665 [hep-th]].

\bibitem{R:2008ub}
  R.~Sini, N.~Varghese and V.~C.~Kuriakose,
  arXiv:0802.0788 [gr-qc].

\bibitem{Shu:2004fj}
  F.~-W.~Shu and Y.~-G.~Shen,
  Phys.\ Rev.\ D {\bf 70} (2004) 084046
  [gr-qc/0410108].

\bibitem{Wang:2009hr}
  C.~-Y.~Wang, Y.~Zhang, Y.~-X.~Gui and J.~-B.~Lu,
  arXiv:0910.5128 [gr-qc].

\bibitem{Eckart:1930zza}
  C.~Eckart,
  Phys.\ Rev.\  {\bf 35} (1930) 1303.

\bibitem{Rosen:1932}
  N.~Rosen and Philip M.~Morse,
  Phys.\ Rev.\  {\bf 42} (1930) 210.

\bibitem{Boonserm:2011}
  P.~Boonserm and M.~Visser,
  JHEP {\bf 03} (2011) 073
  [arXiv:1005.4483 [hep-th]].

\bibitem{Preskill:1991tb}
  J.~Preskill, P.~Schwarz, A.~D.~Shapere, S.~Trivedi and F.~Wilczek,
  Mod.\ Phys.\ Lett.\ A {\bf 6} (1991) 2353.

\bibitem{Koga:1994np}
  J.~-I.~Koga and K.~-I.~Maeda,
  Phys.\ Lett.\ B {\bf 340} (1994) 29
  [hep-th/9408084].

\bibitem{Maldacena:1996ix}
  J.~M.~Maldacena and A.~Strominger,
  Phys.\ Rev.\ D {\bf 55} (1997) 861
  [hep-th/9609026].

\bibitem{Harmark:2007jy}
  T.~Harmark, J.~Natario and R.~Schiappa,
  Adv.\ Theor.\ Math.\ Phys.\  {\bf 14} (2010) 727
  [arXiv:0708.0017 [hep-th]].

\bibitem{Kraus:1995vr}
  P.~Kraus,
  gr-qc/9508007.

\bibitem{Chen:2005rm}
  S.~-B.~Chen and J.~-L.~Jing,
  Class.\ Quant.\ Grav.\  {\bf 22} (2005) 1129.

\bibitem{Cho:2011sf}
  H.~T.~Cho, A.~S.~Cornell, J.~Doukas, T.~R.~Huang and W.~Naylor,
  Adv.\ Math.\ Phys.\  {\bf 2012} (2012) 281705
  [arXiv:1111.5024 [gr-qc]].

\bibitem{Cardoso:2003sw}
  V.~Cardoso and J.~P.~S.~Lemos,
  Phys.\ Rev.\ D {\bf 67} (2003) 084020
  [gr-qc/0301078].

\bibitem{Konoplya:2011qq}
  R.~A.~Konoplya and A.~Zhidenko,
  Rev.\ Mod.\ Phys.\  {\bf 83} (2011) 793
  [arXiv:1102.4014 [gr-qc]].

\bibitem{Nagar:2006eu}
  A.~Nagar, O.~Zanotti, J.~A.~Font and L.~Rezzolla,
  Phys.\ Rev.\ D {\bf 75} (2007) 044016
  [gr-qc/0610131].

\end{thebibliography}
\end{document}